\documentclass{aa}
\pdfoutput=1
\usepackage[varg]{txfonts}
\usepackage[utf8]{inputenc}
\usepackage[english]{babel}
\usepackage{amsmath}
\usepackage{amsfonts}
\usepackage{amssymb}
\usepackage{graphicx}
\graphicspath{{./pictures/}}
\usepackage{subcaption}
\usepackage{textcomp} 

\usepackage{natbib}
\bibpunct{(}{)}{;}{a}{}{,} 
\usepackage{float}
\usepackage{hyperref} 
\usepackage{pdfpages}
\usepackage{siunitx} 
\usepackage{placeins}
\usepackage{afterpage}
\usepackage{rotating} 
\usepackage{ulem} 
\usepackage{xcolor}

\usepackage{array}
\newcolumntype{M}[1]{>{\centering\arraybackslash}m{#1}}
\usepackage{multirow}
\usepackage{amstext}

\begin{document}

\title{ The dipper population of Taurus seen with \textit{K2}}

\author{Noemi Roggero\inst{1}
  \and J\'er\^ome Bouvier \inst{1}
  \and Luisa M. Rebull \inst{2}
  \and Ann Marie Cody \inst{3}
}

\institute{Univ. Grenoble Alpes, CNRS, IPAG, 38000 Grenoble, France \email{noemi.roggero@univ-grenoble-alpes.fr}
 \and Infrared Science Archive (IRSA), IPAC, 1200 E.\
California Blvd., California Institute of Technology, Pasadena, CA
91125, USA
 \and SETI Institute, 189 N Bernardo Ave. Suite 200, Mountain View, CA 94043 USA
}

\abstract {Dippers are typically low-mass, pre-main-sequence stars that display dips in their light curves. These dips have been attributed to dusty warps that form in the inner part of the disk.} {Our goal is to derive the characteristics of dipper stars in Taurus to assess the physical mechanisms that induce dipper light curves.} {We used the light curves of the fourth and thirteenth campaigns of \textit{K2} to select a dipper sample among 179 members and possible members of the Taurus star-forming region based on the light-curve morphology. We studied the periodicities by combining periodograms with wavelet analysis and  derived the stellar parameters of the sample from the photometry. We also studied the morphology of the photometric dips.}{We find a dipper occurrence of $\sim$30\% in disk-bearing stars observed with \textit{K2} that were identified visually by us. This represents a lower limit to their true occurrence, on the one hand because they are ephemeral, and on the other because there are detection limits. About half of the dippers are aperiodic, and most of these are dominated by another type of variability. The chosen sample is of late spectral type (K/M), low mass (mostly $<1\,\mathrm{M_{\odot}}$), and moderate mass accretion rates and has periods of a few days. We observed a transient dipper over a few rotation cycles and observed a dipper with a changing period. The structure of the dips can be complex and varies strongly over timescales of down to one stellar rotation. The corotation radii are located at a few stellar radii, and the temperatures at corotation allow dust survival. Many of the systems are seen at moderate to high inclination. We find that the angular extension of the dusty structure producing the dips is correlated with the stellar period.}{Magnetospheric accretion, which causes an accretion column and its base to occult the star, can explain most of the observed light curves. Although compatible with the model, many of the stellar inclination angles are moderate and do not exclude mechanisms other than the occultation by an inner disk warp to account for dipper light curves.}

\keywords{protoplanetary disks -- stars: pre-main sequence -- stars: variables: T Tauri - accretion, accretion disks -- techniques: photometric}
\maketitle

\section{Introduction}

\label{sec:intro}
Classical T Tauri stars (CTTSs) are still accreting from their circumstellar disk, and their activity is reflected in their complex photometric variability. Many physical mechanisms have been proposed to explain this variability, such as accretion hotspots, accretion bursts, or occultations by dusty structures in the disk  \citep[e.g.,][]{Cody14,Alencar10,Bouvier03}. At the next stage of protostellar evolution, weak-lined T Tauri stars (WTTSs) do not show accretion signatures in their spectral lines. Their magnetic field still produces cold spots, which give a sinusoidal modulation to the light curve.

In the past two decades, more attention has been paid to the low-mass star AA Tau.  Its light curve showed a constant brightness and narrow quasiperiodic occultations that were less pronounced in the infrared (IR) than in the optical. These observations, combined with spectroscopic and spectropolarimetric data, led to the development of an occultation model. The magnetospheric accretion column would cause a dusty warp in the inner disk to obscure the star when it crosses the line of sight of the observer \citep{Bouvier99,Bouvier03,Bouvier07}. 

The possibility that CTTSs are still accreting from a circumstellar disk was proposed by \cite{Bertout88} and had its roots in the work of \cite{LyndenBell74}. Although these authors already suggested that accretion might take place along magnetic field lines, the exact role played by the magnetic field in the accretion process was constrained only later. A rotating magnetosphere of a star with a strong magnetic field (approximately of some kG) disrupts the inner disk at a few stellar radii from the star and accretes the material along the closed field lines; the open field lines induce a strong disk wind \citep{Camenzind90,Koenigl91,Shu94}. The most common magnetic field configuration for T Tauri stars is a strong dipole. Depending on the evolutionary status, other magnetic field topologies are possible, but the dipole in most cases remains the dominant component farther away from the star. Moreover, a misalignment of the dipole axis and the stellar rotation axis seems to be very common \citep{Gregory12}. When there is such a misalignment, the rotation of the disk provokes a twist in the magnetic field lines, which are dragged in the rotation. The magnetic pressure is not the same in the upper and lower part of the disk, and the normal vector of the inner disk becomes tilted with respect to the rotation axis. This induces a disk warp \citep{Lai99,Terquem2000}. From the observational point of view, the strong emission lines observed in the spectra of T Tauri stars are related to accretion and not to winds \citep{Hartmann94}. Radiative transfer modeling provided observational predictions for magnetospheric accretion \citep{Muzerolle01,Kurosawa06,Lima10}.

Magnetohydrodynamical (MHD) simulations confirmed the possibility that magnetospheric accretion might warp the disk \citep[e.g.,][]{Romanova08,Romanova13,Romanova15}. In a stable configuration (stable accretion regime), two broad,  stable, and symmetric matter streams form from the disk and are accreted as funnel flows onto the stellar surface,  producing  a  hotspot  on  each  hemisphere. In this scenario, the magnetosphere truncates the disk at a few stellar radii close to corotation, where the disk material rotates with $\omega \sim \Omega_*$. The base of the accretion column can host warm dust and be optically thick, which explains the photometric behavior of AA Tau.

After the the characterization of AA Tau, additional surveys identified a new class of AA Tau-like YSOs (young stellar objects), the so-called dippers. Dippers have been observed in NGC 2264, Upper Scorpius, $\rho$ Ophiucus, and Orion \citep[e.g.,][]{Alencar10,Cody14,McGinnis15,Ansdell16a,Cody18,Morales11}. The occurrence rate of dippers is estimated to be up to 20\%-30\% of CTTSs. As in the case of AA Tau, the light curve exhibits a brightness continuum interrupted by narrow flux dips, which can be aperiodic or quasiperiodic. Quasiperiodic means that while the occurrence of the dips is periodic, their shape and amplitude are not constant over time. The dips are sharp and irregular, their periodicity is in the range of stellar rotation periods of K/M stars, and they last from 1-2 to  4-5\,d. The amplitudes can range to up to $50-60\%$ in flux. 

 The common consensus about the origin of dippers is that the occultation is produced by dusty structures in the inner disk. The correlation between the dip depth and the mid-infrared excess at 4.6\,$\mu$m supports this hypothesis \citep{Ansdell16a} as IR excess at this wavelength is a tracer for warm dust at the corotation radius. This makes dippers a powerful tool for studying this region, which is difficult to resolve, yet physically complex. The dusty disk warp model \citep{Bouvier07} requires dipper stars to be seen at high inclination, so that the line of sight crosses the base of the accretion column. The generalized magnetospheric accretion model \citep{Bodman17} allows also for moderate inclinations. The amplitude and the shape of the dips depends on parameters such as viewing angle, dust opacity, and the tilt between magnetic field and stellar rotation axis. Another possibility are vertical instabilities in the disk, which should be able to endure several rotation cycles. A star with an aligned magnetic field rather accretes via Rayleigh-Tailor (RT) instabilities \citep[e.g.,][]{Romanova08} in an unstable accretion regime. The so-caused accretion  tongues\ reach the stellar surface at different latitudes and induce a stochastic photometric variability. \cite{McGinnis15} claimed that this mechanism proably lies at the origin of aperiodic dippers, which appear as stochastic occultations of the stellar photosphere, with smaller amplitudes than periodic dippers.
 
However, the high occurrence of dippers among YSOs is difficult to reconcile with the condition of a grazing viewing angle on these systems. It was found that the outer disk inclination of dippers can be low or even close to face-on \citep{Ansdell20}. This either corroborates the possibility of an inner disk that is tilted with respect to the outer disk \citep[e.g.,][]{Alencar18} or requires other mechanisms that are compatible with lower inclination, such as disk winds \citep[e.g.,][]{Bans12}.
 
 The Taurus star-forming region hosts a quite young stellar population $\lesssim 3$\,Myr \citep[e.g.,][]{White01,Kraus09} of $\sim$400 members \citep{Kenyon08} at a distance of $\sim 140$\,pc \citep{Fleming19,Galli18}. \cite{Rebull20} presented the global sample of the Taurus region as covered by the \textit{K2} C13, with a few additions from C4. This work focuses on the dipper population of the same sample. In Sec.~\ref{sec:res} the selection of the dipper sample, the periodicity study, and the stellar parameters are presented. In Sec.~\ref{sec:discussion} the results and the possible scenarios to explain dippers are discussed. Final considerations and future work are presented in Sec.~\ref{sec:conclusion}.

\section{Observations and data reduction }
\label{sec:obs}

The \textit{Kepler} satellite \citep{Borucki2010,Haas2010} was launched in 2009 with the main aim of detecting Earth-like exoplanets. It observed over 170,000 targets simultaneously in its long-cadence mode of 29.4 min. The failure of two reaction wheels led to a substantial change in observing strategy and to a renaming of the mission as \textit{K2} \citep{Howell2014}. The spacecraft used its two remaining reaction wheels in tandem with the solar radiation pressure (plus compensatory periodic thruster firings) to control movement in the third dimension. Instead of staring at one field, as the original \textit{Kepler} mission had done, \textit{K2} was constrained to view fields in the ecliptic plane for about 70 days at a time \citep{Howell2014}.  Within the framework of the \textit{K2} mission (2014 - 2018), 19 fields were observed \citep{Howell2014}. Several discussions of data and reductions of \textit{Kepler} and \textit{K2} data are available. In the context of this work, the two most important artifacts to keep in mind are the periodic thruster firings and the relatively low spatial resolution (pixels of $\sim 4 \times 4 \arcsec$ and the 95\% encircled energy diameter of up to 7.5 px). The thruster firings happened every 0.245 days, and so apparent astrophysical periods near this must be scrutinized carefully to ensure that they are not affected by the spacecraft.  The low spatial resolution means that source confusion can be an issue, and each source (and light curve) must be inspected individually to assess confusion. Customized apertures are sometimes needed to attempt to mitigate this. The effective observed range of \textit{K2} lies between 6\,mag$<K_s<$16\,mag \citep[e.g.,][]{Rebull16a}. Saturation is not a relevant detection bias in Taurus, which does not host a massive star population. However, deeply embedded low-mass stars are likely to remain undetected with \textit{K2}.

A few Taurus members were observed in \textit{K2} C4 (15 February - 24 April 2015), but most of the Taurus members were observed during C13 from 8 March to 27 May 2017\footnote{\url{https://archive.stsci.edu/files/live/sites/mast/files/home/missions-and-data/k2/_documents/KSCI-19116-003.pdf}}. We are building on the analysis presented in \cite{Rebull20}. The \textit{K2} fields do not cover the entire Taurus cluster, but the stellar properties of the cluster are not thought to vary based on the spatial distribution \citep[see, e.g., Fig.~1 in][]{Rebull20}.  We used light curves from several different data reduction pipelines \citep[also see][]{Rebull2017}. First, the pre-search data conditioning (PDC) version from the Mikulski Archive for Space Telescopes (MAST). Second, a position-detrended version with moving apertures as in \cite{Cody18}. Third, the \lq self-flat-fielding\rq\ approach by \cite{Vanderburg2014} and the K2SFF pipeline as obtained from MAST. Last, the EVEREST2 pipeline, based on pixel-level decorrelation \cite{Luger2016}, as downloaded from MAST.
\begin{figure*}
    \centering
    \includegraphics[width=\linewidth]{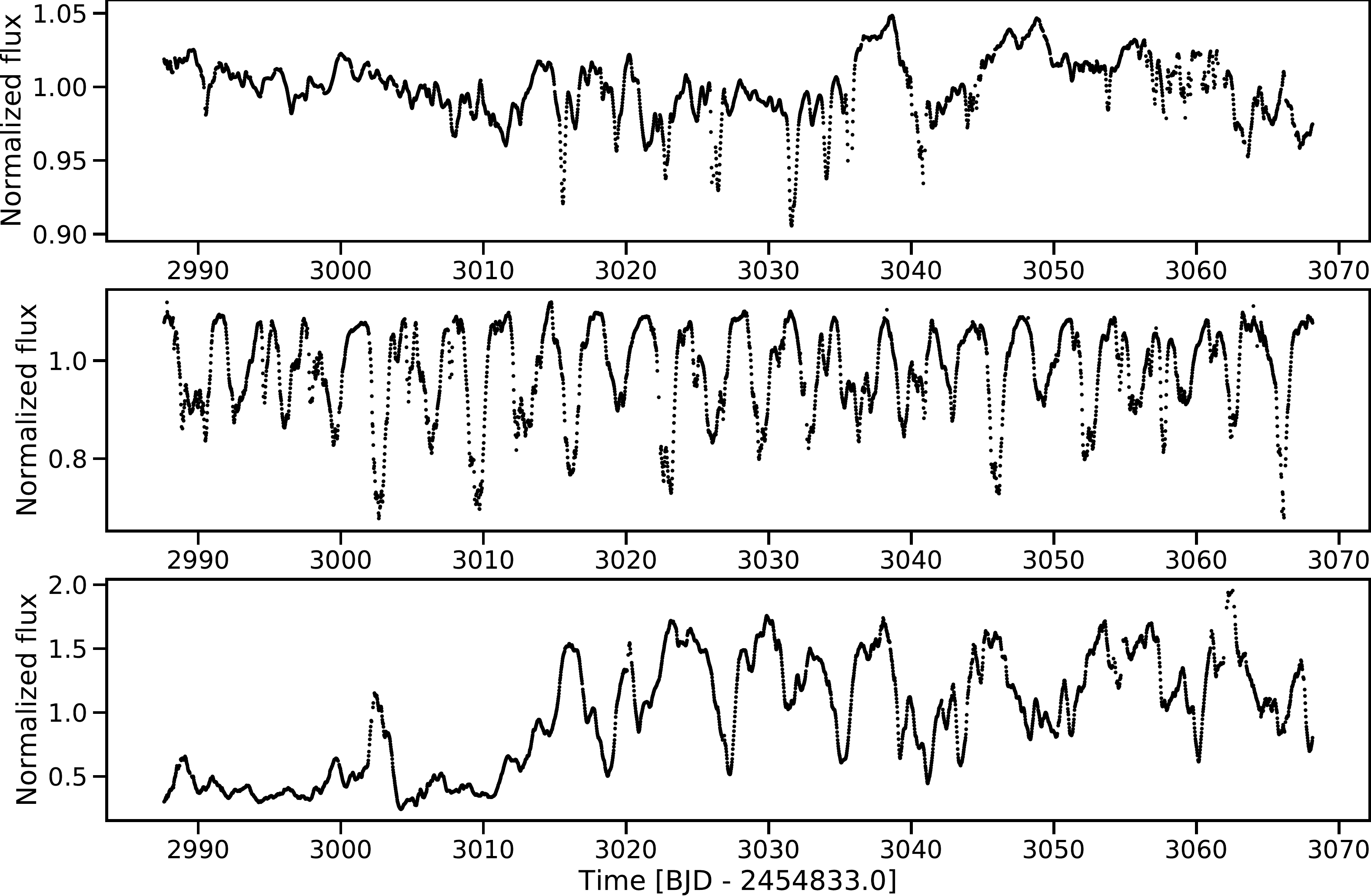}
    \caption{From top to bottom: example of an aperiodic dipper (HD 285893); a quasiperiodic dipper (JH 223); a dipper with a complex light curve (DK Tau). Dipper light curves are characterized by irregularly shaped fading events, which can last to up to a few days.}
    \label{fig:example_lcs_dippers}
\end{figure*}

\begin{figure*}
    \centering
    \includegraphics[width=\linewidth]{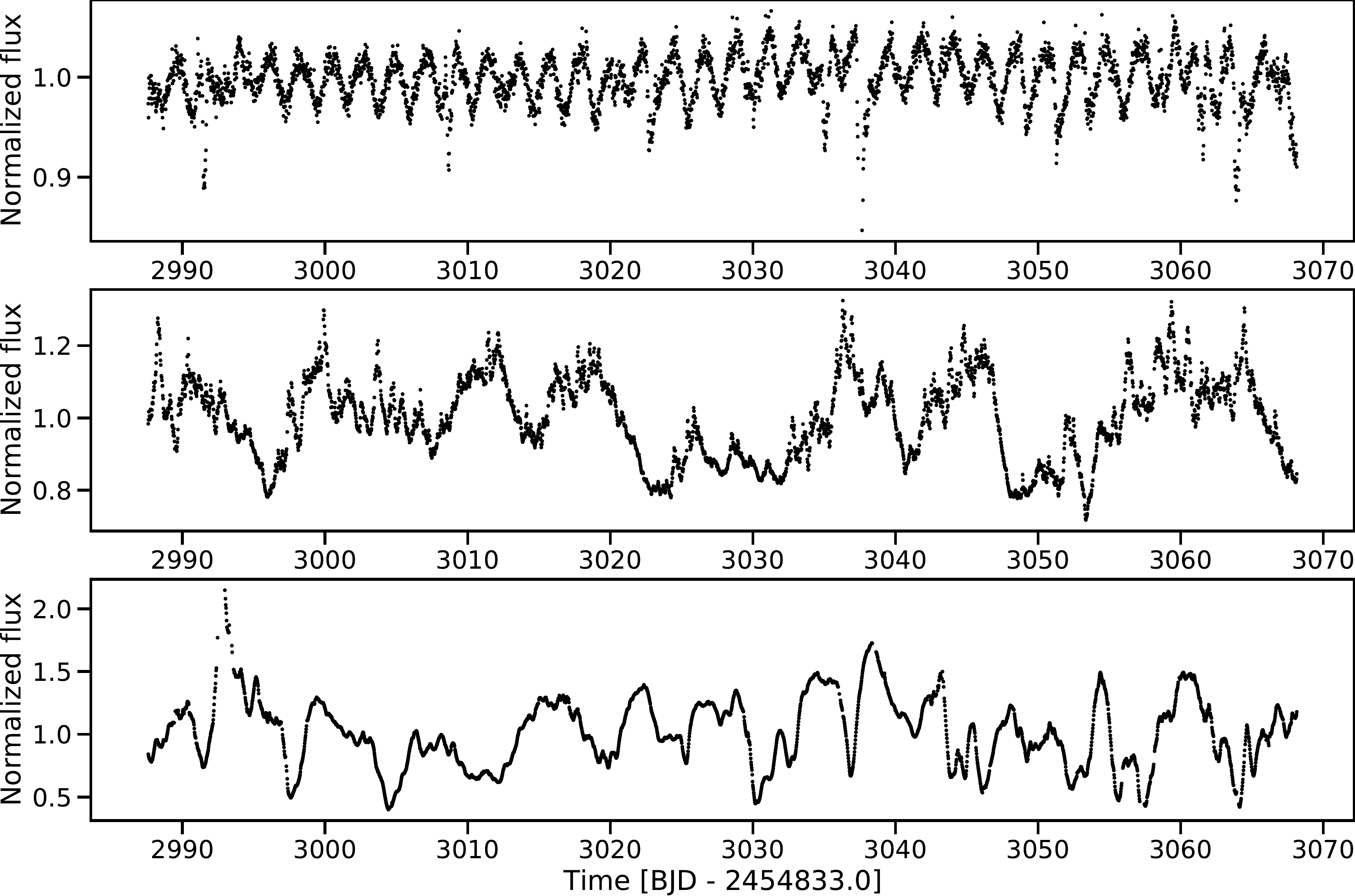}
    \caption{From top to bottom: a spot-dominated light curve with aperiodic dips (EPIC246859790); a burst-dominated light curve with dips (CI Tau); a low-quality dipper candidate (IQ Tau). Although dips as described in Fig.~\ref{fig:example_lcs_dippers} are present, the light curve is dominated by another type of variability.}
    \label{fig:example_lcs_other_dippers}
\end{figure*}
A comparison of all these light-curve versions allows us to constrain the periodicity (or periodicities) and select the  best available periodicity\ \citep[see][and references therein]{Rebull20} to represent that star's light curve. If the light curve does not present particular issues, the version with moving aperture as in \cite{Cody18} was used by default for consistency. This version was also used for the dipper sample presented in this paper.  

As described in \cite{Rebull20}, we started from a more expansive definition of possible Taurus members with \textit{K2} light curves, and then weeded it down to 156 members and 23 possible members. The highest-confidence members are those listed in \cite{Luhman2018}. We searched for dippers in this set of 156+23 Taurus member light curves.
\section{Light-curve analysis}
\label{sec:res}
\subsection{Identification of dippers}
\label{subsec:res:identif}
We identified dippers based on a visual inspection of the light curve (see lists in Table~\ref{tab:SpT_photometry} and Table~\ref{tab:other_dippers}). A common feature in YSOs are sinusoidal patterns, which can be ascribed to cold spots on the stellar photosphere \citep[e.g.,][]{Herbst94}. Irregular, aperiodic increases in the stellar brightness (bursts) are linked to accretion shocks for T Tauri stars \citep{Stauffer14}. Planetary transits and eclipsing binaries do produce dips, but these are strictly periodic and in the first case, have a small amplitude in flux. The criteria for selecting dippers are thus the irregular and sharp shape of the dips, the duration of the dips of up to a few days, and an amplitude of at least $\sim$10\% to $\sim$60\% of the flux. A few examples of dipper light curves are displayed in Fig.~\ref{fig:example_lcs_dippers}. The light curves easiest to classify are those in which a stable brightness continuum is interrupted by irregularly shaped dips (e.g., JH\,223 in Fig.~\ref{fig:example_lcs_dippers}). The bright-faint flux asymmetry is thus high \citep[see, e.g., the M metric in][]{Cody14}. The M metric is a useful tool to investigate a large sample of stars. However, it tends to classify as quasiperiodic symmetric (QPS) light curves with irregular dips that do not have a constant brightness continuum. We consider that from a physical point of view, this should not be a striking selection criterion and therefore preferred a manual classification. Dippers with an irregular brightness continuum are CFHT Tau 12 and ITG 34. More symmetric are HK Tau, GH Tau, GM Tau, and EPIC 247820821 (2MASS J04295950+2433078). The last two are more ambiguous, and we differentiated them from spots because of the sharp shape of their dips. We cannot exclude that the dips are superimposed on spots. GI Tau is contaminated by the neighbor GK Tau, and its light curve is complex. However, its own periodicity can be retrieved, and clear dips are visible in the light curve. DK Tau also exhibits a complex light curve. After $\sim$25\,d of quiescent state, the star becomes a clear dipper. All dipper light curves are shown in Fig.~\ref{fig:app:LCs_dippers}.

In addition to the main dipper sample, two other groups of light curves were taken into account: light curves that show a predominant behavior (e.g., spots, bursts, in Table~\ref{tab:other_dippers}) but also some dips, and low-quality dipper candidates that did not fulfill all of the criteria of the visual inspection and/or were too complex for discerning the different types of their variability (Table~\ref{tab:not_dippers}). In Fig.~\ref{fig:example_lcs_other_dippers} a few example light curves are shown.

As a result, 22 objects were classified as dippers and 12 were classified as dippers dominated by another type of variability. Almost all of the latter exhibit some aperiodic dips superimposed on an otherwise variable light curve (Fig.~\ref{fig:app:LCs_otherdippers}). Eleven additional stars are presented as low-quality candidates in Table~\ref{tab:not_dippers} and Fig.~\ref{fig:app:LCs_candidatedippers} and are not further considered in this study. They are rather listed as visual examples of the selection process. The final sample of 34 dippers represents 19\% of high-confidence and possible Taurus members \citep[156 + 23 as in][]{Rebull20} and 31\% of the disked stars \citep[94 + 7 as in][]{Rebull20} in this sample.

In addition to the dipper prototype AA Tau (which is now in a faint state and appears in Table~\ref{tab:other_dippers}), \cite{Rodriguez17} identified five dippers in Taurus-Auriga with \textit{KELT} light curves, four of which are confirmed in this sample. \cite{Rebull20} counted 21 dippers in this sample. Sixteen of the dippers analyzed in this study are present in \cite{Rebull20}. Four out of the 5 further dippers in \cite{Rebull20} appear here in Table~\ref{tab:other_dippers}. One further light curve (EPIC 247078342) indicated as a dipper in \cite{Rebull20} is discarded in this study because its shape strongly depends on the reduction pipeline. Our classification is therefore not as strict as in \cite{Rebull20}. Further analysis of the Taurus members sample will be presented in Cody \& Hillenbrand (in prep.). As mentioned above, the usage of the M metric or a manual classification can affect the classification for certain stars.
\begin{figure*}
\centering
\includegraphics[width=0.48\linewidth]{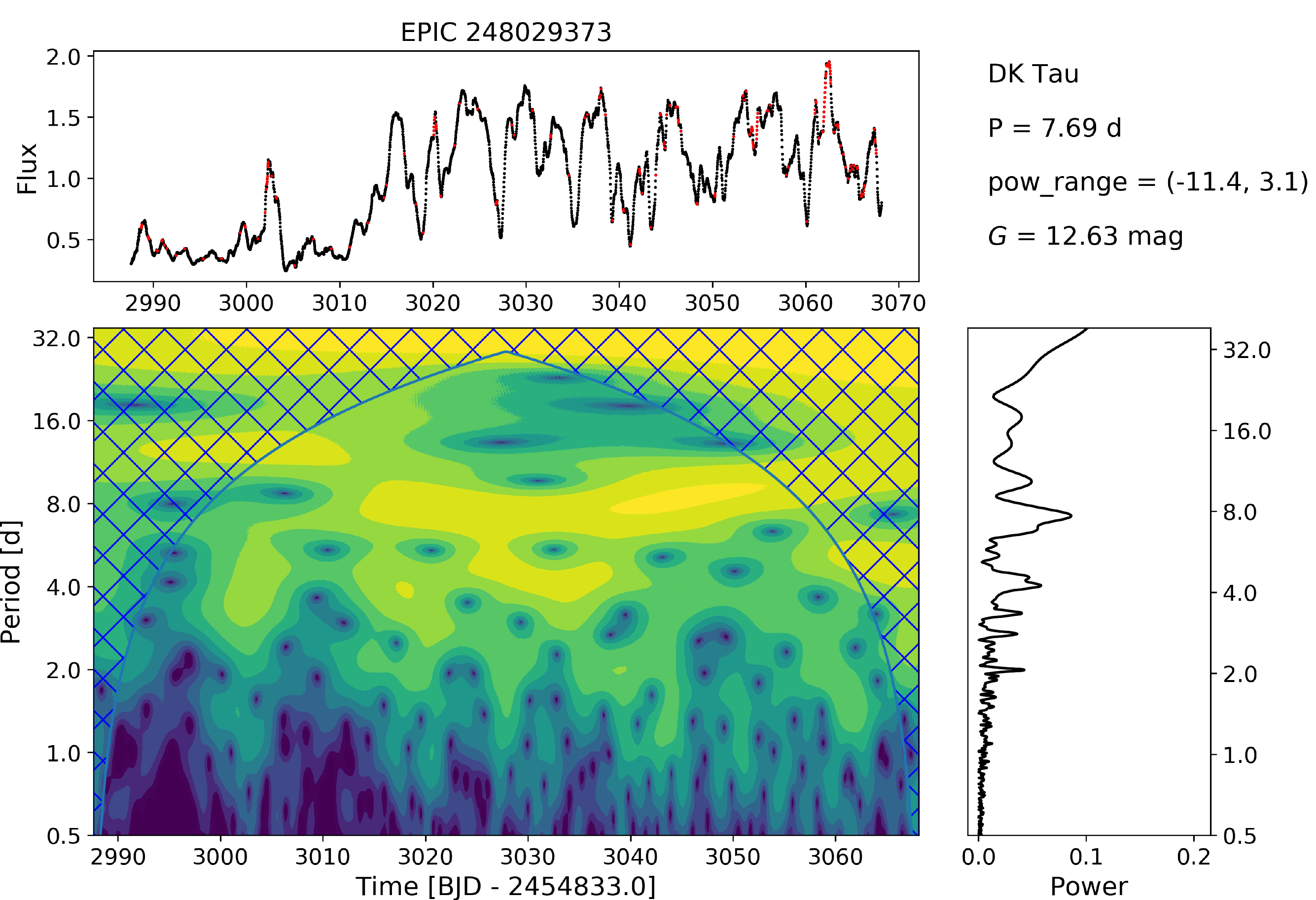}
\includegraphics[width=0.48\linewidth]{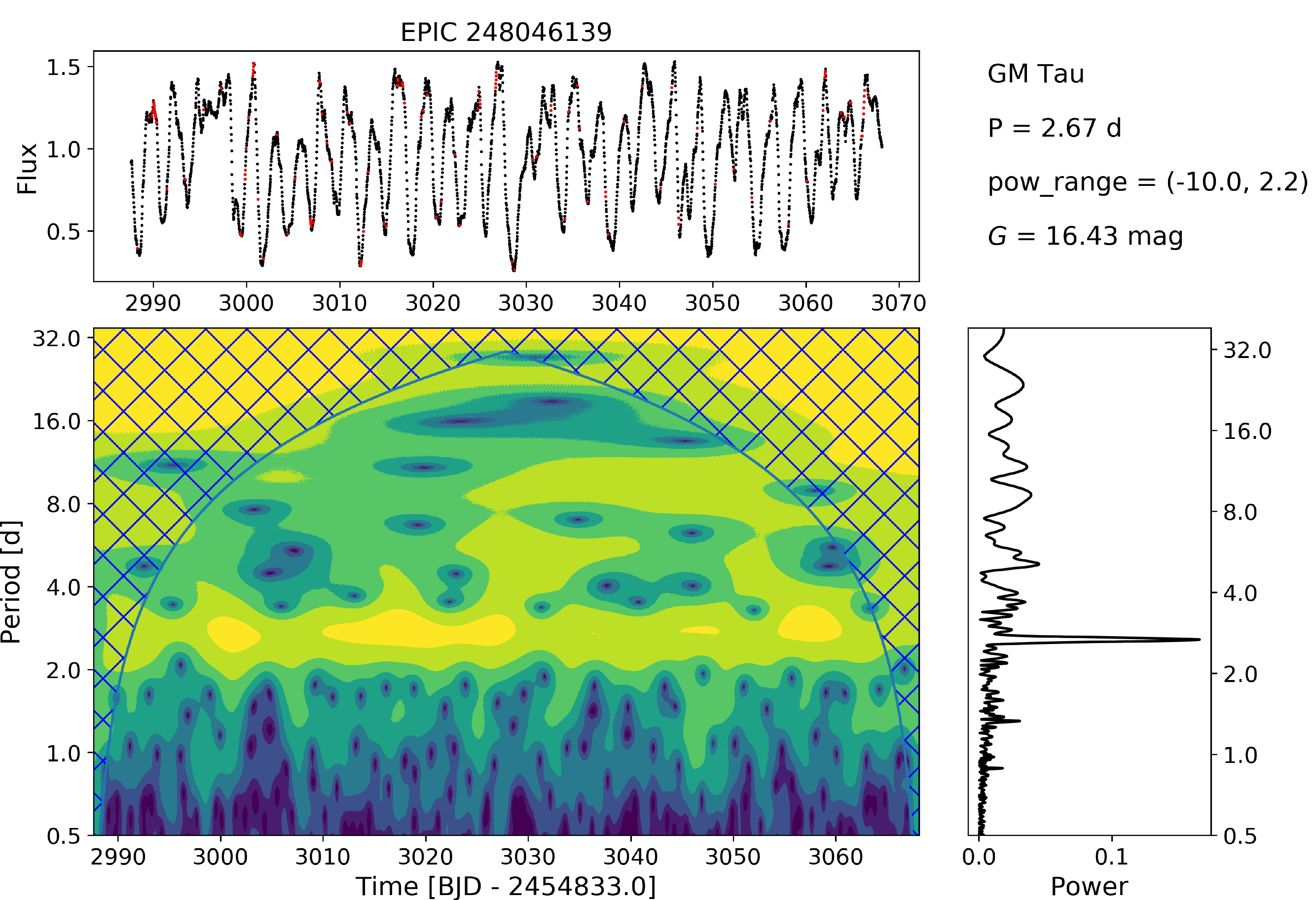}
\caption{Different examples of WPS for a time-resolved changing period (DK Tau) and a periodic dipper (GM Tau). Top panel: Light curves with interpolated points marked in red. Left panel: 2D WPS. The crossed lines mark the COI, where edge effects of the wavelet transform become relevant. The power contours extend from low (blue) to high (yellow). The power range and the \textit{Gaia} magnitude are annotated in the upper right corner. Right panel: CLEAN periodogram. The logarithmic y-axis with the periods is the same as for the WPS. In the case of DK Tau (left), the two peaks in the periodogram at $\sim$8 and $\sim$10\,d can be interpreted as a single period that changes during the observations by means of the WPS. GM Tau displays only one clear period in the WPS and the periodogram. The green stripe at $\sim$10\,d is just a recurring pattern in the light curve: a deeper dip at $t = 3030, 3040, 3050$\,d. This is not a relevant periodicity and does not appear in the periodogram.}
\label{fig:ex_WPS}
\end{figure*}

\subsection{Wavelet analysis of dippers}
\label{subsec:res:wavelets}

Dippers can be quasiperiodic or aperiodic, their shape varying from cycle to cycle. Aperiodic narrow dips might appear in an otherwise periodic light curve, or a quasiperiodic dipper might be a transient phenomenon \citep[e.g.,][]{McGinnis15}. Because the variability is irregular, the use of time-resolved period-search algorithms is of high interest.

Common tools used in astronomy such as the Lomb-Scargle periodogram \citep{Lomb,Scargle} and Fourier analysis are not able to deliver information about transient phenomena because they merely resolve the time series in the frequency domain. The windowed Fourier transform (WFT) convolves the signal most commonly with a Gaussian window that is shifted along the signal before the power spectrum is computed. The resulting spectrograms to some extent allow retrieving a time resolution and were applied to dippers by \cite{Bodman17}. However, the width of the window and the time shift cannot be optimized for the entire frequency range, resulting in poor resolution.

When transient phenomena are to be characterized, the wavelet analysis represents a valid alternative because this method allows keeping track of the time variability of a periodic feature by partly losing frequency resolution \citep[see for a summary][]{TorrenceCompo}. It can be conceived as a time-resolved WFT with a variable window width that allows recognizing both high- and low-frequency features. A wavelet transform produces a power spectrum after the convolution of the signal and the wavelet, which in most cases is a sort of time-confined sinusoidal wave. We used here the complex Morlet wavelet, which is a complex exponential multiplied by a Gaussian envelope \citep{Grossmann84}. The time and frequency resolution are reached by shifting the wavelet along the time series, then stretching or compressing the wavelet, and repeating the procedure again. This delivers a two-dimensional wavelet power spectrum (WPS), which has a power value for each combination of time and frequency (for more information, see Appendix \ref{sec:app}). 

Wavelet analysis has been successfully applied to a number of phenomena in $K2$ light curves: stellar rotation \citep{Mathur14,Garcia14,Bravo14}, planetary transits, stellar variability, pulsation, and binaries \citep{Bravo14}.
The implementation in \texttt{python} runs with the package \texttt{pywt}. 

Because the time series must be evenly sampled for wavelet analysis and some values in the light curve are flagged out during the data reduction, the light curves are linearly interpolated onto even time steps in a preliminary step.

An example of WPS is shown in Fig.~\ref{fig:ex_WPS}, while the full atlas of the sample is presented in Appendix~\ref{sec:app:wps}. The \textit{x}- and \textit{y}-axis present the linear time and the logarithmic periods (equivalent to the scales, as explained in Appendix \ref{sec:app:morlet}). The contours correspond to the power of the wavelet spectrum, and the region omitted from the cone of influence (COI) is cross-hatched. The contours reach from blue (low power) to yellow (high power). To increase the readability of the plots, the contour plot is saturated at the 99th and third percentile of the power. Moreover, the periodogram and the WPS are cut for the plot at a lower limit of 0.5\,d. The WPS is compared to the CLEAN \citep{Roberts87} periodogram (to the right of the WPS). For this sample of light curves, this periodogram shows a slightly higher frequency resolution than the Lomb-Scargle periodogram by better resolving close double-peaked periods. 
The WPS delivers a lower frequency resolution than the periodogram and close peaks are resolved only poorly; on the other hand, it becomes clear when a certain periodicity is present in the time series. High peaks in the periodogram that appear for timescales on the order of one period in the WPS are not real periodicities, but rather parts of the light curve that just highly correlate with the shape of the Morlet wavelet. When two major periodicities are present, the WPS allows us to distinguish which of the two is really predominant in the signal. In case of a changing period, the change can be time-resolved by the WPS, but will appear as several peaks in the periodogram, as is the case for DK Tau (Fig.~\ref{fig:ex_WPS}). Depending on the needs of any given light curve, a higher time or frequency resolution can be obtained by changing the central frequency or the bandwidth of the Morlet wavelet (see Appendix \ref{sec:app:morlet}). This does not affect the physical interpretation of the results. Because of the lower frequency resolution of the wavelet transform, it is better to use the period as retrieved from the periodogram. It is possible to  project\ the WPS onto the y-axis and thus average over time to obtain something similar to a periodogram. However, the lower frequency resolution of the wavelet transform does not provide any particular advantage for the period determination alone compared to a periodogram.

\subsection{Period analysis}
\label{sec:periods}
\begin{figure}[ht]
\begin{subfigure}{.24\textwidth}
  \centering
  \includegraphics[width=\linewidth]{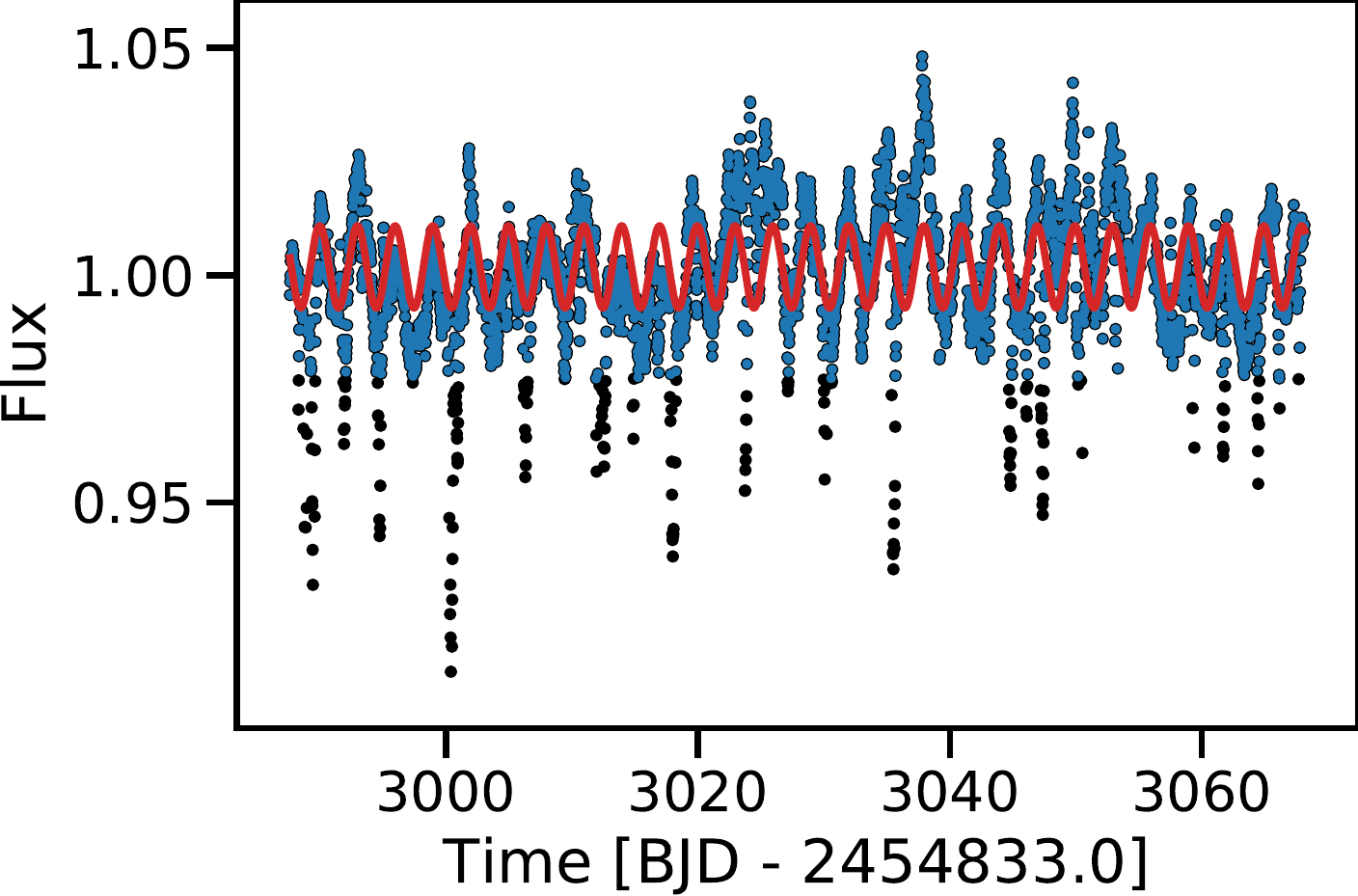}  
  \caption{ }
  \label{fig:sub-first}
\end{subfigure}
\begin{subfigure}{.24\textwidth}
  \centering
  \includegraphics[width=\linewidth]{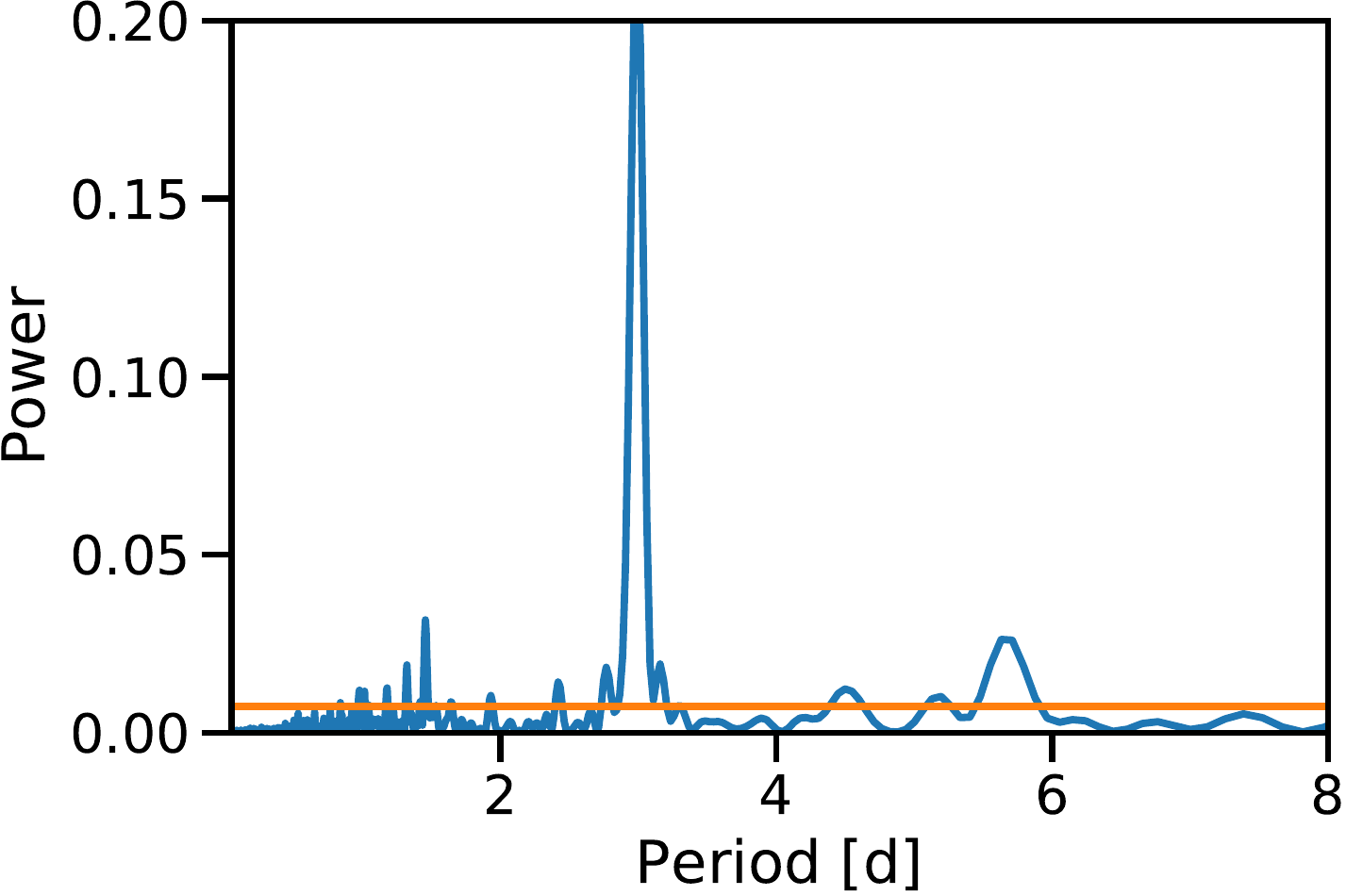}  
  \caption{ }
  \label{fig:sub-second}
\end{subfigure}\\
\begin{subfigure}{.24\textwidth}
  \centering
  \includegraphics[width=\linewidth]{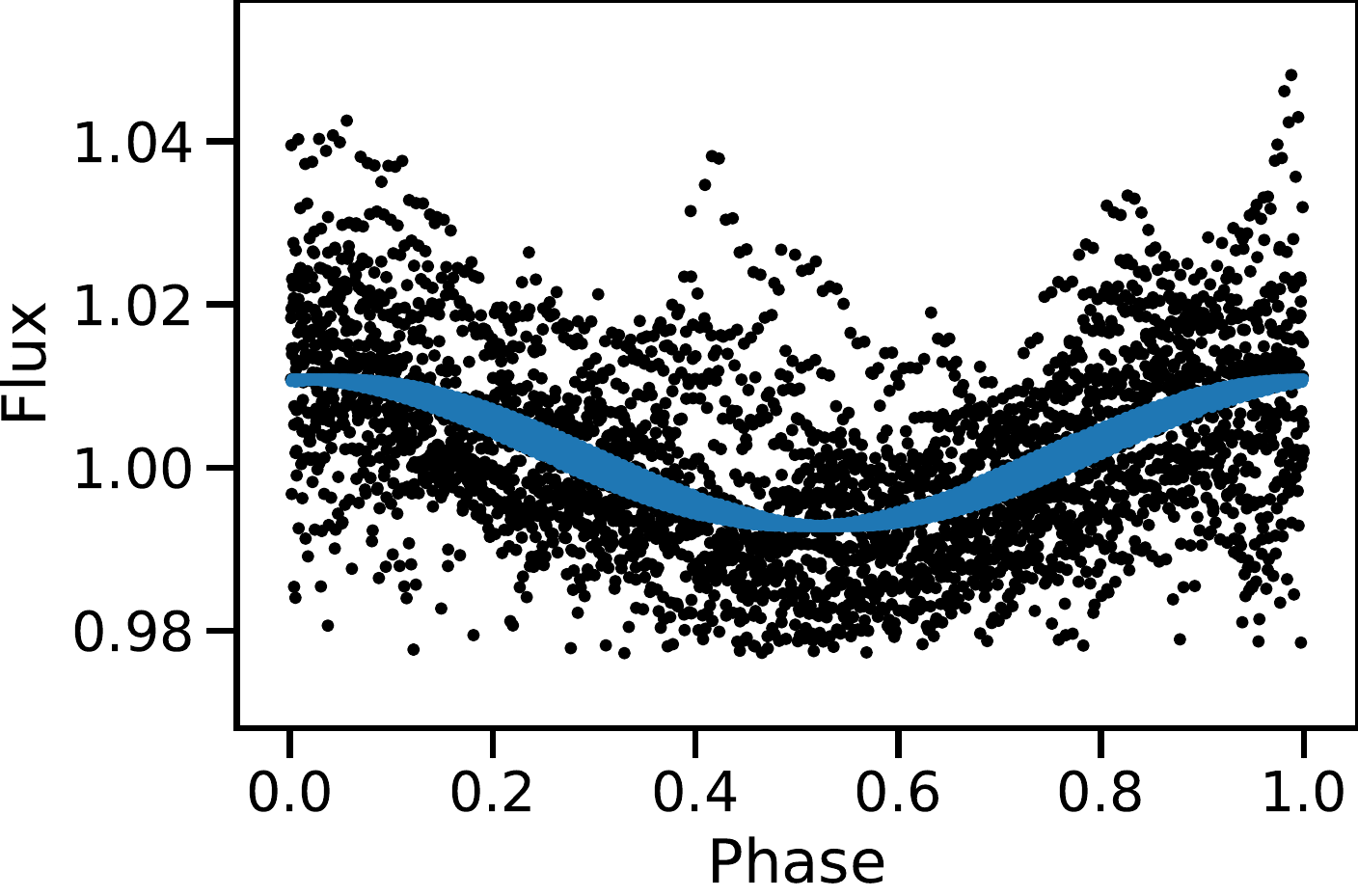}  
  \caption{ }
  \label{fig:sub-second}
\end{subfigure}
\begin{subfigure}{.24\textwidth}
  \centering
  \includegraphics[width=\linewidth]{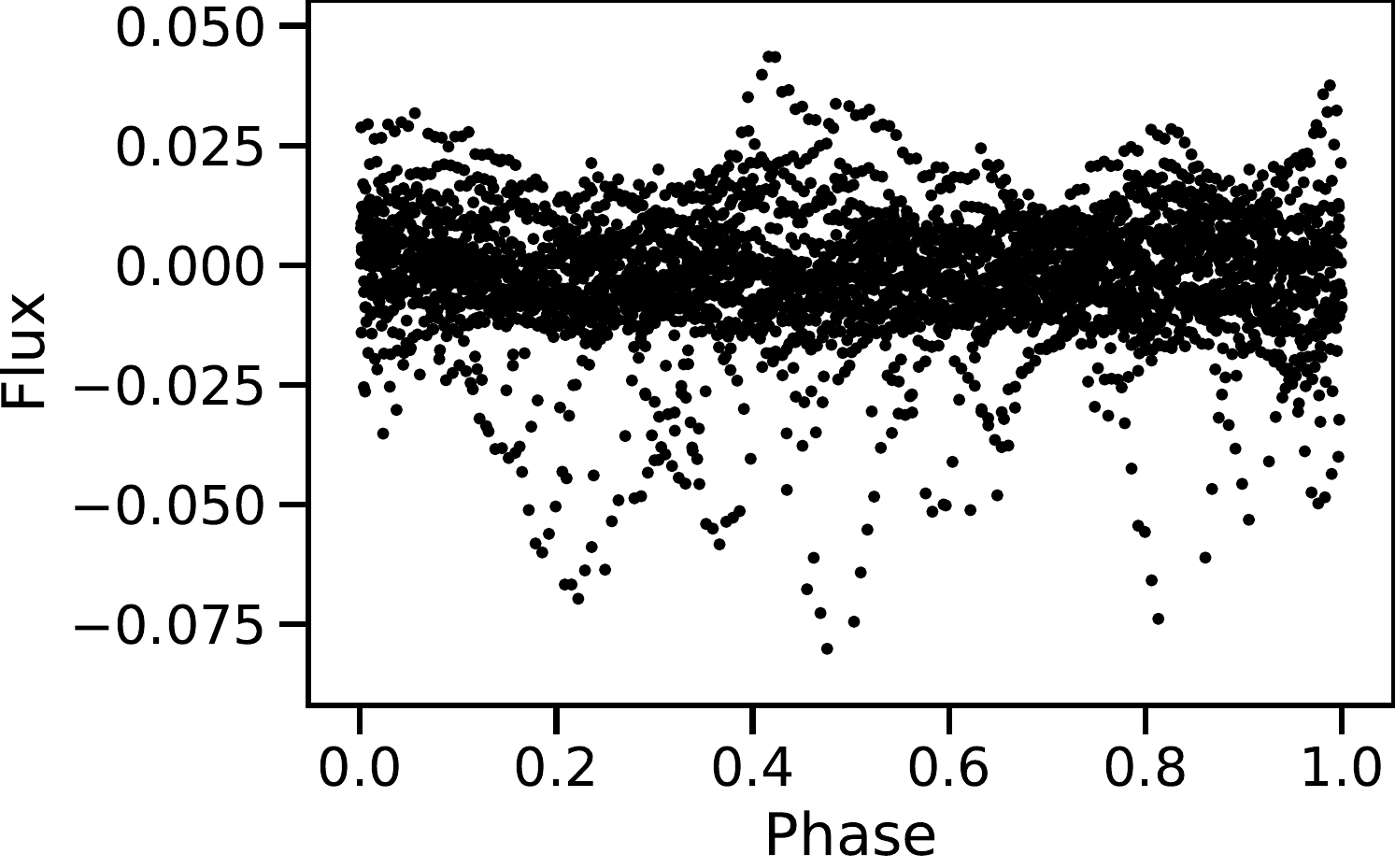}  
  \caption{ }
  \label{fig:sub-second}
\end{subfigure}
\caption{(a) Light curve of EPIC247885481. Red: Sinusoidal fit of the flux above the fifth percentile (blue). (b) Periodogram of the star with the FAP level at 0.05 (orange). (c) Light curve above the fifth percentile folded at 2.99\,d and sinusoidal fit (blue). (d) Residual light curve after subtraction of the fit as in panels (a) and (c), folded at 2.99\,d. The noise is high with respect to the dips, but no evident pattern is present.}
\label{fig:disc:EPIC247885481}
\end{figure}

\begin{figure*}[ht]
\begin{subfigure}{.33\textwidth}
  \centering
  \includegraphics[width=\linewidth]{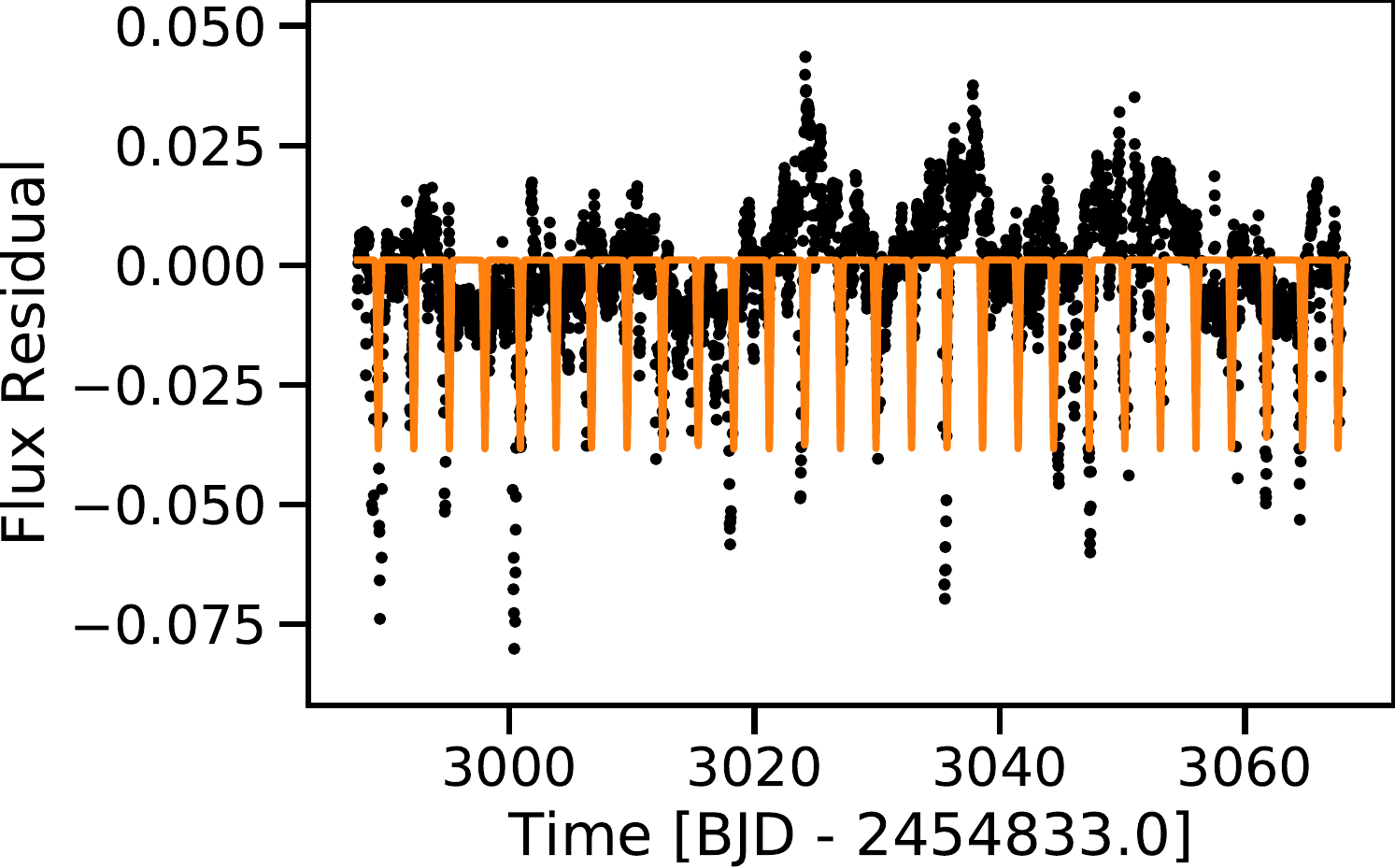}  
  \caption{ }
  \label{fig:sub-first}
\end{subfigure}
\begin{subfigure}{.33\textwidth}
  \centering
  \includegraphics[width=\linewidth]{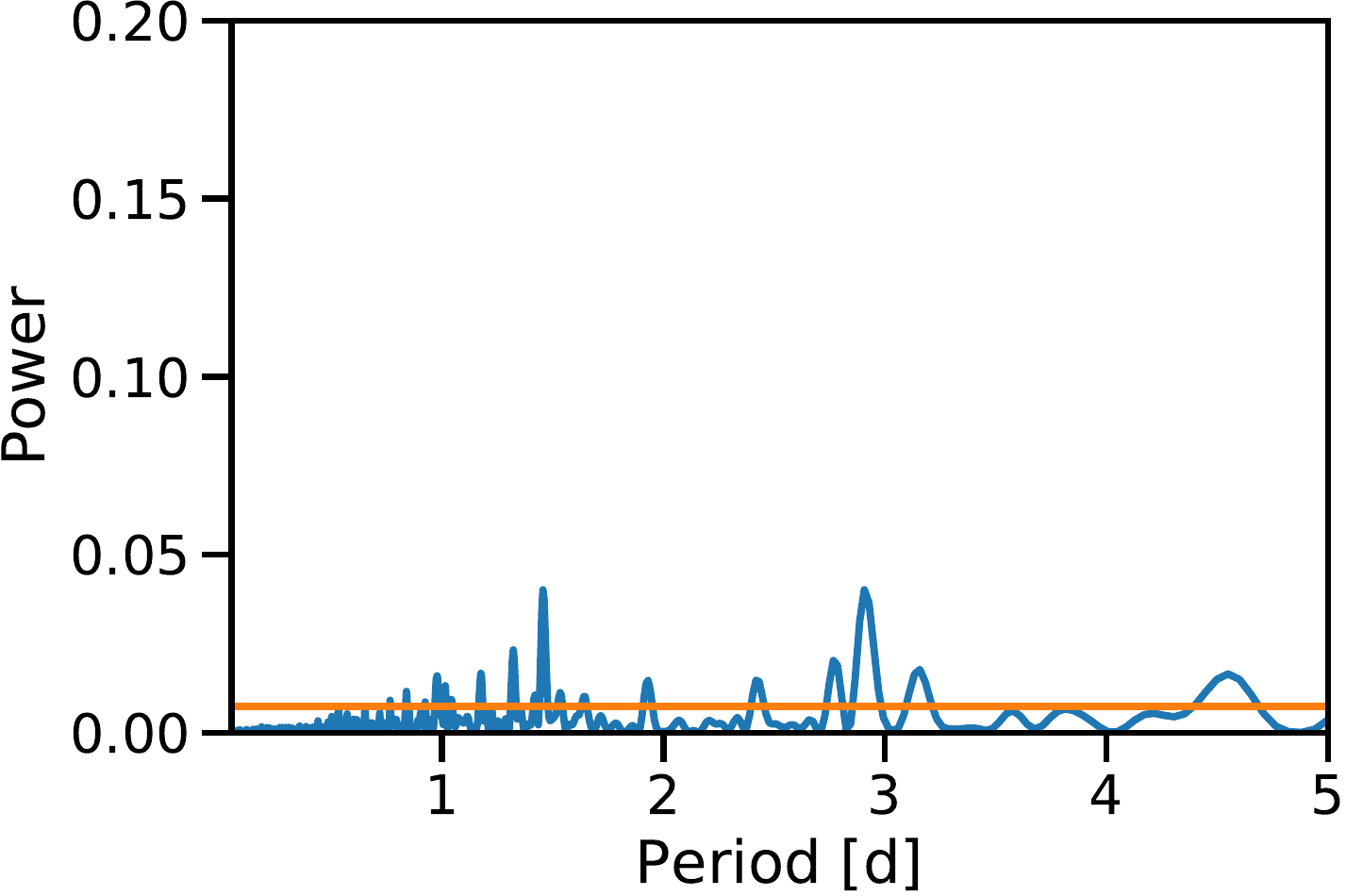}  
  \caption{ }
  \label{fig:sub-second}
\end{subfigure}
\begin{subfigure}{0.33\textwidth}
  \centering
  \includegraphics[width=\linewidth]{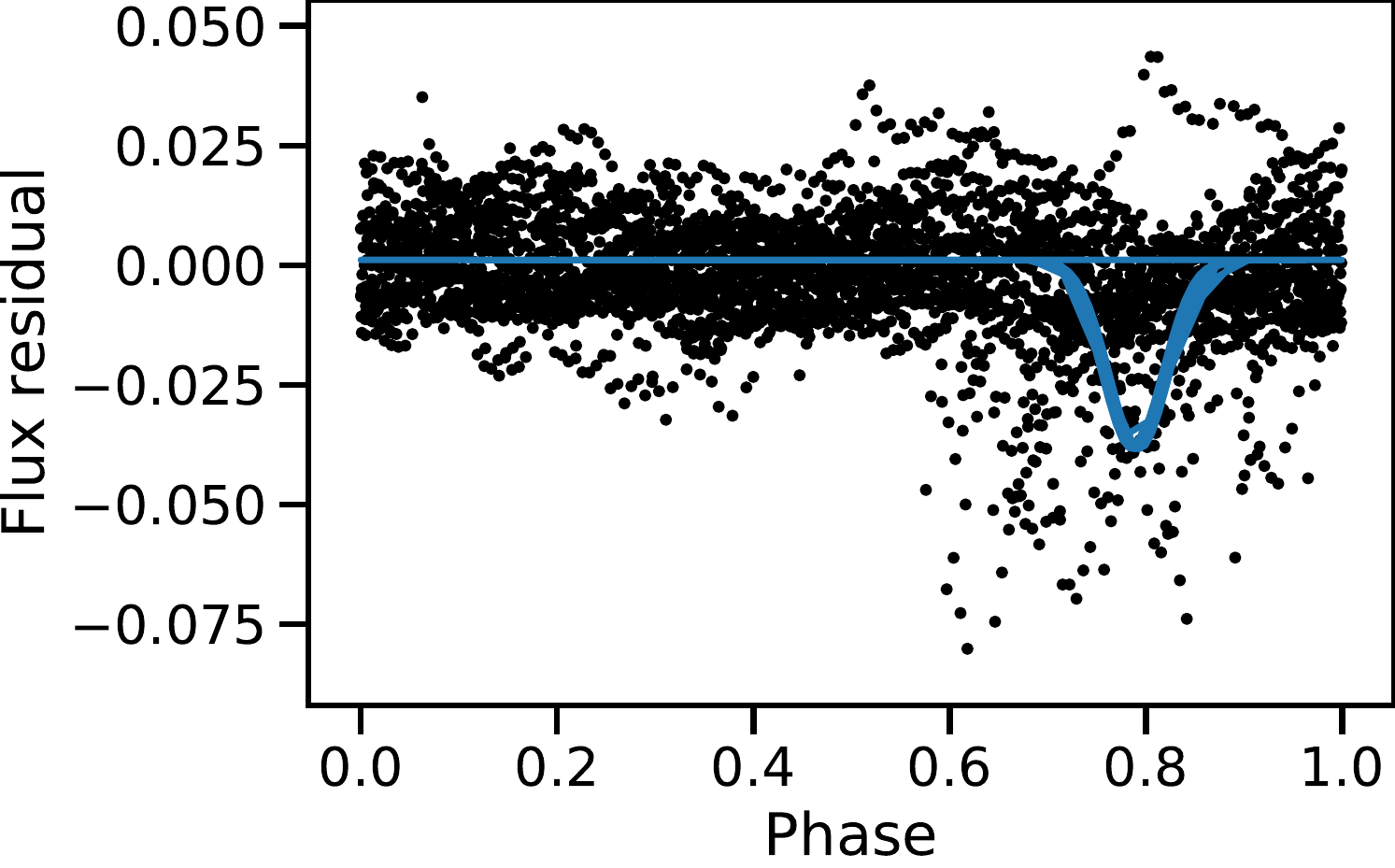}  
  \caption{ }
  \label{fig:sub-second}
\end{subfigure}
\caption{Fitting with periodic Gaussian pulses (a) of the residual light curve of  EPIC247885481 after the removal of the sinusoidal variability. (b) Periodogram of the residual light curve. The peak is far less evident, but still significant. (c) Residual light curve folded at 2.90\,d with the fit Gaussian pulses (blue).}
\label{fig:disc:EPIC247885481_residual}
\end{figure*}
Dippers often appear to be quasiperiodic. Following the disk warp model \citep[e.g.,][]{Bouvier07}, the dusty material obscuring the star must be located close to corotation, which is per definition the region in which circumstellar material rotates with the same angular velocity as the star. \cite{Stauffer15} and \cite{Rebull18} identified some light curves that showed both spots and dips. The overlapping periods of the two phenomena strongly supported the scenario of an inner disk that was rotationally locked to the star because stellar spots follow the stellar rotation. It is thus of interest to discuss the periodicities of dippers compared to stellar rotation.

The variations due to spots are smooth, largely sinusoidal variations. The quasiperiodic dippers produce rather sharp or complex dips, and these events are usually slightly different each time they occur. The distinction between variations originating in spots and dippers can be ambiguous in some unusual cases. Sometimes multiple real periodicities (not harmonics) can show up in the periodogram or in the WPS; in these cases, the physical interpretation of the periodicity can be complicated \cite[see, e.g., discussion in][]{Rebull18,Rebull20}.

In this sample, eight stars exhibit a single clear periodicity (JH 223, EPIC 246989752, CFHT Tau 12, V807 Tau, GK Tau, HK Tau, GM Tau, and IS Tau), three stars are aperiodic (HD 285893, FX Tau, and GO Tau), and for another three stars, the periodicity is unclear (Haro 6-37, St 34, and DK Tau). Two stars are transient quasiperiodic dippers (JH 112 A and DK Tau). The remaining seven dippers are quasiperiodic, but their light curve or periodogram is more complex than that of the other dippers: LkCa 15, EPIC 247885481, HP Tau, GH Tau, GI Tau, EPIC 247820821, and ITG 34. The periods are listed in Table~\ref{tab:periods_ampl}. Those that are uncertain are flagged with a semicolon.

The periods coincide with those published by \cite{Rebull20}, with the exception of that of DK Tau (7.84\,d vs. 7.69\,d derived here; this small difference is due to the complex periodogram and the usage of two different algorithms). The WPS (Fig.~\ref{fig:ex_WPS}) shows that the period increases during the \textit{K2} campaign. The periodogram exhibits two peaks at $\sim$8 and $\sim$10\,d. By means of the WPS, they can be interpreted as a single changing period. This could explain why the period of DK Tau is never constrained precisely in the literature (see Appendix~\ref{sec:app:individuals} for more details). The stars ITG 34 and GH Tau have a second reported period \citep{Rebull20}, which is confirmed here. For GH Tau, a possible third period at 5.09\,d is present. 
An interesting case is EPIC 247885481 (see Sec.~\ref{subsec:disc:diff_period} for a detailed discussion), where the main periodicity of 2.99\,d is most probably caused by a spot, while the dips can be folded in phase with a period of 2.90\,d. \cite{Rebull20} reported 2.99\,d for the main periodicity, while the dipper period is unresolved both in the periodogram here and in \cite{Rebull20}. In the case of GI Tau, the light curve is contaminated by the neighboring GK Tau, and the dominant peak in the periodogram is the period of GK Tau. Only the second period is therefore reported as being of GI Tau. EPIC 247820821 shows a second periodicity on the WPS at 7.00\,d that does not appear in the periodogram. When folded in phase, the pattern appears to be related to a shift of the minimum of the dip. The periodogram of ITG 34 exhibits several double-peaked periods that cannot be harmonics of each other. It is difficult to recognize a dominant structure in the folded light curve. All the folded light curves are presented in Appendix~\ref{sec:app:phasecurves}. The motions of the \textit{Kepler} satellite might create spurious periods at 0.22, 1.75, and 1.97\,d. None of them has been observed here. 

\paragraph{A different period for spot and dips?}
\label{subsec:disc:diff_period}
The light curve for EPIC 247885481 (2MASS J05023985+2459337) shows narrow dips superimposed on a sinusoidal variability. The periodogram shows a clear peak at 2.99\,d with $\sigma=0.08$\,d, but the folded light curve shows that the period refers to the sinusoidal variation (Fig.~\ref{fig:disc:EPIC247885481}). The dips can be folded in phase around $P - \sigma$ at $\sim$2.90\,d. The peak in the periodogram does not appear to have any complex structure. 

In order to better study the periodicity, the spot was fit with a sinusoidal wave with a period of 2.99\,d (Fig.~\ref{fig:disc:EPIC247885481}), which was then subtracted from the light curve (Fig.~\ref{fig:disc:EPIC247885481_residual}). The periodogram before and after the subtraction shows that the main contribution to the periodicity is given by the spot; no substructure appears after the spot subtraction. This can be explained by the fact that the amplitude of the eclipses is very small compared to the noise and that the dip is not clearly present in every period. The shape of the dips is nearly Gaussian, and the dip width is small compared to the period. The residual was fit with a periodic Gaussian pulse such as $f(t) = c + A \sum_{n=-\infty}^{\infty} \exp{-\frac{(t -nT -\delta)^2}{2\sigma^2}}$, with the period $T$ as a free parameter. The retrieved period is thus 2.90\,d for the dips, still very close to $P-\sigma,$ and it cannot be confirmed that they are significantly different. Similar conclusions about other stars were reached by \cite{Stauffer15}. Nevertheless, the fact that the dips cannot be folded in phase with the period of the spot is a clear indication that they must be different. The difference between the two periods could be explained by the differential rotation of the stellar surface, assuming that the corotating material is aligned with the equator and the spot is at higher latitude.

\subsection{Light-curve morphology}
In the following section, the light curve properties of ideal and peculiar dippers are briefly presented and the determination of the dip properties, such as amplitude and width, are explained and discussed.
\begin{figure*}
    \centering
    \includegraphics[width=\linewidth]{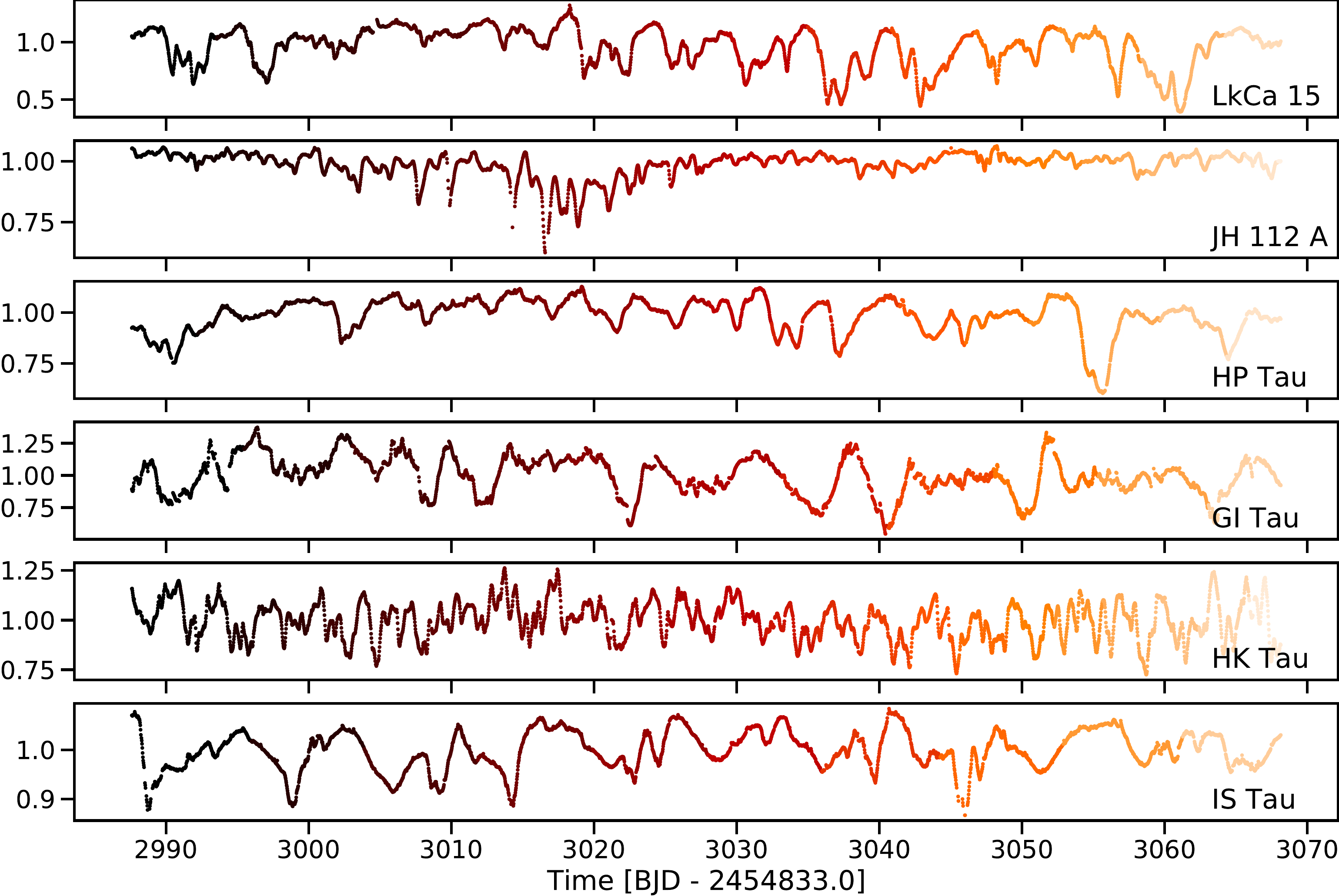}
    \caption{Light curves of the double-peaked periodic dippers. Top to bottom: LkCa 15, JH 112 A, HP Tau, GI Tau, HK Tau, and IS Tau. The color-code is normalized to the length of the observations, one color per phase.}
    \label{fig:disc:double_peaks}
\end{figure*}
\paragraph{Ideal dippers} As mentioned in Sec.~\ref{sec:intro} and in Sec.~\ref{subsec:res:identif}, the main characteristics of dippers are their dips. These are irregular in shape and can either be aperiodic or quasiperiodic. A good example for a quasiperiodic dipper is JH 223 (Fig.~\ref{fig:example_lcs_dippers}): The occurrence of the dips is periodic, their shape is irregular, and their amplitude can almost double in two neighboring dips. No other types of variability in form of bursts or other patterns affect the classification. Its period of 3.31\,d and the dip duration of $\sim$2\,d are in range of the rotation periods of CTTSs and of observed dip properties. Aperiodic dippers tend to present either very narrow dips whose shape is rather simple \citep[e.g.,][]{Stauffer15}, which is the case of HD 285893 (Fig.~\ref{fig:example_lcs_dippers}), or broad and complex dips. 

\paragraph{Transient dippers}
\label{sec:transient}
The only clear example of a quasiperiodic and transient dipper in the sample is JH 112 A. The dips have a period of 2.21\,d, and if directly linked to the stellar rotation, they are present for a timescale of $\text{about ten}$ full rotations. Interestingly, the continuum brightness decreases only as long as the eclipses occur. DK Tau also changes from a fade state to a dipper with a higher brightness continuum, but its light curve is complex, and a strong overall variability has already been observed in former campaigns. The dipper status of quasiperiodic dippers is known to be transient over a few years  (e.g., AA Tau), but it is rare to observe a quasiperiodic dipper on such a short timescale. The explanation of a disk warp would imply that the warp significantly changes its height and becomes not visible for the observer after a few rotations, or that the dust in it is completely dissipated on a short timescale.

\paragraph{Double- and multi-peaked dippers}
\label{subsec:double_peak}
In the sample studied, six periodic dippers exhibit clear double-peaked dips in their light curve:  LkCa 15, JH 112 A, HP Tau, GI Tau, HK Tau, and IS Tau (Fig.~\ref{fig:disc:double_peaks}). These double- or multiple-peaked dippers exhibit at least two well-detached dips in the folded light curve (see App.~\ref{sec:app:phasecurves}). It should be noted that the dips themselves are in general not Gaussian and have a complex shape. The determination of the dip width for these sources depends on the desired information (the width of the primary peak or the total width of the multi-peaked dip) and is handled in Sec.~\ref{sec:res:ampl_width}.
A more detailed description regarding the double dips of individual sources is provided in Appendix~\ref{sec:app:individuals}. In the scenario in which a dusty warp occults the star, the constant presence (or absence) of a double dip delivers an indication about the stability and the shape of the warp. 
In general, it can be remarked that the dips, although periodic, do not occur exactly at the same phase. For HP Tau, the WPS shows that the periodicity slightly varies around a period of 4.33\,d. The light curve displays clear shifts of the minima of the dip for different rotation cycles. Moreover, the variations in amplitude of the different dips do not seem correlated to each other. This suggests that either smaller and independent dusty warps occult the star, or that the shape of the dusty structure is highly unstable.
A quantitative study of this variation will be object of future work.

\subsection{Dip amplitude and dip width}
\label{sec:res:ampl_width}

\begin{figure}[h!]
    \centering
    \includegraphics[width=0.8\linewidth]{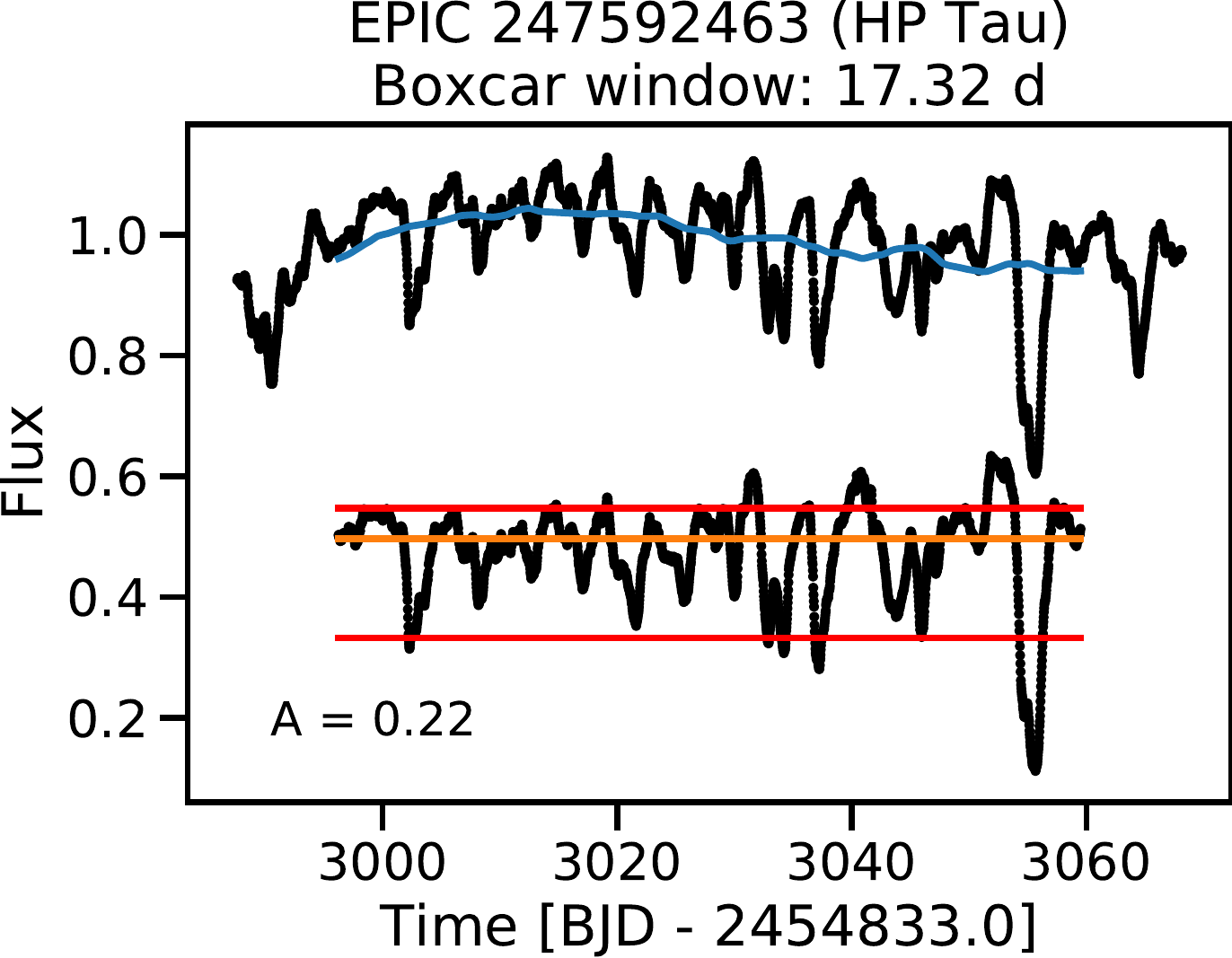}
    \caption{Determination of the dip amplitude with the example of HP Tau. Top: light-curve detrending of the original data (black) with a boxcar of size $4\cdot P$. The blue line is the trend. Bottom: Detrended light curve. The convolution with a boxcar removes a small part of the data at the edges. Orange line: Flux median. Red lines: 90th and fifth percentiles of flux. The usage of flux percentiles allows us to consider the global variability of the light curve.}
    \label{fig:LC_morph:HPTau}
\end{figure}

Among all other sources of variability for YSOs, the long-term trends particularly affect the determination of the dip amplitude because the continuum brightness is unstable.
To determine the dip amplitudes, the light curves were therefore
detrended using a boxcar filler with a width chosen to be four times the period. This removed the long-term variability and did not interfere with the dips (Fig.~\ref{fig:LC_morph:HPTau}). For nonperiodic dippers, a standard window of 7\,d was chosen. The amplitudes were computed as the difference between the 90th and fifth percentile and peak-to-peak and are listed in Table~\ref{tab:periods_ampl}. The light curves of Haro 6-37, JH 223, IS Tau, and DK Tau (for this star, the faint state of the first $\sim$30\,d has not been included for this and the following analysis) did not need detrending because the continuum brightness is stable.

\begin{figure}[h!]
    \centering
    \includegraphics[width=0.75\linewidth]{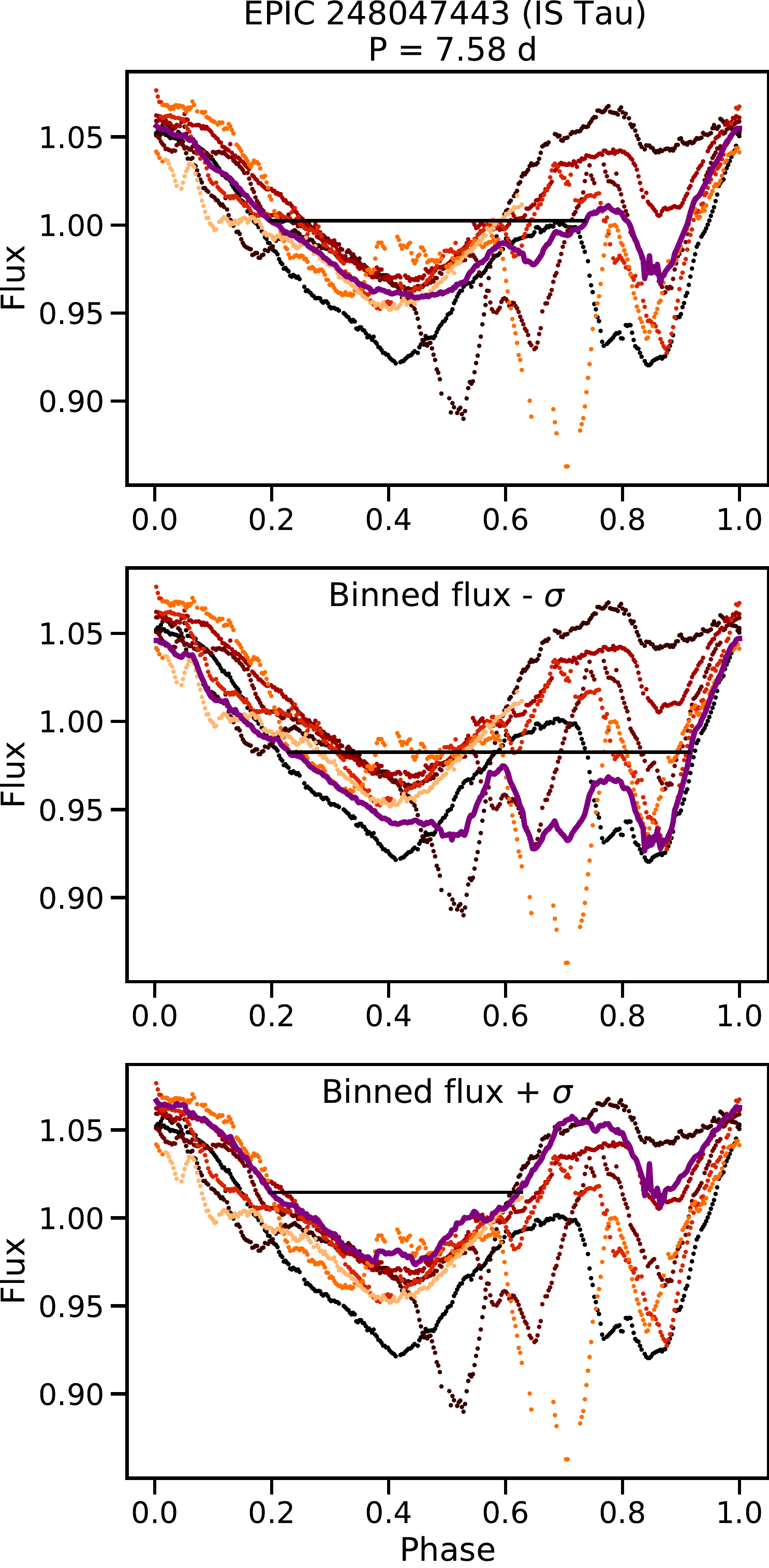}
    \caption{Dip width measured as the FWHM of the binned light curve (purple line, top). For the folded light curve, each color represents a different cycle. By adding or subtracting the standard deviation of the flux for each phase, it is possible to distinguish between the width of the main dip (bottom) and the double peak (center).}
    \label{fig:LC_morph:ISTau:width}
\end{figure}

The detrending was not applied for the study of the periodicity because it might affect the result and interfere with the physical phenomena that cause variability. It is only used in a pragmatic approach to study the dip morphology, without the effect of other stellar variability.

To define an eclipse width, it is useful to bin, that is, to average, the detrended and folded light curve (purple line in the upper panel of Fig.~\ref{fig:LC_morph:ISTau:width}). The dip width (black horizontal line in the upper panel of Fig.~\ref{fig:LC_morph:ISTau:width}) is then defined at the full width at half maximum (FWHM) of the resulting binned curve. The results are listed in Table~\ref{tab:periods_ampl} and shown in Appendix~\ref{sec:app:phasecurves}. In general, the dip width is about half of the period. The largest dip widths are linked to double- or multi-peaked dippers, with the exception of JH 112 A and HK Tau, which both exhibit a relatively small dip width ($\lesssim 0.5~P$) and the shortest periods in this group (2.21 and 3.3\,d, respectively). The discussion of the correlation between dip width and period is presented in Sec.~\ref{subsec:width}. 

We estimate an uncertainty on the dip width as follows. For each phase bin, the standard deviation $\sigma(\mathrm{phase})$ of the flux between the different cycles  was first computed (purple line in the upper panel of Fig.~\ref{fig:LC_morph:ISTau:width}). For every point of the binned light curve, the standard deviation of the phase bin was then added in one case and subtracted in the other (purple line on center and bottom panel in Fig.~\ref{fig:LC_morph:ISTau:width}). Finally, the dip width was computed again as FWHM of these modified binned light curves. Upper and lower error bars are defined as difference between the dip width of the modified binned light curve and the original one. Because the shape of the dips may vary strongly after this operation, these error bars can be large. 

For double-peaked dippers, the line drawn at the FWHM can cross the binned light curve in more than two points: the error bars $w + \sigma$ and $w - \sigma$ were then chosen to represent the primary peak width and the total width of the dip, which is very large in these cases. This can result in strongly asymmetric error bars (see Sec.~\ref{subsec:width}). In Table~\ref{tab:periods_ampl}, some width values are flagged because they are not reliable. This can happen when the folded light curve is very noisy and the binned light curve does not show the dips as they can be seen by eye. These plots appear in Fig.~\ref{fig:app:phased_lc}. 

Another possible representation of the uncertainty on the dip width would be to compute the dip width at different thresholds of the same binned light curve, for example, at 0.3, 0.5 (FWHM), 0.7 of the flux maximum. However, the information about the width of the main dip and the total dip in double-peaked dips (see, e.g., IS Tau) would be lost.

\begin{table*}[h]
\centering
\caption{Spectral types, effective temperatures, and \textit{VJHK} photometry for the presented sample of 22 best dippers. \textit{JHK} measurements are from \textit{2MASS}.}
\tiny
\begin{tabular}{llllllllllll}
\hline \hline
EPIC     &      2MASS    &      Name     &      SpT      &      Refs.    &       $T_{\mathrm{eff}}$       &      $A_V$    &      \textit{V}       &       Refs.    &      \textit{J}       &      \textit{H}       &      \textit{K}          \\
& &  & & & [K] & [mag] \\
\hline
246929818                        &      J04465897+1702381                        &       Haro 6-37                &      K8                                   &   1,2     &       3940                             &      2.1        &     12.98            & 7         &  9.24             &      7.99             &       7.31     \\
246942563                        &      J04542368+1709534                        &       St 34                    &      M3                           &  3       &         3360                             &      0.5        &    14.59            & 7         &   10.69            &      10.08            &      9.79     \\
246989752                        &      J04384725+1737260                        &        \ldots                  &      M5.5                         &  1,2     &   2920\tablefootmark{**}   &      0.0        &       \ldots        & \ldots &      12.75            &      12.11            &      11.75    \\
247103541                        &      J04363081+1842153                        &       HD 285893                &      F8                   &  4       &         6100                             &      0.3        &    10.01            & 8         &   8.76             &      8.37             &      7.99     \\
247520207                        &      J04391779+2221034                        &       LkCa 15                  &      K4                           &  5       &         4492\tablefootmark{***}  &      0.4        &    12.14            & 9         &   9.42             &      8.60             &      8.16     \\
247575958                        &      J04330945+2246487                        &       CFHT Tau 12              &      M6                                   &   1       &       2860\tablefootmark{**}   &      3.0        &         \ldots       & \ldots  &    13.15            &      12.14            &       11.54    \\
247589612                        &      J04324911+2253027                        &       JH 112 A                 &      K5.5                         &  1,2     &   4080                             &      2.9        &    14.44            & 7         &   10.24            &      8.99             &      8.17     \\
247592463                        &      J04355277+2254231                        &       HP Tau                   &      K4                                   &   1,2     &       4330                             &      3.2        &     13.78            & 9         &  9.55             &      8.47             &       7.62     \\
247763883                        &      J04330622+2409339                        &       GH Tau                   &      M2                                   &   1,2     &       3490                             &      0.4        &     12.87            & 7         &  9.11             &      8.23             &       7.79     \\
247764745                        &      J04330664+2409549                        &       V807 Tau                 &      K7\tablefootmark{*}  &  1,2     &         3970                             &      0.4        &    11.31            & 7         &   8.15             &      7.36             &      6.96     \\
247791801                        &      J04333456+2421058                        &       GK Tau                   &      K6.5                         &  1,2     &   3995                             &      1.0        &    12.67            & 9         &   9.05             &      8.11             &      7.47     \\
247792225                        &      J04333405+2421170                        &       GI Tau                   &      M0.4                         &  1,2     &   3714                             &      1.3        &    13.39            & 9         &   9.34             &      8.42             &      7.89     \\
247799571                        &      J04315056+2424180                        &       HK Tau                   &      M1\tablefootmark{*}  &  1,2     &         3630                             &      2.4        &    15.12            & 7         &   10.45            &      9.25             &      8.59     \\
247805410                        &      J04302961+2426450                        &       FX Tau                   &      M2.2                         &  1,2     &   3464                             &      1.0        &    13.39            & 7         &   9.39             &      8.40             &      7.92     \\
247820821                        &      J04295950+2433078                        &        \ldots                  &      M5                                   &   1       &       2880                             &      0.0        &     15.97\tablefootmark{****}                & 10        &  11.68            &       10.54            &      9.81     \\
247885481                        &      J05023985+2459337                        &        \ldots                  &      M4.25                        &  6       &         3090                             &      0.0        &    16.04            & 7         &   11.78            &      11.09            &      10.78    \\
247935061                        &      J04430309+2520187                        &       GO Tau                   &      M2.3                         &  1,2     &   3451                             &      1.6        &    14.43            & 7         &   10.71            &      9.78             &      9.33     \\
248006676                        &      J04404950+2551191                        &       JH 223                   &      M2.8                         &  1,2     &   3386                             &      1.4        &    15.66            & 7         &   10.75            &      9.92             &      9.49     \\
248015397                        &      J04411078+2555116                        &       ITG 34                   &      M5.5                         &  1       &         2920\tablefootmark{**}   &      2.2        &       \ldots          & \ldots     &         13.19            &      12.12            &       11.45    \\
248029373                        &      J04304425+2601244                        &       DK Tau                   &      K8.5                         &  1,2     &   3910                             &      0.7        &    12.58            & 9         &   8.72             &      7.76             &      7.10     \\
248046139                        &      J04382134+2609137                        &       GM Tau                   &      M5                                   &   1,2     &       2880                             &      2.1        &     17.83            & 7         &  12.80            &      11.59            &       10.63    \\
248047443                        &      J04333678+2609492                        &       IS Tau                   &      M0\tablefootmark{*}  &  1,2     &         3770                             &      2.4        &    14.94            & 7         &   10.32            &      9.29             &      8.64     \\
\hline
\end{tabular}
\tablefoot{
\tablefoottext{*}{Spectral type by \cite{Esplin19} differs by 0.5 subclasses or more in comparison with \cite{Herczeg14}.\\}
\tablefoottext{**}{For spectral types later than M5, the temperature conversion by \cite{Herczeg14} is used. The extrapolation for M6 in \cite{Pecaut13} would deliver 3038\,K.\\}
\tablefoottext{***}{$T_{\mathrm{eff}}$ derived spectroscopically by \cite{Alencar18}.\\}
\tablefoottext{****}{The provided amplitude and $<V>$ were combined to derive a more representative $V_{\mathrm{min}}$.}
}
\tablebib{
(1) \citet{Esplin19}; (2) \citet{Herczeg14}; (3) \citet{Dahm11}; (4) \citet{Nesterov95}; (5) \citet{Alencar18}; (6) \cite{Esplin14};
 (7) \citet{Lasker08}; (8) \citet{Henden16}; (9) \citet{Grankin07}; (10) \citet{Drake14}.
}
\label{tab:SpT_photometry}
\end{table*}

\begin{table*}[htbp] 
\centering 
\caption{Additional dippers that exhibit dips in their light curves as secondary variability.} 
\tiny
\begin{tabular}{llllll}
\hline \hline 
EPIC     &      2MASS    &      Name     &      Type     &      Period   &       \textit{G}       \\ 
&&&& [d] & [mag] \\
\hline 
210725857                &      J04285053+1844361                &      \ldots                  &      Sd                       &      2.06                     &       15.19            \\ 
246859790\tablefootmark{*}               &      J04440164+1621324                &       \ldots                   &      Sd                       &      2.16                  &      17.30            \\ 
247575425                &      J04331907+2246342                &      IRAS 04303+2240               &      Td                       &      7.5                       &      16.09            \\ 
247584113                &      J04335200+2250301                &      CI Tau                      &      db                       &      8.5;                      &      12.15            \\ 
247585465                &      J04322415+2251083                &      \ldots                  &      Bd?                      &      2.48                     &       15.25            \\ 
247788960                &      J04323058+2419572                &      FY Tau                      &      B?d                      &      6.94;                     &      13.77            \\ 
247810494                &      J04345542+2428531                &      AA Tau                      &      d?                       &      \ldots                    &      14.49            \\ 
247827638\tablefootmark{*}               &      J04293606+2435556                &       XEST 13-010              &      Sd                       &      3.9                          &      14.85            \\ 
247915927                &      J04442713+2512164                &      IRAS 04414+2506               &      S?d                      &      4.42                      &      15.48            \\ 
248009353                &      J04324282+2552314                &      UZ Tau                      &      Bd                       &      \ldots                    &      12.39            \\ 
248018164\tablefootmark{*}               &      J04413882+2556267                &       Haro 6-33                &      Td?                      &      \ldots                  &      15.94            \\ 
248030407\tablefootmark{*}               &      J04394488+2601527                &       ITG 15                   &      S?d                      &      3.45                  &      15.00            \\ 
\hline 
\end{tabular} 
\tablefoot{
The variability type has to be interpreted as: capital letter = dominant variability; d = dipper; s = spot; b = burster; t = long-term trend. The periods are derived with the CLEAN periodogram, and the listed value is the dominant period of the light curve, i.e., of the spot, when present. Only CI Tau and EPIC 247585465 exhibit quasiperiodic dips. The brightness is obtained from the \textit{Gaia G} band.
\tablefoottext{*}{Star reported as a dipper by \cite{Rebull20}.} 
}
\label{tab:other_dippers}
\end{table*}
\begin{table*}[htbp] 
\centering 
\caption{Variable YSOs that are dipper candidates of lower quality.} 
\tiny
\begin{tabular}{llllll}
\hline \hline 
EPIC     &      2MASS    &      Name     &      Type     &      Period   &       \textit{G}       \\ 
&&&& [d] & [mag] \\
\hline 
210689083                &      J04313747+1812244                &      HH 30                       &      t                        &      7.51                      &      \ldots           \\ 
210690735                &      J04300399+1813493                &      UX Tau A+C                  &      t?                       &      3.6;                      &      11.29            \\ 
210690913                &      J04313843+1813576                &      HL Tau                      &      t?                       &      \ldots                    &      \ldots           \\ 
246923113                &      J04470620+1658428                &      DR Tau                      &      Bd?                      &      14.71                     &      11.65            \\ 
247837468                &      J04293008+2439550                &      IRAS 04264+2433               &      ?                        &      11.9                      &      \ldots           \\ 
247923794                &      J04423769+2515374                &      DP Tau                      &      t                        &      3.66;                     &      13.51            \\ 
247992574                &      J04392090+2545021                &      GN Tau                      &      t                        &      5.75;                     &      14.56            \\ 
248017479                &      J04410826+2556074                &      ITG 33A                      &      t                        &      6.6;                      &      17.60            \\ 
248040905                &      J04295156+2606448                &      IQ Tau                      &      t                        &      \ldots                    &      13.24            \\ 
248055184                &      J04335470+2613275                &      IT Tau                      &      t                        &      2.74                      &      \ldots           \\ 
248058354                &      J04334465+2615005                &      \ldots                  &      ?                        &      \ldots                   &       16.50            \\ 
\hline 
\end{tabular} 
\tablefoot{
The variability type has to be interpreted as: capital letter = dominant variability; d = dipper; s = spot; b = burster; t = long-term trend. The periods are derived with the CLEAN periodogram and the listed value is the dominant period of the light curve. Most objects in this list are dominated by a long-term trend. The brightness is obtained from the \textit{Gaia G} band.
}
\label{tab:not_dippers}
\end{table*}
\begin{table*}[htbp] 
\centering 
\caption{Morphological properties of the dips.} 
\tiny
\begin{tabular}{lllllllll}
\hline \hline 
EPIC     &      Name     &      $P$      &      $\sigma_P$       &      $W$          &      $W_{-\sigma}$    &      $W_{+\sigma}$    &      $A$      &       $p2p$    \\ 
&& [d] & [d] & [$P$] & [$P$] & [$P$] & [mag] & [mag] \\
\hline 
246929818\tablefootmark{a}               &      Haro 6-37                &       10.63;                   &      0.72             &      0.59                      &      0.65                     &      0.50                     &       0.75                     &      1.03             \\ 
246942563                &      St 34                    &      5.23;                     &      0.20             &      0.65\tablefootmark{b}                     &      0.59                     &      0.76                     &       0.14                     &      0.33             \\ 
246989752                &      \ldots                   &      1.62                      &      0.03             &      0.22                     &       0.18                     &      0.37                     &      0.16                  &      0.46             \\ 
247103541                &      HD 285893                &      \ldots                    &      \ldots           &      \ldots                   &       \ldots                   &      \ldots                   &      0.05                  &      0.13             \\ 
247520207                &      LkCa 15                  &      5.78                      &      0.20             &      0.64\tablefootmark{c}                     &      0.47                     &      0.62                     &       0.54                     &      1.05             \\ 
247575958                &      CFHT Tau 12              &      3.48                      &      0.07             &      0.59                     &       0.60                     &      0.70                     &      0.23                  &      0.50             \\ 
247589612                &      JH 112 A                 &      2.21                      &      0.07             &      0.20\tablefootmark{c}                     &      0.18                     &      0.34                     &       0.11                     &      0.53             \\ 
247592463                &      HP Tau                   &      4.33                      &      0.17             &      0.55\tablefootmark{c}                     &      0.51                     &      0.64                     &       0.24                     &      0.65             \\ 
247763883                &      GH Tau                   &      2.49                      &      0.06             &      0.44                     &       0.54                     &      0.48                     &      0.37                  &      0.74             \\ 
247764745                &      V807 Tau                 &      4.39                      &      0.12             &      0.39                     &       0.40                     &      0.35                     &      0.11                  &      0.18             \\ 
247791801                &      GK Tau                   &      4.61                      &      0.15             &      0.43                     &       0.41                     &      0.48                     &      0.62                  &      1.28             \\ 
247792225                &      GI Tau                   &      7.13                      &      0.40             &      0.67\tablefootmark{c}                     &      0.34                     &      0.68                     &       0.49                     &      0.97             \\ 
247799571                &      HK Tau                   &      3.3                       &      0.07             &      0.50                     &       0.53                     &      0.44                     &      0.29                  &      0.55             \\ 
247805410                &      FX Tau                   &      \ldots                    &      \ldots           &      \ldots                   &       \ldots                   &      \ldots                   &      0.18                  &      0.36             \\ 
247820821                &      \ldots                   &      2.38                      &      0.04             &      0.41                     &       0.43                     &      0.36                     &      1.18                  &      1.77             \\ 
247885481                &      \ldots                   &      2.90\tablefootmark{d}                  &      0.08             &      0.60\tablefootmark{b}                     &      0.32                     &      0.89                     &       0.05                     &      0.14             \\ 
247935061                &      GO Tau                   &      \ldots                    &      \ldots           &      \ldots                   &       \ldots                   &      \ldots                   &      0.17                  &      0.36             \\ 
248006676\tablefootmark{a}               &      JH 223                   &       3.31                     &      0.09             &      0.36                      &      0.36                     &      0.39                     &       0.32                     &      0.54             \\ 
248015397                &      ITG 34                   &      3.91                      &      0.17             &      0.51                     &       0.46                     &      0.58                     &      0.53                  &      1.22             \\ 
248029373\tablefootmark{a}               &      DK Tau                   &       7.69;                    &      0.45             &      0.57                      &      0.65                     &      0.35                     &       1.08                     &      1.91             \\ 
248046139                &      GM Tau                   &      2.67                      &      0.05             &      0.44                     &       0.49                     &      0.40                     &      1.30                  &      2.00             \\ 
248047443                &      IS Tau                   &      7.58                      &      0.41             &      0.54\tablefootmark{c}                     &      0.68                     &      0.43                     &       0.13                     &      0.24             \\ 
\hline 
\end{tabular} 
\tablefoot{
Columns: the photometric period $P$, the uncertainty estimated from the Gaussian fitting of the periodogram peak $\sigma_P$, the dip width $W$ measured as the FWHM of the folded and binned light curve in units of phase, the dip width estimated considering the binned light curve $\pm \sigma$ of the flux ($W_{-\sigma}, W_{+\sigma}$), the dip amplitude with the 90 and 5 flux percentiles $A$, the peak-to-peak amplitude $p2p$. The stars are detrended before the width was determined, with the exception of those marked \tablefoottext{a}{.} \tablefoottext{b}{Value discarded because it was not representative of the real dip width.} \tablefoottext{c}{Double-peaked dip.} \tablefoottext{d}{The peak of the periodogram is at 2.99\,d and is linked to a cold spot. A more thorough analysis is needed to retrieve the dipper period at 2.90\,d.}
}
\label{tab:periods_ampl}
\end{table*}

\section{Stellar parameters}
\label{sec:stellar_params}
\begin{table*}[htbp]
\centering
\caption{Stellar properties of the dipper sample.}
\tiny
\begin{tabular}{llllllllllll}
\hline \hline
EPIC     &      Name     &      $L_*$    &      $M_*$    &      $R_*$    &       $R_\mathrm{cor}$         &      $T_\mathrm{cor}$         &      $v\sin{i}$          &      Ref.     &      $i_*$ & $\log{\dot{M}_{\mathrm{acc}}}$ & Refs. \\
         &               &       [L$_{\odot}$]   & [M$_{\odot}$]         &        [R$_{\odot}$]   &       [R$_*$]         &       [K]     &       [km~s$^{-1}$]    &               &      [$\degr$] & [$\mathrm{M_{\odot}~yr^{-1}}$]  &       \\
\hline
246929818                &      Haro 6-37                &      $0.79_{-0.37}^{+0.40}$                  &      $0.60_{-0.05}^{+0.05}$                   &      $1.91_{-0.48}^{+0.54}$                  &      8.98                     &      $930_{-158}^{+179}$                          &      $12.1\pm 1.2$                            &       1                        &      $>47$   &  -7.00/-8.12   &  6,7,8    \\
246942563                &      St 34                    &      $0.18_{-0.03}^{+0.02}$                  &      $0.25_{-0.05}^{+0.05}$                   &      $1.25_{-0.13}^{+0.14}$                  &      6.38                     &      $941_{-107}^{+118}$                          &      \ldots                                   &       \ldots                   &      \ldots            &  \ldots   &  \ldots    \\
246989752                &      \ldots                   &      $0.023_{-0.002}^{+0.004}$                  &      $0.071_{-0.011}^{+0.014}$                &      $0.59_{-0.04}^{+0.08}$                  &      4.08                     &      $1022_{-91}^{+122}$                          &      \ldots                                   &       \ldots                   &      \ldots            &  \ldots   &  \ldots    \\
247103541                &      HD 285893                &      $1.99_{-0.25}^{+0.25}$                  &      $1.15_{-0.02}^{+0.02}$                   &      $1.26_{-0.10}^{+0.08}$                  &      \ldots                   &      \ldots                                    &      \ldots                                   &       \ldots                   &      \ldots          &   \ldots  &   \ldots   \\
247520207                &      LkCa 15                  &      $0.96_{-0.17}^{+0.19}$                  &      $1.12_{-0.01}^{+0.01}$                   &      $1.62_{-0.18}^{+0.19}$                  &      8.69                     &      $1077_{-95}^{+99}$                          &      $13.9\pm 1.2$                            &       1                        &      $79_{-24}^{+11}$    &  -8.83/-8.87/-9.17   & 8,6,9        \\
247575958                &      CFHT Tau 12              &      $0.038_{-0.013}^{+0.013}$                  &      $0.065_{-0.005}^{+0.010}$                &      $0.79_{-0.14}^{+0.14}$                  &      4.90                     &      $913_{-112}^{+100}$                          &      \ldots                                   &       \ldots                   &      \ldots            &   \ldots  &   \ldots   \\
247589612                &      JH 112 A                 &      $0.90_{-0.17}^{+0.19}$                  &      $0.700_{-0.015}^{+0.020}$                &      $1.90_{-0.32}^{+0.28}$                  &      3.33                     &      $1581_{-217}^{+172}$                          &      \ldots                                   &       \ldots                   &      \ldots            &  \ldots   &  \ldots    \\
247592463                &      HP Tau                   &      $2.30_{-0.46}^{+0.50}$                  &      $0.9_{-0.2}^{+0.3}$                      &      $2.70_{-0.54}^{+0.53}$                  &      4.00                     &      $1531_{-336}^{+294}$                          &      $15.4\pm 1.6$                            &       2                        &      $29_{-11}^{+11}$  &   \ldots  &   \ldots   \\
247763883                &      GH Tau                   &      $0.62_{-0.13}^{+0.14}$                  &      $0.325_{-0.025}^{+0.025}$                &      $2.16_{-0.30}^{+0.30}$                  &      2.46                     &      $1572_{-168}^{+166}$                          &      $30.3\pm 0.7$                            &       1                        &      $44_{-10}^{+10}$    &  -7.92/-8.02/-8.90   & 7,10,6,8     \\
247764745                &      V807 Tau                 &      $1.39_{-0.36}^{+0.38}$                  &      $0.60_{-0.05}^{+0.05}$                   &      $2.50_{-0.38}^{+0.38}$                  &      3.81                     &      $1437_{-161}^{+154}$                          &      $13.6\pm 0.7$                            &       1                        &      $28_{-7}^{+7}$      &  -8.40/<-8.68   & 8,10         \\
247791801                &      GK Tau                   &      $0.94_{-0.10}^{+0.11}$                  &      $0.65_{-0.05}^{+0.05}$                   &      $2.03_{-0.19}^{+0.16}$                  &      4.98                     &      $1266_{-117}^{+93}$                          &      $18.7\pm 3.5$                            &       2                        &      $57_{-22}^{+23}$    &  -8.19   &   6,7,8,11       \\
247792225                &      GI Tau                   &      $0.74_{-0.12}^{+0.12}$                  &      $0.45_{-0.05}^{+0.05}$                   &      $2.08_{-0.23}^{+0.23}$                  &      5.75                     &      $1095_{-118}^{+118}$                          &      $12.7\pm 1.9$                            &       1                        &      $59_{-24}^{+24}$      &  -8.00/-8.02/-8.08   & 11,6,7,8       \\
247799571                &      HK Tau                   &      $0.37_{-0.06}^{+0.05}$                  &      $0.40_{-0.05}^{+0.05}$                   &      $1.54_{-0.16}^{+0.16}$                  &      4.47                     &      $1214_{-115}^{+115}$                          &      $21.8\pm 2.5$                      &    3                          &      $68_{-21}^{+21}$    &  -7.65   &   12   \\
247805410                &      FX Tau                   &      $0.56_{-0.16}^{+0.16}$                  &      $0.300_{-0.025}^{+0.050}$                &      $2.09_{-0.35}^{+0.35}$                  &      \ldots                   &      \ldots                                    &      $9.61\pm 0.19$                           &       1                        &      \ldots            &  -8.65   &  6,7    \\
247820821                &      \ldots                   &      $0.045_{-0.004}^{+0.004}$                  &      $0.075_{-0.003}^{+0.015}$                &      $0.85_{-0.07}^{+0.04}$                  &      3.70                     &      $1058_{-110}^{+39}$                          &      $18.4\pm 1.0$                            &       4                        &      $>64$     &   \ldots  &  \ldots           \\
247885481                &      \ldots                   &      $0.065_{-0.005}^{+0.005}$                  &      $0.15_{-0.03}^{+0.03}$                   &      $0.89_{-0.07}^{+0.07}$                  &      5.23                     &      $956_{-100}^{+100}$                          &      \ldots                                   &       \ldots                   &      \ldots            &   \ldots  &   \ldots   \\
247935061                &      GO Tau                   &      $0.25_{-0.08}^{+0.08}$                  &      $0.300_{-0.025}^{+0.050}$                &      $1.41_{-0.26}^{+0.26}$                  &      \ldots                   &      \ldots                                    &      $17.5\pm 9.7$                            &       2                        &      \ldots            &  -7.93/-8.33/-8.42   & 6,9,8     \\
248006676                &      JH 223                   &      $0.21_{-0.03}^{+0.03}$                  &      $0.275_{-0.025}^{+0.025}$                &      $1.34_{-0.14}^{+0.14}$                  &      4.52                     &      $1126_{-103}^{+107}$                          &      \ldots                                   &       \ldots                   &      \ldots            &   \ldots  &   \ldots   \\
248015397                &      ITG 34                   &      $0.033_{-0.010}^{+0.011}$                  &      $0.075_{-0.015}^{+0.015}$                &      $0.71_{-0.13}^{+0.15}$                  &      6.23                     &      $827_{-128}^{+143}$                          &      \ldots                                   &       \ldots                   &      \ldots            &   \ldots  &    \ldots  \\
248029373                &      DK Tau                   &      $1.14_{-0.18}^{+0.16}$                  &      $0.55_{-0.05}^{+0.05}$                   &      $2.33_{-0.24}^{+0.27}$                  &      5.78                     &      $1150_{-112}^{+131}$                          &      $17.5\pm 1.2$                            &  1                     &      $>60$     &  -7.42   & 6,7,11            \\
248046139                &      GM Tau                   &      $0.033_{-0.005}^{+0.006}$                  &      $0.065_{-0.005}^{+0.015}$                &      $0.73_{-0.09}^{+0.06}$                  &      4.49                     &      $961_{-126}^{+60}$                          &      $9.0\pm 2.0$                             &       5                        &      $41_{-15}^{+17}$    &  -8.60/-8.70   &  13,14       \\
248047443                &      IS Tau                   &      $0.63_{-0.13}^{+0.12}$                  &      $0.50_{-0.05}^{+0.05}$                   &      $1.86_{-0.23}^{+0.24}$                  &      6.93                     &      $1013_{-110}^{+116}$                          &      \ldots                                   &       \ldots                   &      \ldots            &  -7.91/-8.01   & 10,8      \\
\hline
\end{tabular}
\tablefoot{Stellar inclinations reported as lower limits have $\sin{i}>1$.}
\tablebib{
(1)~\cite{Nguyen12}; (2) \cite{Guedel07}; (3) \cite{Hartmann86}; (4) \cite{Kraus17}; (5) \cite{Mohanty05}; (6) \cite{Hartmann98}; (7) \cite{Muzerolle98}; (8) \cite{White01}; (9) \cite{Isella09}; (10) \cite{Hartigan03}; (11) \cite{Gullbring98}; (12) \cite{White04}; (13) \cite{White03}; (14) \cite{Herczeg08}.
}
\label{tab:dipper_prop}
\end{table*}
Dipper stars are most commonly low-mass T Tauri stars. Their periods of a few days are consistent with the range of rotational periods found in general for young, low-mass stars. There is consensus that the dips in the light curve are caused by dust; in the dusty disk warp scenario, the inner disk warp is located at the corotation radius. Moreover, regardless of the position of the occulting dusty structure, the temperature must be low enough for dust to be able to be present, that is, not to sublimate.

In the following section, the derivation of the different stellar parameters for dipper stars is discussed. In order to verify the conditions and to compute the radius and temperature at corotation, the masses and radii of the star are required. The discussion about each parameter follows in the next paragraphs.
\subsection{Effective temperatures and spectral types} Effective temperatures are listed in Table~\ref{tab:SpT_photometry} and were derived according to the SpT-$T_{\mathrm{eff}}$ conversions of \cite{Pecaut13}. The main sources for the spectral types are \cite{Esplin19} and \cite{Herczeg14}, who agree for almost all objects. When spectral subclasses were not explicitely listed in the conversion tables, the value of $T_{\mathrm{eff}}$ was linearly interpolated between the two closest subclasses. For spectral types later than M5, the temperature conversions by \cite{Herczeg14} were used.  

The uncertainty on $T_{\mathrm{eff}}$ follows the uncertainty on the spectral type: for spectral types up to K9, the uncertainty is one spectral subclass; for spectral types between M0 and M4, 0.4 subclasses; and for spectral types later than M4, 0.25 subclasses \citep{Herczeg14}. As a consequence, asymmetric error bars were computed from \cite{Pecaut13} for an earlier and later spectral type, respectively. If the later spectral type is missing at the end of the table, a symmetric error bar is produced. The systematic uncertainty linked to the choice of the models of up to 150\,K is not taken into account here. A discussion of the consequences on the derived stellar parameters is given in Sec.~\ref{subs:inclination}.\\

\subsection{\textit{VJHK} photometry} The light curves of young, accreting stars might be affected by contamination (e.g., by hotspots) in the blue band. Photometry in the \textit{V} and \textit{JHK} bands is therefore preferred. The dipper list was cross-matched with the Two Micron All-Sky Survey \citep[\textit{2MASS},][]{Skrutskie06} to obtain \textit{JHK} photometry. 

Because a photometric monitoring is the best way to determine the magnitude of variable stars, values reported in the photometric monitoring by \cite{Grankin07} are preferred for the \textit{V} band, where available. An inspection of the light curves presented for the objects studied here led to the choice of their $\overline{V}_{m}$ as best estimate for the brightness continuum. However, many of the dipper stars are absent from that collection, and other sources have to be included as well. Three photometric catalogs are common to most sources and include \textit{V}: GSC2.3.2 \citep{Lasker08}, NOMAD-1 \citep{Zacharias05}, and APASS \citep{Henden16}.\footnote{The entries in NOMAD are from the YB6 catalog, which consists of scanned plates and is thus considered a lower-quality source.} For individual stars, other additional measurements were retrieved from the literature. All available values for each star were inspected to verify whether any strong inconsistencies existed between the collections. More entries are available in the GSC catalog than in any other collection. If the star was not already reported in \cite{Grankin07} and there were no particular problems with the photometry, the entry of the GSC catalog was set as $V$ value in Table~\ref{tab:SpT_photometry}, when available.

In the few cases where these single measurements were highly different ($\Delta V > 1$\,mag) between the catalogs because of the intrinsic variability of young stars, special attention was given to those stars. One way to assess the plausibility of a $V$ band measurement is to verify that it does not lead to a strongly negative extinction. In order to verify that the individually selected visual magnitudes from different sources were self-consistent, the values of $V$ were compared with the \textit{Gaia} $G$ \citep{Gaia16,Gaia18} converted into $V$. No offset was found.

\subsection{Visual extinction} Most stars in Taurus are subject to at least some reddening. Many stars have moderate or strong infrared excesses, which can affect even the near-IR bands. At the same time, the fainter stars are harder to detect at shorter wavelengths. We therefore calculated the visual extinction by comparing the (V-J) colors as observed to those for the corresponding spectral type in \cite{Pecaut13}. \cite{Herczeg14} measured a quite high veiling at 7510\,{\AA} for some objects in this sample. For those with high veiling, extinction derived with photometry is not as accurate as with spectroscopy. For the stars with $r>0.1$, the $A_V$ computed by \cite{Herczeg14} from spectroscopy is preferred. 

We wished to minimize the risk of inconsistent values that is attached to very few, or just one, photometric measurement. For this reason, the available measurements of $A_V$ in the optical in the literature were collected and compared (see Table~\ref{tab:Av}). This allowed us to filter out several clear outliers that are flagged in Table~\ref{tab:Av}. For stars with lower veiling, the final $A_V$ is an average of the extinction computed here using $(V - J)$ and the other values in the literature  (a comparison of the values we derived and those in the literature is shown in Fig.~\ref{fig:app:Av}). The uncertainty was set to the rms of the different values, with a reasonable minimum of $0.3\,$mag. For measurements of $A_V < 0, A_v$ was set to 0 in order to be physical, as was the uncertainty if $A_V\ll 0$.

\subsection{Luminosities} The stellar luminosity was computed from the bolometric magnitude $M_{bol}$ as in
\begin{equation}
    L_*/L_{\odot} = 10^{-\frac{M_{bol} - M_{bol,\odot}}{2.5}}
,\end{equation}
where $M_{bol,\odot} = 4.25$\,mag. Because the brightness and bolometric corrections are mostly available for the \textit{J} band, the bolometric magnitude was derived according to
\begin{equation}
    M_{bol} = m_J - 5\log(p^{-1}) + 5 - A_J + \mathrm{BC}_J
,\end{equation}
where $m_J$ is the $J$-band magnitude, $p$ is the parallax (here from \textit{Gaia}), $A_J$ is the extinction in the $J$ band, and BC$_J$ is the bolometric correction for the $J$ band.
The bolometric corrections were retrieved from \cite{Pecaut13} and interpolated for intermediate spectral types. In case of spectral types later than M5 (not presented in the tables), the BC$_J$ was extrapolated. The computation of $A_J$ from $(J - K)$ is affected by IR excess. Thus, we preferred to convert $A_v$ into $A_J$ as $\frac{A_J}{A_v}=0.282$ \citep{Cardelli89}. 

The \textit{Gaia} parallax was not available for three objects (GH Tau, IS Tau, and FX Tau). We used the distance distributions in Taurus by \cite{Fleming19} for these stars. FX Tau and GH Tau are located in the B18 cloud, which contains members of the near population at $127.4\pm3.8$\,pc. A standard deviation of 7.9\,pc was applied as uncertainty for unknown parallaxes. IS Tau lies on the filament L1495, which contains members of both populations. Its luminosity was computed for the two mean distances and is shown with an arrow in Fig. \ref{fig:HR_diag}.  IS Tau appears to more likely be a member of the far population when we assume that it has the same age as the other dippers. In the case of Haro 6-37, the 2MASS measurement in the \textit{J} band is corrupted and no other measurements are available. The visual magnitude and the corresponding BC$_V$ were considered instead. 

The procedure for deriving uncertainties on the bolometric correction is the same as for the effective temperature.
A major issue for the determination of the luminosity uncertainty is the intrinsic photometric variability of dipper stars, which cannot be included here in absence of an extensive observation campaign. Nevertheless, for stars on their Hayashi tracks, the derivation of the mass depends much more on a precise spectral type than on luminosity (see Fig.~\ref{fig:HR_diag}). The uncertainty on the luminosity affects the derivation of age far more; the isochrones in Fig.~\ref{fig:HR_diag} accordingly are approximations. 

\subsection{Masses and radii} 
\begin{figure}[h]
    \centering
    \includegraphics[width = \linewidth]{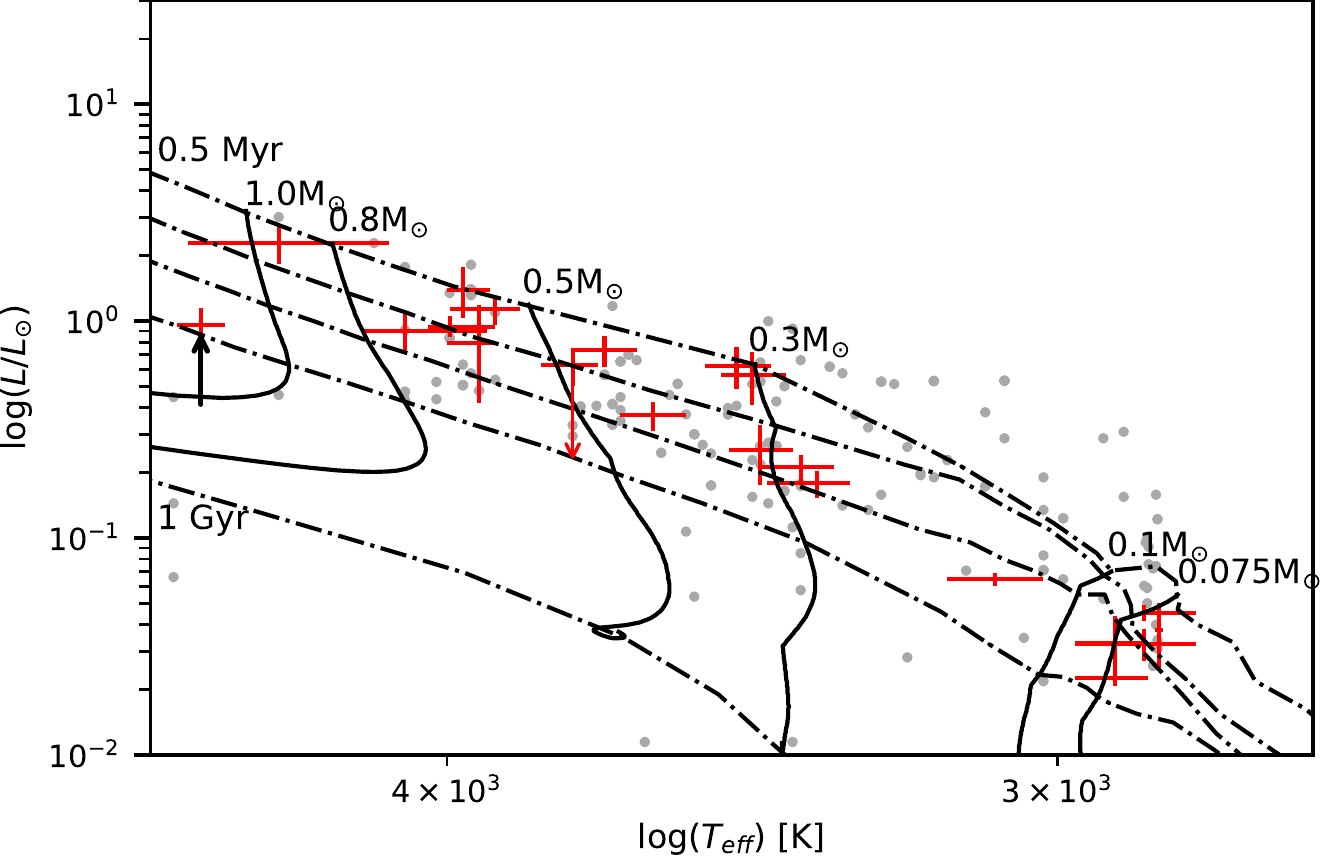}
    \caption{Evolutionary tracks (solid lines) and isochrones (dashed lines) from \cite{Baraffe15}. The brown dwarf limit lies at 0.073\,M$_{\odot}$. Isochrones from top to bottom: 0.5, 1, 2, 5\,Myr, and 1\,Gyr. The red points are the Taurus members classified as dippers, while the gray points are Taurus members as in \cite{Herczeg14}. HD 285893 (SpT F8) does not appear in this plot for the sake of readability because its derived $T_{\mathrm{eff}}$ is much higher than that of the rest of the sample. The stellar ages scatter around 1\,Myr, and with the exception of LkCa15 (marked with a black arrow) and HD 285893, they are on their Hayashi tracks and are still fully convective. }
    \label{fig:HR_diag}
\end{figure}

In order to derive the mass of the dipper stars in this sample, evolutionary models that reach the lowest end  of stellar formation are needed. We therefore preferred the tracks of \cite{Baraffe15} (Fig.~\ref{fig:HR_diag}) as they include M dwarfs. 

The aim here is to be self-consistent, therefore we derived all masses from the same model. The effective temperature is the dominant term for identifying the corresponding track for this sample. Thus, an identification of the corresponding tracks along the x-axis is the most efficient, with the only exception of the latest M dwarfs, for which the luminosity also plays an important role. Because the tracks are quantized, with a resolution of $0.1\,\mathrm{M}_{\odot}$ down to $m = 0.2\,\mathrm{M}_{\odot}$, $0.02\,\mathrm{M}_{\odot}$ for $0.1 < m \le 0.2\,\mathrm{M}_{\odot}$, $0.01\,\mathrm{M}_{\odot}$ for $m \le 0.1\,\mathrm{M}_{\odot}$, the error bars represent the distance between the upper or lower values of the data point and the closest track. The small error bars on the masses listed in Table~\ref{tab:dipper_prop} reflect the small error bars on the effective temperatures. The Hertzsprung-Russell diagram with some of the evolutionary tracks and isochrones is shown in Fig. \ref{fig:HR_diag}.

The dippers are uniformly distributed around 1\,Myr. With the exception of LkCa 15 and HD 285893, the stars identified as dippers have a mass lower than $1\,\mathrm{M}_{\odot}$, with a few M dwarfs close to the brown dwarf limit of $\sim 0.072\,\mathrm{M}_{\odot}$. The stars in \cite{Herczeg14} are cross-matched with the members and possible members of Taurus in \cite{Rebull20} and appear as gray points on the HR diagram. The spectral types are converted into temperatures with the same conversion as we applied to dippers. For spectral types later than K4, the conversion provided by \cite{Herczeg14} exhibits a systematic temperature offset of about $+60$\,K and up to $+130$\,K, while for earlier spectral types, the temperatures proposed by \cite{Pecaut13} are higher. When several luminosities are provided for a star, the values were averaged in the HR diagram. Because dippers are associated with circumstellar extinction events and their $V$ measurement might have been taken in the fainter state, it might be expected that they are less luminous than non-dipper stars. This does not seem to be the case here. Interestingly, many stars with a mass between 0.1 and 0.3\,M$_{\odot}$ lie well above the youngest isochrone. On the one hand, this might be an indication of the difficulty of providing precise evolutionary models for stars with a very low mass. On the other hand, for faint and accreting brown dwarfs, the accretion luminosity might dominate the faint stellar luminosity.  At the other extreme, with a $T_{\mathrm{eff}} = 6100$\,K and $L = 1.99\,\mathrm{L}_{\odot}$, HD 285893 almost lies on the ZAMS on the HR diagram. A more detailed discussion is given in the individual notes in Appendix~\ref{sec:app:individuals}. This \lq old\rq\ dipper might belong to a recently identified class of debris disk dippers \citep{Gaidos19,Tajiri20}, for which aperiodic extinction events have been attributed to the disruption of planetesimals. 

The stellar radii are derived according to the Stefan-Boltzmann law as 
\begin{equation}
    \frac{R}{R_{\odot}} = \left( \frac{L}{L_{\odot}} \right)^{\frac{1}{2}} \left(\frac{T_{\odot}}{T_{\mathrm{eff}}} \right)^2
\end{equation}
and are presented along with the masses in Table~\ref{tab:dipper_prop}. The uncertainties on $T_{\mathrm{eff}}$ and the luminosity are  propagated to the stellar radius.
\subsection{Radius and temperature at corotation} In the 
 accretion scenario, the dusty warp that obscures the star is located at corotation. The corotation radius defines the distance from the star at which the material of the Keplerian disk rotates with the same angular velocity as the star.  Because in a stable orbit, the gravitational potential of the star is equal to the centrifugal force, the corotation radius is derived as
\begin{equation} \label{eq:Rcor}
    R_{\mathrm{cor}} = \left( \frac{P}{2\pi} \right)^{\frac{2}{3}} (G M_*)^{\frac{1}{3}}
,\end{equation}
where $P$ is the stellar period.
Considering the simple approximation of no energy transfer between a dust grain situated at $R_{\mathrm{cor}}$ and the remaining disk, which reduces the effective irradiation by a factor 4 as in \cite{Bouvier99}, it is possible to derive the temperature at corotation as
\begin{equation} \label{eq:Tcor}
    T_{\mathrm{cor}} = 2^{-\frac{1}{2}}T_{\mathrm{eff}} \left( \frac{R_*}{R_{\mathrm{cor}}} \right)^{\frac{1}{2}}
.\end{equation}

The corotation radii and temperatures computed in this way are listed in Table~\ref{tab:dipper_prop}. The corotation radii extend to a few stellar radii, between 2.5 and 9\,$R_*$. The temperatures at corotation are $\sim 1000$\,K, and a few stars reach up to $\sim 1500$\,K. The error bars on this quantity are large because the uncertainties on the stellar radii are about 10\%-20\% for young stars.

\subsection{Inclinations}
\label{par:inclination}
In order to probe the capacity of the magnetospheric accretion model to account for dippers, the angle under which we observe the system is required because dippers cannot be seen close to face-on if the dusty part of the accretion column has to cross the observer's line of sight to produce dips in the light curve \citep[e.g.,][]{McGinnis15,Bodman17}. A more detailed discussion of this point is provided in Sec.~\ref{subs:inclination}.
 The stellar inclination is derived according to the formula
\begin{equation}
    v \sin{i} = \sin{i} \frac{2\pi R_*}{P}
\label{eq:vsini}
.\end{equation}
The convention we used to define the inclination is $0\degr$ for the star seen pole-on and $90\degr$ for edge-on. The $v\sin{i}$ values were retrieved from the literature (see Table~\ref{tab:dipper_prop}): the main sources were \cite{Nguyen12} and  \cite{Guedel07}, who retrieved their values for this sample from \cite{Rebull04}.

The uncertainty on $i$ grows with $\sin{i}$, thus higher inclination angles have larger uncertainties. Some stars (Haro 6-37, EPIC 247820821, and DK Tau) exhibit a  $\sin{i} > 1$, which probably indicates that something is amiss with the stellar parameters. In these cases, the minimum inclination angle is indicated as a lower limit.
The results are listed in Table~\ref{tab:dipper_prop}. For Haro 6-37, the period is uncertain. It is possible that the observed period is not correct or not directly related with the rotation period. For DK Tau, different periods have been reported in the literature because the light curve is complex, and the WPS shows an increasing period during the \textit{K2} observations. Because the stellar rotation period is not assumed to vary much, other physical phenomena might affect the dips. For instance, \cite{Grankin07} reported a long-term variability amplitude of $\Delta V \sim 1.8$\,mag for this star, which propagates to any inclination angle between 30$\degr$ and 90$\degr$ if taken into account as the uncertainty on $V$.

\begin{table*}[h]
\centering
\caption{Inclination of the stellar axis $i_*$ from this study compared to observations of the outer disk $i_{disk}$ from the literature.}
\tiny
\begin{tabular}{llllll}
\hline \hline
EPIC     &      Name     &      $i_*$    &      \multicolumn{2}{l}{\hspace{7mm}$i_{disk}$}   \\
&&[$\degr$] & \multicolumn{2}{l}{\hspace{7mm}[$\degr$]}    \\
&&& (1) & (Other)  \\
\hline
246929818                &      Haro 6-37                &      $>47$                             &      \ldots                                   &       \ldots                   \\
246942563                &      St 34                    &      \ldots                                    &      \ldots                                   &       \ldots                   \\
246989752                &      \ldots                   &      \ldots                                    &      \ldots                                   &       \ldots                   \\
247103541                &      HD 285893                &      \ldots                                    &      \ldots                                   &       \ldots                   \\
247520207                &      LkCa 15                  &      $79_{-24}^{+11}$                          &      \ldots                                   &       55 (2); $50^{+4}_{-6}$ (3)             \\
247575958                &      CFHT Tau 12              &      \ldots                                    &      \ldots                                   &       \ldots                   \\
247589612                &      JH 112 A                 &      \ldots                                    &      \ldots                                   &       \ldots                   \\
247589803                &      JH 112 B                 &      \ldots                                    &      \ldots                                   &       \ldots                   \\
247592463                &      HP Tau                   &      $29_{-11}^{+11}$                          &      $18.3_{-1.4}^{+1.2}$                     &       \ldots                   \\
247763883                &      GH Tau                   &      $44_{-10}^{+10}$                          &      \ldots                                   &       \ldots                   \\
247764745                &      V807 Tau                 &      $28_{-7}^{+7}$                          &      \ldots                                   &       \ldots             \\
247791801                &      GK Tau                   &      $57_{-22}^{+23}$                          &      $40.2_{-6.2}^{+5.9}$                     &       $73.0_{-59}^{+59}$  (4); $71_{-5}^{+5}$ (5)      \\
247792225                &      GI Tau                   &      $59_{-24}^{+24}$                          &      $44.0_{-2.0}^{+2.0}$                     &       \ldots                  \\
247799571                &      HK Tau                   &      $68_{-21}^{+21}$                          &      $56.9_{-0.5}^{+0.5}$                     &       $46.0_{-29}^{+29}$  (4); $51_{-2}^{+2}$ (5); $57.0_{-4}^{+4}$ (6)          \\
247805410                &      FX Tau                   &      \ldots                                    &      \ldots                                   &       $40_{-4}^{+4}$ (5)               \\
247820821                &      \ldots                   &      $>64$                             &      \ldots                                   &       \ldots                   \\
247885481                &      \ldots                   &      \ldots                                    &      \ldots                                   &       \ldots                   \\
247935061                &      GO Tau                   &      \ldots                                    &      $53.9_{-0.5}^{+0.5}$                     &       \ldots                   \\
248006676                &      JH 223                   &      \ldots                                    &      \ldots                                   &       \ldots                   \\
248015397                &      ITG 34                   &      \ldots                                    &      \ldots                                   &       \ldots                   \\
248029373                &      DK Tau                   &      $>60$                             &      $12.8_{-2.8}^{+2.5}$                     &       $41.0_{-11}^{+11}$  (4); $20_{-5}^{+5}$ (5); $27.0_{-9}^{+9}$ (6)          \\
248046139                &      GM Tau                   &      $41_{-15}^{+17}$                          &      \ldots                                   &       \ldots                   \\
248047443                &      IS Tau                   &      \ldots                                    &      \ldots                                   &       \ldots                   \\
\hline
\end{tabular}
\tablebib{
(1)~\cite{Long19}; (2) \cite{vanderMarel15}; (3) \cite{Thalmann14}; (4) \cite{Akeson14}; (5) \cite{Simon17}; (6) \cite{Harris12}.
}
\label{tab:incl}
\end{table*}
\subsection{Mass accretion rates and accretion regime}
The mass accretion rates of the dippers were collected from the literature. When available, they are listed in the order of magnitude of $10^{-8}\,\mathrm{M_{\odot}yr^{-1}}$ (Table~\ref{tab:dipper_prop}). This is in agreement with the observed mass accretion ranges for T Tauri stars \citep{Gregory06}. Exceptions are Haro 6-37, HK Tau, and DK Tau, which are stronger accretors of about $10^{-7}\,\mathrm{M_{\odot}yr^{-1}}$. LkCa 15 has a very low mass accretion rate for its mass, but it is also a transition object, and a lower mass accretion rate is expected at this stage. In general, quasiperiodic dippers are expected to be generated in a stable accretion regime (see Sec.~\ref{subsec:width}) and aperiodic dippers instead in an unstable regime \citep{McGinnis15}, which could be caused by a phase of enhanced mass accretion. The complex light curve of DK Tau, one of the strongest accretors, might be an example of unstable accretion. However, for the dipper sample presented in this paper, we found no correlation between mass accretion rate and periodicity.

\section{Discussion}
\label{sec:discussion}
Several possible mechanisms for the origin of dippers have been presented in the literature, including dusty disk warps, dusty winds, and disk vortices. In the study of the protoypical dipper AA Tau, \cite{Bouvier99} proposed that a magnetic dipole, tilted with respect to the stellar rotation axis, might contribute to an inner disk distortion. This can produce a dusty, optically thick disk wall that can account for the photometric variation, if observed at high inclination. This model was further explored by \cite{McGinnis15}, using as free parameters the stellar inclination and location, height, and azimuthal extension of the dusty wall. In the dipper sample of NGC 2264, which has a similar size as that of  Taurus, inclinations down to $\sim$50$\degr$ could be measured, with large uncertainties. \cite{McGinnis15} were able to fit the light curves with a dusty warp with a mean inclination of $\sim$70$\degr$. Dippers close to edge-on are assumed to be undetectable in photometry because the disk would block the line of sight toward the stellar photosphere.

The first parameter we discuss for the compatibility with the model therefore is the inclination of the star and the disk (Sec.~\ref{subs:inclination}). For stable accretion, the magnetospheric truncation radius is close to corotation, and the photometric periods of dippers provide evidence for the dusty warp to be located at corotation. One other condition for the validity of the model is thus that the environment at corotation must be cold enough to avoid dust sublimation (Sec.~\ref{subsec:disc:t_cor}). It would be helpful to derive the truncation radius to compare it with the corotation radius; unfortunately, not enough precise magnetic field measurements are available for this purpose. \cite{Bodman17} provided a rough estimate for the mass accretion rate and the magnetic field strength. However, this estimate has a very large uncertainty.

This scenario was challenged by the discovery of dippers seen in millimeter wavelengths in a full range of inclination angles \citep{Ansdell2016b,Ansdell20}, on which the hypothesis was founded that the inner and outer disk might be misaligned. Binarity can be invoked as a cause of misalignement in a circumbinary disk \cite[e.g.,][]{Facchini13,Facchini18,Franchini19}. This opens the possibility of finding a dipper with a low outer disk inclination. Nonetheless, the inclination of the inner region must be reasonably high to detect dips. Two examples of well-studied dippers with a misaligned inner disk are LkCa 15 and RX J1604.3-2130A \citep{Alencar18,Sicilia20}.

Another way of lifting dust away from the disk midplane is through dusty disk winds. These are spatially more extended than the base of accretion funnel flows, and might shield the stellar brightness at lower viewing angles. Magnetic field lines inclined with respect to the disk symmetry axis ($\theta \geq 30\degr$) are able to accelerate matter away from the disk. Dust, dragged away with gas, is able to survive in a disk-driven wind at temperatures of several $10^3$\,K because heating the dust through collisions with gas particles and sputtering is highly inefficient \citep{Tambovtseva08}. 
A high mass-accretion rate and strong magnetic field (kGauss) are required to launch such a disk wind \citep{Miyake16,Labdon19}, and only the small dust grains are dragged by the wind, while the large-size grains remain in the disk midplane \citep{Miyake16}. The resulting floating dust in the disk is able to produce optical fading events and NIR brightening. \cite{Bans12} modeled the NIR $3\,\mathrm{\mu m}$ bump as produced by absorbed stellar radiation in a dusty wind.
However, it seems that such disk winds are observed under a rather high inclination $\sim70\degr$ \citep{VInkovic20}, much higher than what has been invoked to explain low-inclination dippers \citep[e.g.,][]{Ansdell20}.

It is also possible to lift dust from the midplane through vortices in the disk that are caused, for example, by Rossby waves. However, the occultations caused in this way would have a very small amplitude and might thus account only for a small fraction of aperiodic dippers \citep{Stauffer15}.

\subsection{Inclination of star and disk}
\label{subs:inclination}

In this section, we investigate whether the stellar inclination agrees with the inclination of the outer disk and if this is compatible with an AA Tau-like star. For most of the objects considered here, a moderate to high inclination is clear despite the large error bars.

GH Tau and GM Tau are compatible with an inclination of $\sim$50$\degr$, which is at the lowest end of the magnetospheric accretion scenario. This suggests that the magnetic field axis is highly tilted with respect to the stellar spin axis. 
In the case of HP Tau and V807 Tau, the proposed dusty warp scenario for the production of dippers fails to explain fading events seen at such a low inclination. We tested whether the photometric variability might significantly affect the estimation of the luminosity and propagate up to the derived stellar inclination. None of these stars shows an amplitude in the \textit{K2} light curve that would be large enough to explain the discrepancy. Nevertheless, for HP Tau, a long-term amplitude variation of $\Delta V \sim$1$\,$mag was reported by \cite{Grankin07}. This can be roughly converted into an upper limit of $i \sim$50$\degr$, which is still at the lowest limit for the magnetospheric accretion model. For V807 Tau, which is a triple system, an overestimation of the brightness of the primary is possible. \cite{Schaefer12} estimated that V807 A could be 0.5\,mag fainter than the total system in the $J$ band and fit an extinction of 0 to the system. From these parameters, a higher stellar inclination of $40\degr$ can be derived, which is more plausible. This is not sufficient to explain the discrepancy, however. A stellar radius of $\sim 1.5\,\mathrm{R}_{\odot}$ would be required to reach an inclination angle of at least $50\degr$. 

It is thus more probable that some other phenomenon contributes to the quasiperiodic extinction events for these stars that are inconsistent with a sufficiently high inclination. Dusty disk winds also need high inclination angles to be observable \citep{VInkovic20}, and the mass accretion rate of V807 Tau is too low (Table~\ref{tab:dipper_prop}). The small amplitude of $\sim 0.1$\,mag might indicate vortices caused by Rossby wave instabilities. However, given the clear periodicity of the light curve, this mechanism can be discarded.  Moreover, the almost constant shape of the dips means that the structure occulting the star is probably stable, not very large, optically thick, and it might be that just a small fraction of it is seen by the observer under this low inclination. 

One  more possibility in the magnetospheric accretion scenario would be that the dust high in the accretion column increases its optical depth immediately before sublimation because evaporation and free-fall timescales are similar \citep{Nagel20}. This might imply that an optically thick part of the accretion column could cross the observer's line of sight, thus requiring a lower inclination to see dips. 

We compared our dipper sample to inclinations of outer disks that were directly measured at millimeter wavelengths (Table~\ref{tab:incl}). \cite{Akeson14} derived the disk inclination in the image plane by fitting the clean continuum maps with 2D Gaussians. This results in larger error bars than with the methods used by other authors because the continuum map is reconstructed from the visibilities. In general, the inclination of the outer disk is slightly lower than the inclination derived for the inner region, with the exception of DK Tau, which seems strongly misaligned with the outer disk, although this value has to be considered with caution (see Sec.~\ref{sec:stellar_params}). 

\cite{Appenzeller_2013} pointed out that the inclination of CTTSs derived from $v\sin{i}$ are not precise for $i>30\degr$ because the uncertainty grows with $i$. The inclination might also be overestimated when not all line broadening effects are properly considered. For a sample of known CTTSs in different regions, $i_*$ and $i_{disk}$ are correlated, but the inclination derived from rotation is $19\degr$ higher on average \citep{Appenzeller_2013}. This trend is also observed in Taurus for HP Tau, GK Tau, GI Tau, and HK Tau. Nevertheless, the recent observations of misaligned inner disks support the possibility that this systematic also has a physical basis. 

Another systematic uncertainty on the stellar inclination is produced by the use of conversion tables between spectral type and effective temperature, which are still not precise for young stars. The use of different conversions can lead to discrepancies of about 150\,K, which for a late-type star with $T_{\mathrm{eff}} \sim$ 3000-3500\,K means an additional uncertainty of $\sim$5\%. Propagated to $R_*$ and $\sin{i}$, this final uncertainty of $\sim$10\% corresponds to about $5\degr$ for $i_* < 30\degr$ or $10\degr$ for $i_* > 30\degr$. Thus, the stellar inclinations might be higher than derived by about 10$\degr$  .

No extreme cases of face-on dippers, as reported by \cite{Ansdell2016b}, are observed in this sample. The stellar inclination was derived for 11 dippers (50\%), and inclinations of the outer disk are available for 8 (36\%) dippers. The dippers for which it was possible to derive a stellar inclination are compatible with the disk warp scenario (thus 9/11), with the exception of HP Tau, whose stellar parameters might be significantly affected by a strong overall photometric variability, and V807 Tau, for which a different scenario is required, even after decomposing the primary and secondary brightness of the system. In general, the stellar inclinations of Taurus dippers are lower than those of dippers in NGC 2264 \citep{McGinnis15}, for instance.

\subsection{Temperature at corotation and dust survival}
\label{subsec:disc:t_cor}
Given the assumption that dusty material at corotation causes the dips, it is interesting to compare the temperature at cototation with dust sublimation temperatures in order to confirm or exclude certain characteristics of the dust grains. The sublimation temperature, $T_{\mathrm{sub}}$, of the grains depends on the gas pressure, latent heat, and molecular weight more than on the stellar parameters \citep{Kobayashi11}. The full range of $T_{\mathrm{cor}}$ in Table~\ref{tab:dipper_prop} extends from 800\,K to 1600\,K, with error bars of about 100\,K. The only materials able to withstand these temperatures are olivine, pyroxene, and iron \citep[see their Table~3]{Pollack94}. Moreover, the sublimation temperature increases with the gas density; stars with $T_{\mathrm{cor}}>1400$\,K exclude gas densities $<10^{-8}\,\mathrm{g~cm^{-3}}$. The stars with the highest temperatures at corotation in Table~\ref{tab:dipper_prop} can only host dust for even higher gas densities $>10^{-6}\,\mathrm{g~cm^{-3}}$. Because it is currently difficult to resolve the inner disk rim for low-mass YSOs, models for this region have only been developed for intermediate-mass stars \citep[e.g.,][]{Isella05,Tannirkulam07}. It is thus of interest to better constrain these properties for CTTSs in future studies to verify the conditions for dust survival.

\begin{figure}
    \centering
    \includegraphics[width=\linewidth]{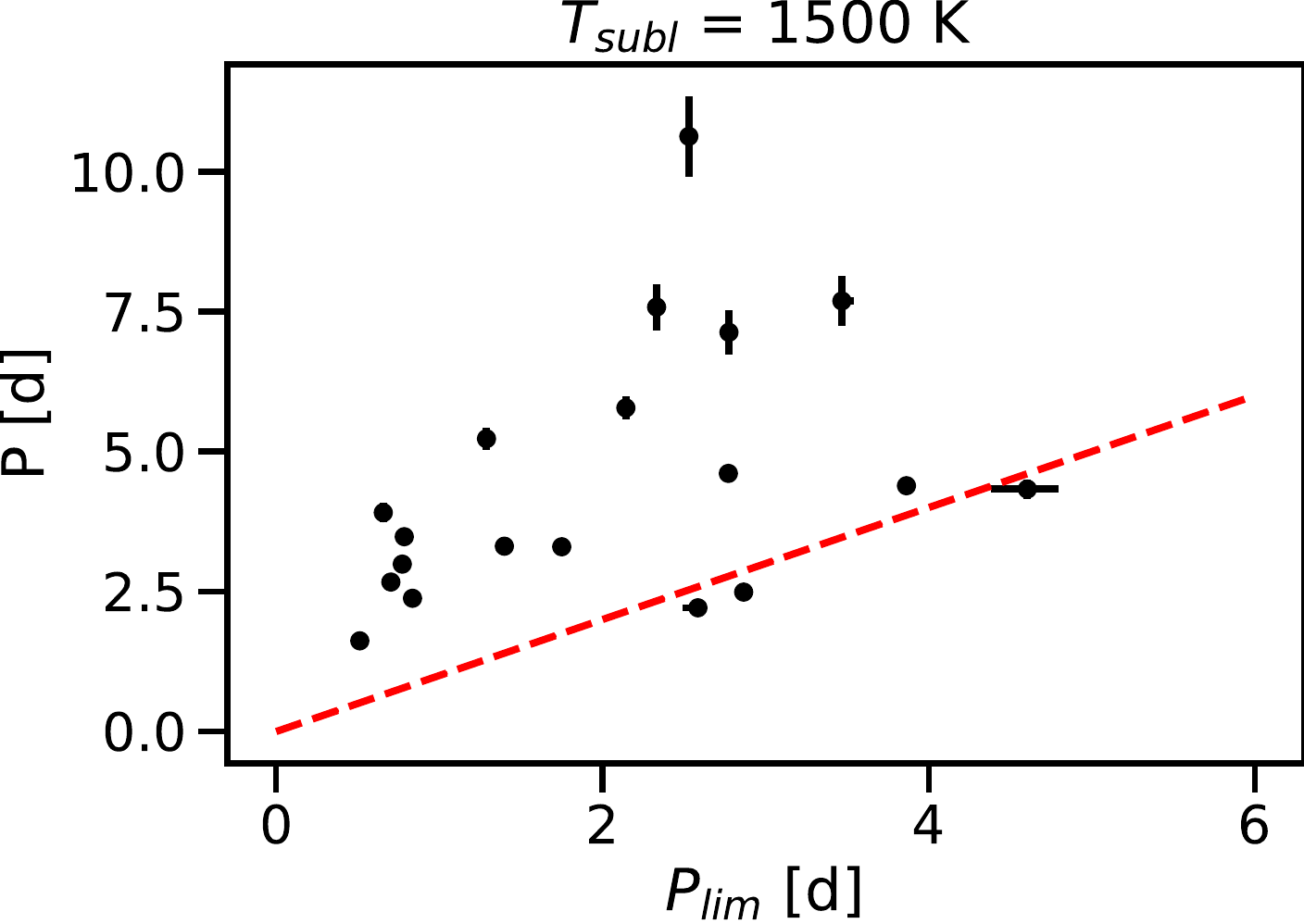}
    \caption{Period of the dippers in comparison to the minimum rotation period (red line) required for a temperature $\leq$1500\,K at corotation. If the star rotates faster, the inner disk is purely gaseous at corotation.} 
    \label{fig:p_lim_tcor}
\end{figure}

Setting a critical temperature for dust $T_{\mathrm{subl}} = 1500\,K$, it is possible to derive from Eq.~\ref{eq:Tcor} and Eq.~\ref{eq:Rcor} the minimum stellar rotation period for a given mass and radius that permits a sufficiently cold environment for dust to survive at corotation as
\begin{equation} \label{eq:Plim}
    P_{lim} = 2^{-\frac{1}{2}} \pi \left( \frac{T_{\mathrm{eff}}}{T_{\mathrm{subl}}} \right)^3 R_*^{\frac{3}{2}} (G M_*)^{-\frac{1}{2}}
.\end{equation}
Fig.~\ref{fig:p_lim_tcor} shows that all periods of dipper stars are well above this limit. For the few stars close to the limit, the condition is satisfied if a $T_{\mathrm{subl}} = 1600$\,K is assumed. Previous observations of dippers in $\rho$~Oph and Upper Sco \citep{Bodman17} also indicate that it is possible for dust to survive at the corotation radius of dipper stars.

\subsection{Dippers in Taurus and in other clusters}
\label{subsec:other_clusters}
The star-forming regions that are known so far to host dippers include NGC 2264, which is 2-3\,Myr old \citep{Alencar10,Cody14,McGinnis15,Stauffer15}, Orion, which is~3\,Myr old \citep{Morales11}, $\rho$ Oph and Upper Sco, which are 1 and ~10\,Myr old, respectively \citep{Ansdell16a,Hedges18,Cody18,Rebull18}, and Taurus \citep{Rebull20}. A feature that is shared by all of these regions is the late spectral type, K to M, of the identified dippers. A common explanation for this is the longer pre-main-sequence phase of late-type stars, which allows the disk to be present for a longer time, thus increasing the probability of observing a dipper. Moreover, the lower surface temperature enables dust to be present close to the stellar magnetosphere. In Taurus, as in $\rho$ Oph and Upper Sco, the occurrence of M-type dippers is higher than that of K-type dippers because stars of lower mass always exist in a larger number.

For the fractional rate at which dippers are found among YSOs, the statistics differ from region to region. The main reasons for this are the different counting methods (i.e., fraction of disk-bearing stars vs. all members of the cluster), the different selection techniques, and the target selection process of the survey, which is not necessarily optimized for finding dippers. For $\rho$~Oph and Upper Sco, dippers represent about 20\% of disked YSOs \citep{Hedges18}, which was increased to 30\% by \cite{Cody18}. These authors also pointed out that in the older Upper Sco, a higher dipper fraction is observed among disked stars. \cite{Alencar10} claimed that 30\% to 40\% of disked stars in NGC 2264 might be dippers, depending on whether a sample of thick or anemic inner disks is considered. This amount was found to be 20\% by \cite{Cody14}, who also studied NGC 2264. An exception is Orion, where only 5\% of the stars are dippers. This might be due to the photometric accuracy of the survey and a biased target selection \citep{Morales11}, as well as the fact that they did not consider a disked-star sample. In Taurus, considering the additional 12 dippers in Table~\ref{tab:other_dippers}, the occurrence rate is 19\% of all the members observed with $K2$, and 31\% of the members hosting a disk. 
The ratio of quasiperiodic to aperiodic dippers is about 1:1 for NGC 2264, $\rho$ Oph, and Upper Sco. In Orion, only one-third of the dippers is quasiperiodic. In contrast, most dippers appear to be quasiperiodic in Taurus. The sample analyzed in this study might be unusual in this sense (only 3 aperiodic and 3 uncertain dippers out of 22 in Table~\ref{tab:periods_ampl}); it must be noted that for only 2 of the 12 additional dippers in Table~\ref{tab:other_dippers} the reported periods are most probably linked to quasiperiodic dips. The predominant periodicity in the light curve for the remainder is caused by cold stellar spots, while the extinction events are stochastic. Only 17 out of the 34 presented dippers are strictly quasiperiodic, which brings the ratio to about 1:1. The periods marked as uncertain in Table~\ref{tab:periods_ampl} are not counted as quasiperiodic here.

\subsection{Dip width and period}
\label{subsec:width}
Previous studies investigated whether dip width, dip amplitude, and/or dip width are correlated for dippers. No such correlation has been found so far \citep[e.g.,][]{Bodman17}, except for the class of short-period narrow dips presented by \cite{Stauffer15}. For dip width and amplitude, a dependence would be a constraint on the geometrical properties of the dusty warp as presented by \cite{Bouvier99}. The amplitude of the eclipses depends not only on the viewing angle, but also on the vertical extent of the dusty material occulting the star. On the other hand, the dip width delivers an estimate of its azimuthal extent. 

\begin{figure}
    \centering
    \includegraphics[width=\linewidth]{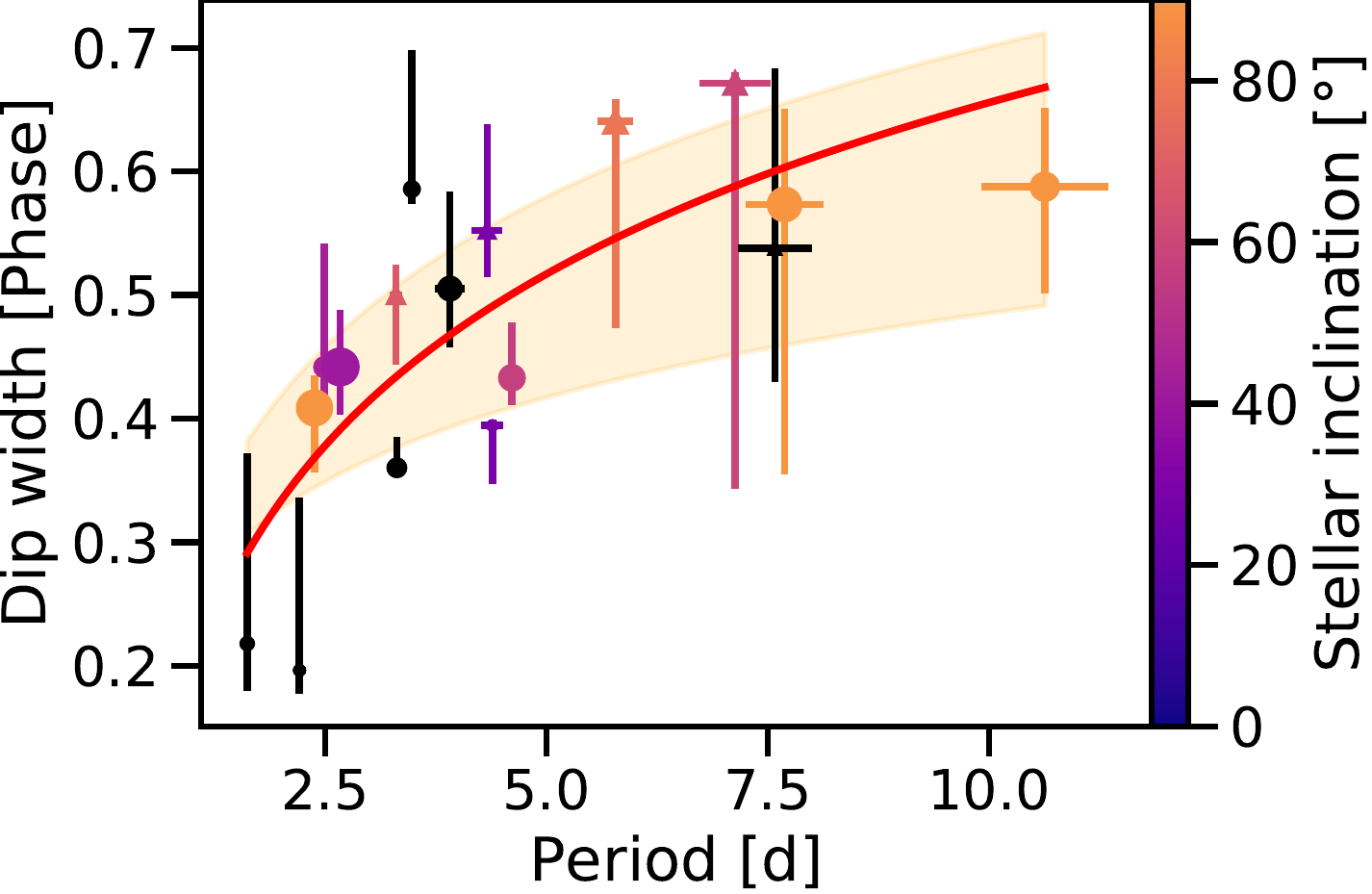}
    \caption{Dependence of dip width in phase units and period of the quasiperiodic stars in the dipper sample. The error bars on dip width and period correspond to those listed in Table~\ref{tab:periods_ampl}. A logarithmic fit $a\log{x} + b$ is performed just on the data points (red line), and for upper and lower limits of the dip width (orange interval). The double-dipped stars are represented as triangles. The size of the markers is proportional to the square root of the dip amplitude (between 0.05 and 1.3\,mag). Two stars for which the dip width derived in this way is not reliable are not represented in this plot. The stellar inclination in degrees is coded as colors of the data points, when available. Although the exact coefficients of the fitting function cannot be constrained from this small sample, $W$ correlates with $\ln{P}$ with Pearson's $r=0.77$. No significant correlation with the dip amplitude can be measured.}
    \label{fig:disc:width_period}
\end{figure}

For our sample of Taurus dippers, the quasiperiodic dippers for which it was possible to derive a reliable width are shown in Fig.~\ref{fig:disc:width_period}. The data suggest a logarithmic trend of the dip width and the period, with a high scatter due to the intrinsic variability of dippers. It is remarkable that many dippers are obscured $\gtrsim$50\% of the time, suggesting that the azimuthal extent of the dusty structure must be larger than 180$\degr$ around the star.
According to the geometrical constraints of a dusty warp, the observed dip width should grow with the inclination. The inclination values listed in Table
~\ref{tab:incl} are represented in Fig.~\ref{fig:disc:width_period} as a color code of the data points. The amplitude of the dips is proportional to the size of the data points as $2 \sqrt{\frac{A}{A_{min}}}$, where $A_{min}$ is the smallest amplitude of the sample. No correlation is observed within the inclination and the dip width, nor between inclination and dip amplitude.
Toward the higher end of the widths, the  dip is often double- or multi-peaked, and its shape and position vary from phase to phase. This is the case for HP Tau, LkCa 15, GI Tau, JH 112 A, HK Tau, and IS Tau (see Sec.~\ref{subsec:double_peak}). The large error bars on the dip width for these double- or multi-peaked stars reflect the difference between considering the width of the full occultation and the width of the main dip.

\begin{figure}
    \centering
    \includegraphics[width=\linewidth]{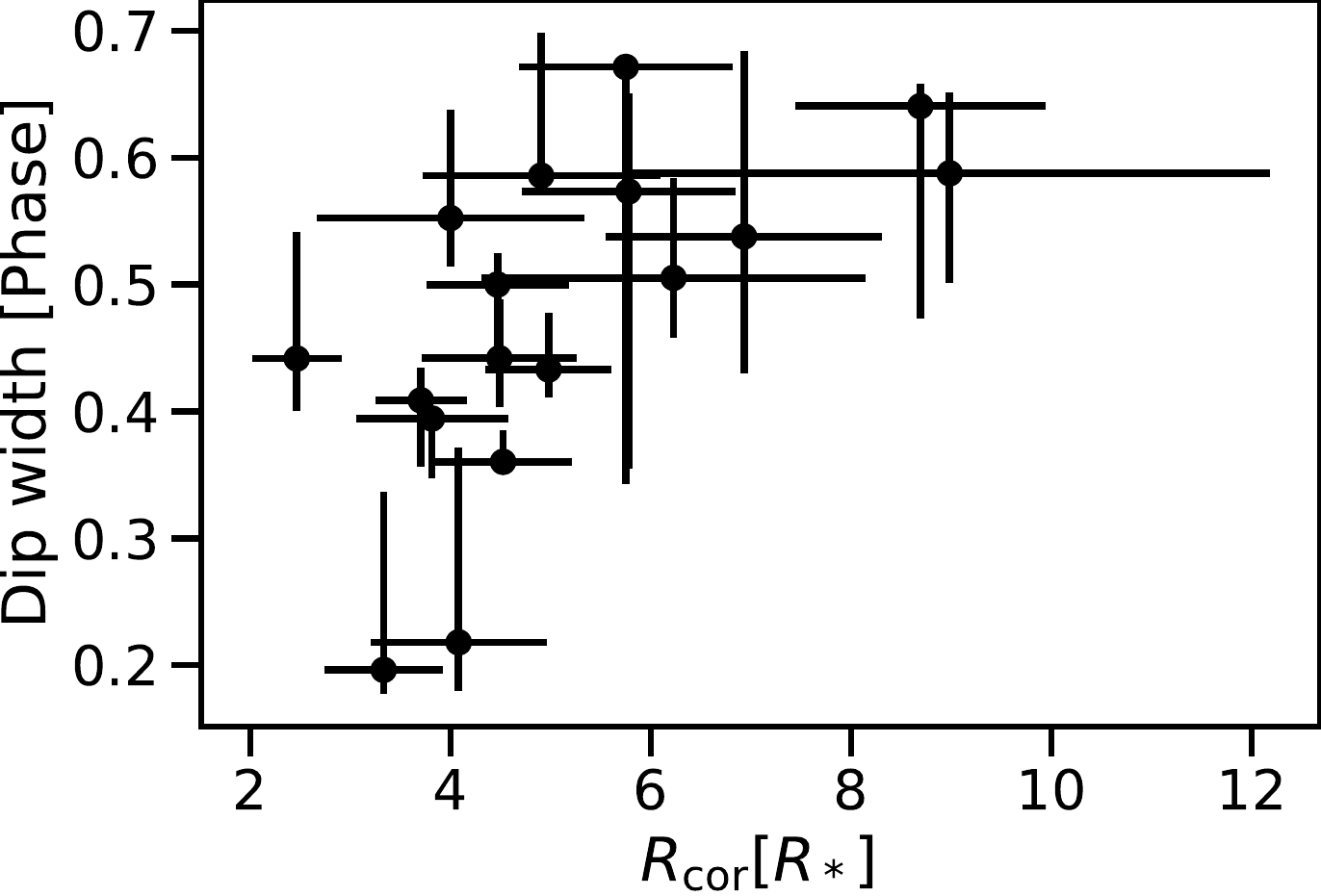}
    \caption{Dependence of the dip width in units of phase and the corotation radius. Although weaker (Pearson's $r$ = 0.63), a correlation of $W$ and $\ln{R_{\mathrm{cor}}}$ is also evident here. The lower correlation can be explained by the fact that $R_{\mathrm{cor}}$ is a function of both $P$ and $M_*$.}
    \label{fig:disc:widths_rcor}
\end{figure}

Although the dependence can be best described with a logarithmic curve, the size of the sample is too small to draw a solid conclusion on the nature of the best-fitting curve. It is also possible that two different regimes are present, such as a linear trend for short periods, followed by a plateau for longer periods. The two quantities dip width ($W$) and the natural logarithm of the period ($\ln{P}$) are correlated with a Pearson correlation coefficient $r = 0.77$ and a p-value of $2.8\cdot 10^{-4}$. A logarithmic, rather than linear, trend is expected because the star cannot be eclipsing close to 100\% of the time, thus flattening the curve for longer periods. The two stars in the lower left corner are crucial for this correlation, however. EPIC246989752 and JH 112 A (see Fig.~\ref{fig:app:phased_lc} for the dip width) fall in the category of short-duration quasiperiodic dips \citep{Stauffer15}, with the interesting case of JH 112 A being a transient dipper. The derived dip width describes the dip as observed in the folded light curve well. Moreover, the upper error bar is very generous because the binned light curve $+\sigma$ is almost flat. We can therefore exclude that the dip width for these two stars is overestimated. \cite{Stauffer15} identified a particular class of Gaussian-shaped, narrow, and short-periodic dips. The definition of ``narrow'' is set where the ratio of the FWHM of the fit Gaussian and the period is lower than 0.25, while the narrow dips in NGC 2264 have a FWHM-to-period ratio of 0.15 on average. In Taurus, EPIC 246989752, and JH 112 A would correspond to this category, with a FWHM-to-period ratio of 0.22 and 0.20, respectively. Both exhibit a Gaussian shape. The short-period dippers in NGC 2264 tend to have narrow eclipses for later spectral types \citep{Stauffer15}. Here, this is true for EPIC 246989752, which is an M dwarf, while JH 112 A has a spectral type of K5.5. We could not find a correlation between dip width and spectral type in the dipper sample. 

Another star that can be associated with the population defined by \cite{Stauffer15} is EPIC 247885481 (Sec.~\ref{subsec:disc:diff_period}). It does not appear in Fig.~\ref{fig:disc:width_period} because the width of the dips cannot be constrained reliably in a standardized way. The amplitude is very small compared to the noise, and the dips do not clearly appear in each phase (see Sec.~\ref{subsec:disc:diff_period}). The star is an M dwarf and is therefore compatible with the group identified in NGC 2264. 

A correlation between $R_{\mathrm{cor}}$ and the dip width in units of phase is presented in Fig.~\ref{fig:disc:widths_rcor}. Although weaker (Pearson's $r=0.63$), a behavior similar to that of Fig.~\ref{fig:disc:width_period} is evident. This is expected because the corotation radius as derived in Eq.~\ref{eq:Rcor} is a function of the stellar period. In this case, the correlation is evident even without considering the two data points in the lower left corner, which only argue for a logarithmic instead of a linear dependence.

It is of interest to verify whether this correlation of dip width with period is an implication of existing models and if the observed dip widths are compatible with the mechanisms that have been proposed to explain dippers. In order to generate a quasiperiodic dipper, the most stable configuration of the magnetic field seems to be an inclined dipole in the framework of magnetospheric accretion \citep{Romanova13}. Two broad, stable, and symmetric matter streams form from the disk and are accreted as funnel flows onto the stellar surface, producing an hotspot in each hemisphere. The observer can see the star obscured either by the dusty warp or even by the funnel flow itself. In this case, the extent of the magnetosphere nearly coincides with the corotation radius, $R_\mathrm{T} \approx R_{\mathrm{cor}}$. In case of $R_{\mathrm{cor}} < R_{\mathrm{T}}$, the inner region of the disk rotates with $\omega > \Omega_K$ (where $\Omega_K$ is the Keplerian velocity at corotation) and the star is in the propeller regime. The magneto-centrifugal forces lead to the ejection of disk material into an outflow \citep{Romanova15}. During phases of enhanced mass accretion, in which the dipole component of the magnetic field decreases, the disk compresses the magnetosphere and reduces $R_\mathrm{T}$ \citep{Romanova13}. For $R_{\mathrm{T}}<R_{\mathrm{cor}}$, the star is in the accretion regime and the magnetosphere rotates more slowly than the inner disk ($\omega < \Omega_K$).
In case of a small or no misalignment between the stellar spin axis and the magnetic moment, the matter from the inner disk accumulates at $R_T$ and the accretion instead occurs through unstable accretion tongues \citep{Romanova13}, which are taller and thinner than funnel flows that reach the stellar surface close to equator. They can be observable as irregular hotspots in the \textit{UV}. As a result, the light curve is rather stochastic and bursting. An unstable accretion regime has been invoked to explain aperiodic dippers \citep[e.g.,][]{McGinnis15}.

The extent of the corotation radius directly depends on the stellar rotation period. For slow rotators, the corotation radius is located at a larger distance from the star, and the inner disk therefore rotates faster than the star. This facilitates the accretion through instabilities \citep{Romanova13,Kulkarni08}. 

An observer would expect broad dips to be quasiperiodic because the unstable accretion tongues are rather narrow and tall and would not be able to obscure the star for a significant fraction of time \citep{Romanova08}. The aperiodic dippers observed in Taurus indeed exhibit rather narrow dips in their light curves.

The quasiperiods observed in dippers might also be linked to a configuration in the accretion regime.
In this case, trapped density waves in the inner disk can produce a warp that rotates more slowly than the star and beyond the corotation radius, with a period that might vary over time. In this case, the simulations show that the inner disk is tilted and also rotates more slowly \citep{Romanova13}. This could explain the changing, rather long period observed in the WPS of the \textit{K2} light curve of DK Tau, which is associated with an outer disk seen at low inclination. However, this would not suffice to explain why the star is obscured in the first part of the light curve. The slowly rotating warp should also have a smaller height than a fast-rotating warp, and DK Tau has an amplitude of 1\,mag.

A correlation of period and dip width thus suggests that slow rotators are more probably occulted by more azimuthally extended dusty structures. This aspect has not been directly considered in simulations to date. In the scenario of magnetospheric accretion, this might imply a broader funnel flow for slow rotators. 

For T Tauri stars of late spectral type, the internal structure is either fully convective or has small radiative cores \citep{Gregory12}. The HR diagram of Taurus dippers shows that with the exception of LkCa 15, all stars are on their Hayashi tracks and fully convective. In this case, the magnetic field is expected to be axisymmetric and the dipole component dominates, as is the case for AA Tau.

\cite{Vidotto14} found a correlation of magnetic flux and rotation period for accreting PMS stars. This implies that stars with a simple magnetic field (i.e., dipole) and with the strongest magnetic fields are also the slowest rotators and truncate the disk at larger distances $R_T$. The star-disk interaction therefore affects the magnetic field, which in turn affects the stellar rotation. We speculate that the large-scale magnetic field topology might cause the correlation of dip width and period. A stronger magnetic field might have a stabilizing effect on a large warp.

\section{Conclusion}
\label{sec:conclusion}
We have studied a sample of 179 YSOs in Taurus, which are members or possible members of the cluster \citep{Rebull20}. We identify a total of 34 dippers, 22 of which are dippers not dominated by another type of variability. This makes up $\sim$\,20\% of the high-confidence and possible Taurus members and $\sim$\,30\% of the disk-bearing Taurus members as observed with $K2$. 

The observations in Taurus highlight again how ephemeral dippers are. A striking example is the dipper prototype AA Tau, which is now in a fainter state and is no longer classified as a quasiperiodic, prototypical dipper. Dipper light curves can persist over timescales of a few years or even a few days. This strongly suggests that the observed occurrence rates are a lower limit to their true occurrence among CTTSs.

The ratio of quasiperiodic to aperiodic dippers is 1:1, although most of the dippers studied in detail are quasiperiodic. The large majority of aperiodic dippers observed in Taurus is dominated by another type of variability, mainly that due to cold spots. As found in other surveys, dipper stars are of late spectral type K or M because the lower surface temperature allows dust to survive in the inner disk and create the dips. The dipper stars of the sample are fully convective, low-mass ($<1\,\mathrm{M_{\odot}}$) stars down to the brown dwarf limit, with the presence of a probable debris disk dipper.

Supporting previous dipper surveys, the observed periods are in the range of the rotation periods of low-mass CTTSs. In the sample, a transient period, a changing period, and a dipper period very close to the period of a stellar spot are also identified. The temperatures derived at corotation are also compatible with dust survival, and the mass accretion rates are typical for CTTSs.

Many of the dippers for which it was possible to derive a stellar inclination are seen at a rather moderate inclination angle, and the outer disk inclination is, when available, slightly lower than the stellar inclination. Whether the stellar inclination is systematically higher than the outer disk inclination cannot be constrained accurately in this small sample, but it would be consistent with recently observed stars with a tilted inner disk. 

Magnetospheric accretion is able to explain most but not all Taurus dippers. However, dusty disk winds and Rossby wave instabilities do not seem convincing as explanations for the dippers seen at low inclination in this sample. 

A fraction of the quasiperiodic dippers exhibits double or complex dips, whose structure varies with time and whose minima are shifted from one cycle to the next. These examples are worth exploring in a future more precise dynamical analysis.

Finally, a correlation is found between the dip width in units of phase (i.e., the equivalent of the angular extent of the occulting structure) and the rotational period of the dippers. We speculate that this might indicate that the accretion columns of more slowly rotating stars have a larger base, possibly reflecting a different magnetic topology. This dependence needs to be investigated for dippers in other clusters in future work.

\begin{acknowledgements}
We wish to thank the anonymous referee for constructive comments which have improved the clarity of this paper, and Didier Fraix-Burnet and G\'erard Gr\'egoire for the helpful discussions. This project has received funding from the European Research Council (ERC) under the European Union's Horizon 2020 research and innovation programme (grant agreement No 742095; {\it SPIDI}: Star-Planets-Inner Disk-Interactions). 

Some of the data presented in this paper were obtained from the
Mikulski Archive for Space Telescopes (MAST). Support for MAST for
non-HST data is provided by the NASA Office of Space Science via grant
NNX09AF08G and by other grants and contracts.

This research has made use of data products from the Two Micron All-Sky Survey (2MASS), which
is a joint project of the University of Massachusetts and the Infrared
Processing and Analysis Center, funded by the National Aeronautics and
Space Administration and the National Science Foundation. The 2MASS
data are served by the NASA/IPAC Infrared Science Archive, which is
operated by the Jet Propulsion Laboratory, California Institute of
Technology, under contract with the National Aeronautics and Space
Administration. 

This work has made use of data from the European Space Agency (ESA) mission
{\it Gaia} (\url{https://www.cosmos.esa.int/gaia}), processed by the {\it Gaia}
Data Processing and Analysis Consortium (DPAC,
\url{https://www.cosmos.esa.int/web/gaia/dpac/consortium}). Funding for the DPAC
has been provided by national institutions, in particular the institutions
participating in the {\it Gaia} Multilateral Agreement.

\end{acknowledgements}

\bibliographystyle{aa}
\bibliography{Roggero_paper}

\begin{appendix}

\section{Wavelets}
\label{sec:app}
The Fourier transform (FT) is an extension of the Fourier series, which expresses periodic functions as sums of sinusoids. It decomposes the signal in the frequency domain, and its power spectrum represents each frequency that is present in the original signal. Fourier analysis requires (nearly) evenly sampled data as a condition.  Fourier analysis is not able to provide information about the time localization of a given frequency because it has no time resolution in the frequency domain. This is because the sinusoids employed to transform the signal are infinite in time, thus the time information is lost, and only the frequencies are retrieved.

The windowed Fourier transform can help in solving this issue by considering a  window\ (i.e., a slice) of the signal \textit{f} and transforming only that section, then translating the window along the signal. Given the window \textit{g(t)}, the windowed signal is defined as
\begin{equation}
 f_t(s) \equiv g(s-t)f(s) 
.\end{equation}
Its Fourier transform becomes 
\begin{equation}
 \mathcal{F}f(\omega, t) = \hat{f}_t(\omega) = \int_{-\infty}^{\infty} ds f_t(s) e^{-i \omega s} 
= \int_{-\infty}^{\infty} ds f(s) g(s-t) e^{- i \omega s} 
.\end{equation}
In physics, $g(t)$ is commonly chosen to be a Gaussian. The window is multiplied with a chop of the signal, transformed separately, and then shifted along the signal. If $g$  is centered in time and frequency, $\hat{f}_t(\omega)$  corresponds to the information of $f$ around time $t$ and frequency $\omega$. The WFT underlies the uncertainty principle: the lower bound for the product of time and frequency resolution is

\begin{equation}
\Delta f \cdot \Delta \tau \geq \frac{1}{4\pi}
.\end{equation}

For this method, the quality of the resulting analysis strongly depends on the chosen size $\tau$ of the window function. This determines the time and the frequency resolution a priori and might be inconvenient if both a good time and good frequency resolution are required.

\label{sec:app:morlet}
The windowed Fourier transform is \textit{\textup{constrained by}} the fixed size of the window, while the Fourier transform delivers no time resolution at all. This issue can be solved by wavelets, by adapting the time-width of the window to the frequency that is to be investigated,

\begin{equation}
\text{CWT}_f (s, \tau) = \frac{1}{\sqrt{s}} \int_{-\infty}^{\infty} f(t) \psi^* \left( \frac{t-\tau}{s}\right) dt
,\end{equation}
where CWT is the continuous wavelet transform.
The \textit{\textup{mother wavelet}} $\psi(t)$ is in most cases a sinusoidal wave confined in time, stretched and compressed according to the \textit{\textup{scale}} $s$ and shifted with a \textit{\textup{time step}} $\tau$ along the time series. The normalization factor $\frac{1}{\sqrt{s}}$ ensures that the transformed signal has the same energy at each scale. The 2-dimensional wavelet power spectrum (WPS) is produced by a convolution between the wavelet and the signal, which is recomputed for each scale along the time series.

The varying size of the wavelet allows us to identify both high- and low-frequency features in the signal. Both the discrete (DWT) and continuous wavelet transform (CWT) exist; here only the CWT is presented.
As for the WFT, the wavelet transform is subject to the uncertainty principle. Based on its shape, a certain wavelet has either a higher time or frequency resolution. Being confined in the time domain results in a broader peak in the frequency domain, which would be a $\delta$-function in the case of the sinusoid of the FT. The complex Morlet wavelet is a good compromise in this respect; moreover, its form is similar to the sought-for signal. 

The complex Morlet wavelet used in this work is defined as a superposition of a complex wave and a Gaussian, such as
\begin{equation}
\psi(x)=\frac{1}{\sqrt{\pi T_b}} e^{-\frac{x^2}{T_b}}e^{i2\pi F_c x}
\label{eq:morlet_general}
,\end{equation}
where the parameters $F_c$ and $T_b$ are the central frequency and the bandwidth parameter, respectively, and  $x$ is a unitless time parameter. This formulation differs from that in \cite{TorrenceCompo}, as they assumed a Gaussian with unit variance, which can be varied here, and a frequency $\omega_0 = 6$. 
The advantage of freely setting the  bandwidth and the central frequency is that we can customize the time and frequency resolution of the resulting wavelet plot, in order to highlight the position of a periodicity or to focus on the frequencies. A broader Gaussian in the time domain, thus a lower time resolution,  is represented by a narrower peak in the frequency domain, translating into higher frequency resolution.

Increasing the bandwidth $T_b$, that is, the width of the Gaussian, in the time domain decreases the bandwidth in the Fourier domain, that is, delivers a higher frequency resolution by losing time resolution. The central frequency should be close to the frequency of the signal of interest. 
In order to create an atlas of \textit{K2} light curves, the parameters were set to $F_c = 1.0$ and $T_b = 1.5$ as default. For individual cases of interest, these parameters may be changed.
In the WPS, a certain scale $s$ corresponds to a frequency $F_s$ as
\begin{equation}
    F_s = \frac{F_c}{s\cdot dt}
    \label{eq:freq_scales}
.\end{equation}

The set of scales of the wavelet transform is equivalent to the custom frequency resolution of a periodogram. They can be set linearly to extend from $s_0$ to $s_{max}$, where $s_{max}$ corresponds to the longest period of interest according to Eq.~\ref{eq:freq_scales}. For a faster and more efficient computation of the WPS, it is better to choose the set of scales according to a power law,
\begin{equation}
    s = s_0 2^{j\delta j}, \quad j = 0,1,...,J, \\
    \delta j = \frac{\log_2{s_{max}}}{J}.
\end{equation}
Here $J$ is the total number of scales and $s_0$ should be chosen close to $2dt$.

The cone of influence (COI) defines the range within which the WPS is affected by edge effects, which grow with the scale, as the wavelet is more and more stretched in time. The power spectrum contained in this region should not be regarded as significant. Considering that for the Morlet wavelet the Fourier period $\lambda$ is almost equivalent to the scale, as $\lambda = 1.03 s $ \citep{TorrenceCompo}, the COI is plotted as $\sqrt{2} p$, where p is the period corresponding to a frequency. A  central frequency different than 1.0 affects the COI as $\sqrt{2} p F_c$.

\section{Light curves of the dipper sample}
\label{sec:append:LCs_dippers}
\begin{figure*}
\centering
\includegraphics[width=0.45\linewidth]{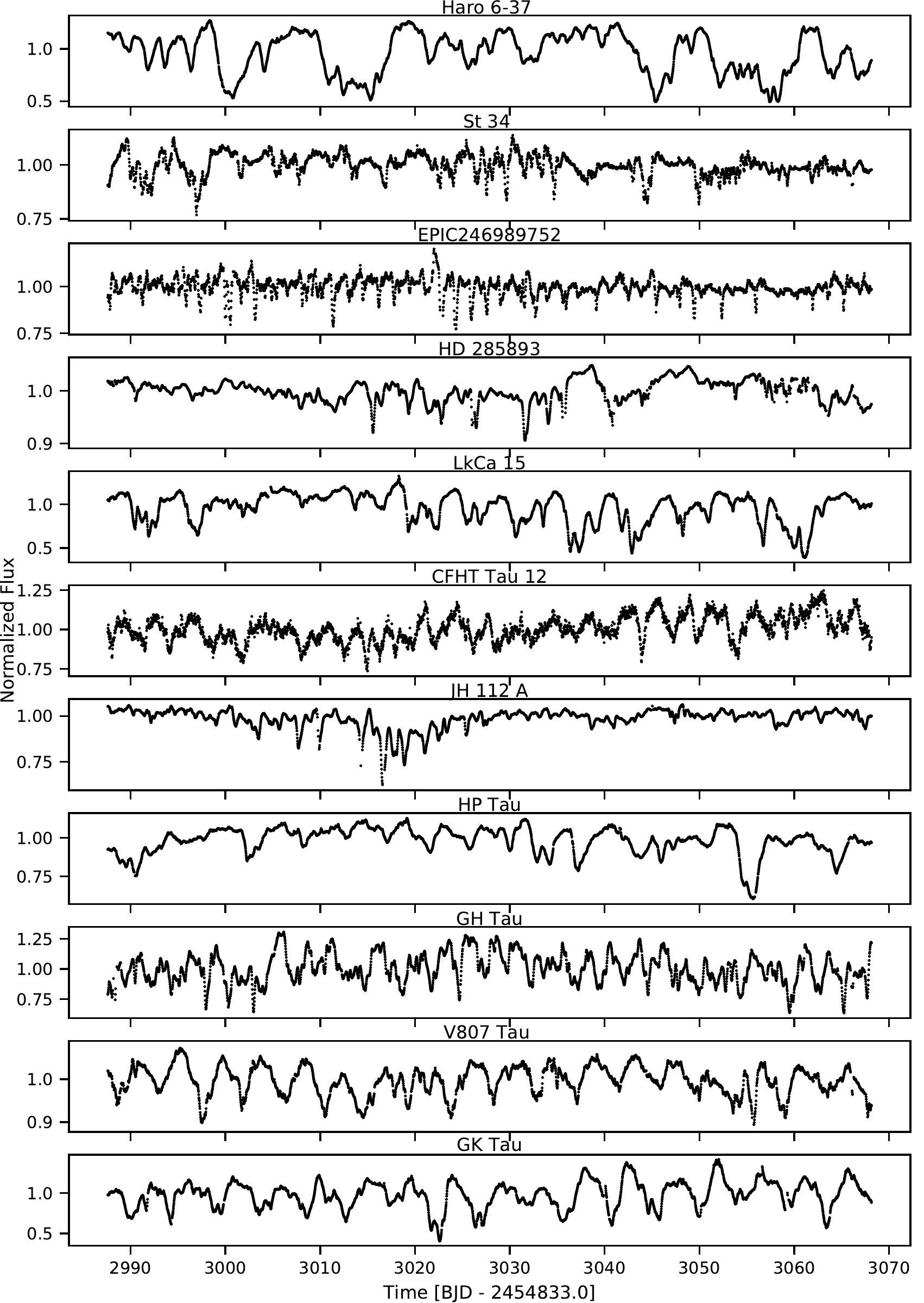}
\includegraphics[width=0.45\linewidth]{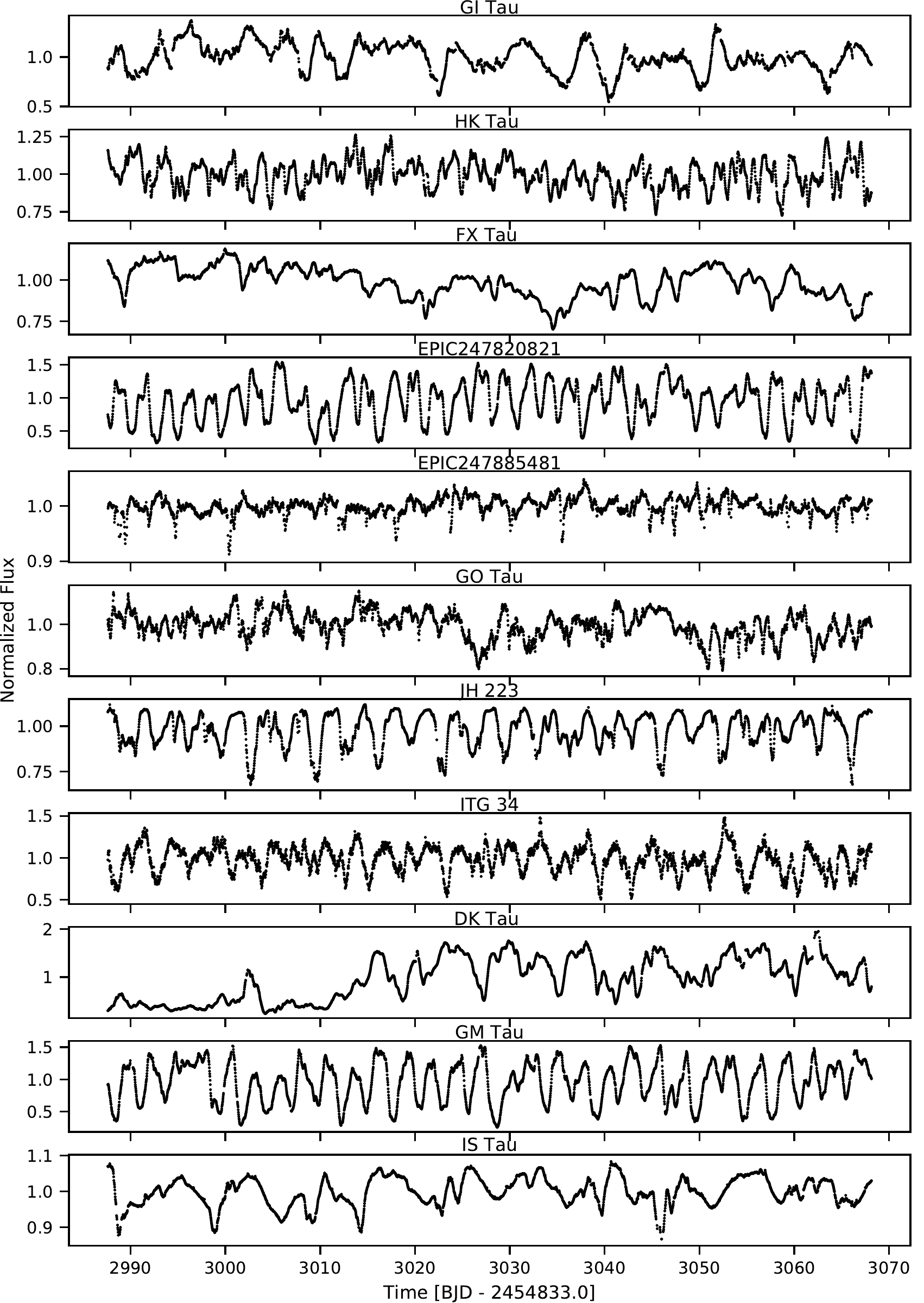}
\caption{Light curves for the dipper sample.}
\label{fig:app:LCs_dippers}
\end{figure*}
\begin{figure}
\centering
\includegraphics[width=\linewidth]{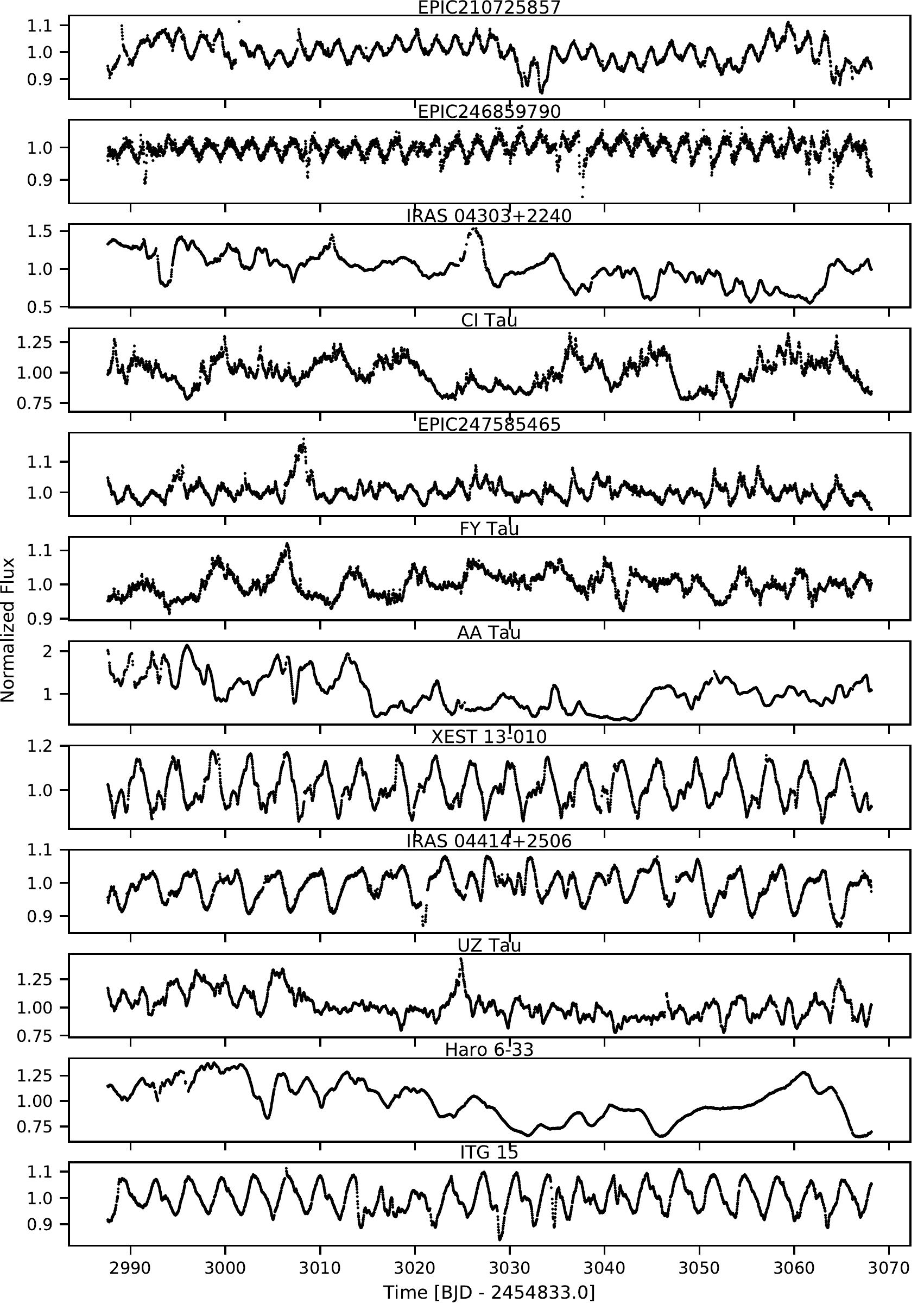}
\caption{Light curves of the additional 12 dippers with another dominant variability in addition to the dips.}
\label{fig:app:LCs_otherdippers}
\end{figure}
\begin{figure}
\centering
\includegraphics[width=\linewidth]{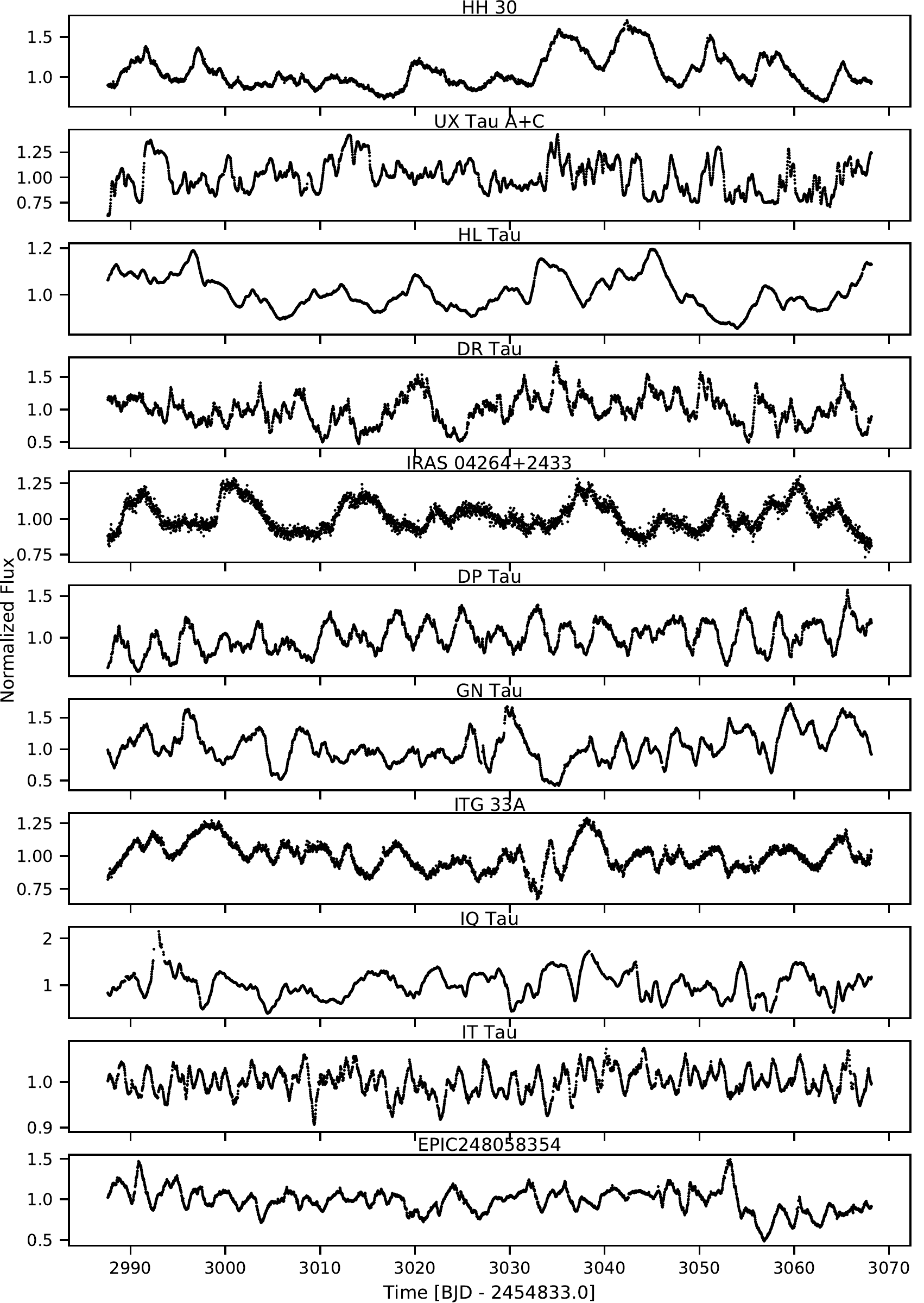}
\caption{Light curves of the 11 low-quality dipper candidates.}
\label{fig:app:LCs_candidatedippers}
\end{figure}
\section{Extinction}
\begin{figure}
    \centering
    \includegraphics[width=0.9\linewidth]{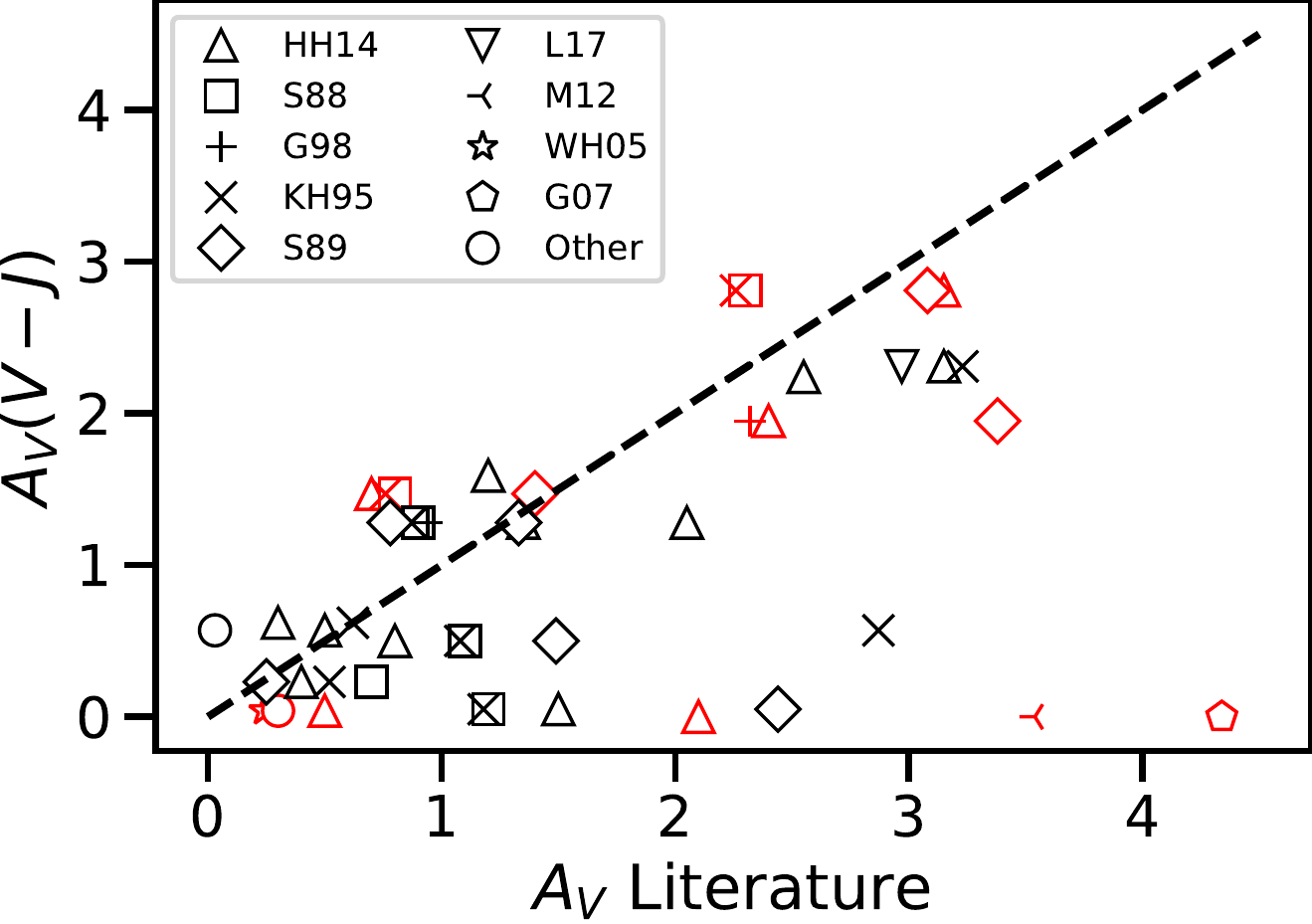}
    \caption{Comparison of the $A_V$ derived here from the $(V-J)$ colors compared to the literature values derived also from the optical. References appear in Table~\ref{tab:Av}. Red points show stars with high veiling. For these, the spectroscopic measurement by \cite{Herczeg14} is preferred. The dashed line represents $A_V(V-J) = A_V(\mathrm{lit})$. The stars with the largest scatter are strongly veiled, and the extinction measured with photometry is less reliable. For GO Tau (four black symbols at the bottom, close to $A_V(V-J) = 0$), the $A_V(V-J)$ value is discarded because it is inconsistent with the literature. Another outlier is the $A_V$ of V807 Tau as measured by KH95; this value is discarded for the same reason.}
    \label{fig:app:Av}
\end{figure}

Table~\ref{tab:Av} presents the extinction values computed from $(V - J)$ compared with different $A_V$ from the literature. Using $(J-K)$ leads to an increase in $A_V$ up to a factor 5 because of the infrared excess. The uncertainty on $A_V$ is set as the root-mean-square (rms) of the $A_V$ derived in this study compared with the literature. A minimum reasonable uncertainty of 0.3\,mag is applied. Empirically, the uncertainty does not grow with extinction.
Fig.~\ref{fig:app:Av} shows that veiling affects the derivation of the extinction from photometry. For stars not affected by veiling, the $A_V$ computed from $(V -J)$ is consistent with optical measurements in the literature.

\begin{table*}[htbp]
\centering
\caption{$A_V$ as measured with \textit{V-J} compared to the literature.}
\tiny
\begin{tabular}{lllllllllllllll}
\hline \hline
EPIC     &      Name     &      $A_V$    &      $r_{7510}^a$     &      $A_V^a$          &      $A_V^b$          &      $A_V^c$          &      $A_V^d$           &      $A_V^e$          &      $A_V^f$          &      $A_V^g$          &       $A_V^h$          &       $A_V^i$         &      $A_V^j$          &       $\sigma_{A_V}$   \\
& & \textit{(V-J)}  & & & & & & & & & & & & \\
\hline
246929818                &      Haro 6-37                &      \ldots           &       0.33                     &      2.05                     &      1.8                          &      \ldots                   &      2.12                      &      2.44                     &      \ldots                   &       \ldots                   &      \ldots                   &      \ldots                  &      2.1                      &      0.3              \\
246942563                &      St 34                    &      0.04             &       0.14                     &      0.5                      &      \ldots                  &      \ldots                   &      \ldots                   &       \ldots                   &      \ldots                   &      \ldots                  &      0.24                     &      0.3$^1$                  &       0.5                      &      0.3              \\
246989752                &      \ldots                   &      \ldots           &       \ldots                   &      \ldots                   &      \ldots                  &      \ldots                   &      \ldots                   &       \ldots                   &      0.0                      &      \ldots                  &      \ldots                   &      \ldots                   &       0.0                      &      0.3              \\
247103541                &      HD 285893                &      0.32             &       \ldots                   &      \ldots                   &      \ldots                  &      \ldots                   &      \ldots                   &       \ldots                   &      \ldots                   &      \ldots                  &      \ldots                   &      \ldots                   &       0.3                      &      0.3              \\
247520207                &      LkCa 15                  &      0.62             &       0.04                     &      0.3                      &      \ldots                  &      \ldots                   &      0.62                     &       \ldots                   &      \ldots                   &      \ldots                  &      \ldots                   &      \ldots                   &       0.5                      &      0.3              \\
247575958                &      CFHT Tau 12              &      \ldots           &       \ldots                   &      \ldots                   &      \ldots                  &      \ldots                   &      \ldots                   &       \ldots                   &      \ldots                   &      2.64                  &      \ldots                   &      3.44$^2$                 &       3.0                      &      0.6              \\
247589612                &      JH 112 A                 &      2.31             &       0.0                      &      3.15                     &      \ldots                  &      \ldots                   &      3.23                     &       \ldots                   &      2.97                     &      \ldots                  &      \ldots                   &      \ldots                   &       2.9                      &      0.4              \\
247592463                &      HP Tau                   &      2.81             &       0.16                     &      3.15                     &      2.3                          &      \ldots                   &      2.26                      &      3.08                     &      \ldots                   &       \ldots                   &      \ldots                   &      \ldots                  &      3.2                      &      0.3              \\
247763883                &      GH Tau                   &      0.23             &       0.0                      &      0.4                      &      0.7                          &      \ldots                   &      0.52                      &      0.25                     &      \ldots                   &       \ldots                   &      \ldots                   &      \ldots                  &      0.4                      &      0.2              \\
247764745                &      V807 Tau                 &      0.57             &       0.05                     &      0.5                      &      \ldots                  &      \ldots                   &      2.87\tablefootmark{*}                  &      \ldots                   &      \ldots                   &       \ldots                   &      \ldots                   &      0.03$^3$                  &      0.4                      &      0.3              \\
247791801                &      GK Tau                   &      1.28             &       0.08                     &      1.35                     &      0.9                          &      0.94                     &      0.87                      &      0.78                     &      \ldots                   &       \ldots                   &      \ldots                   &      \ldots                  &      1.0                      &      0.2              \\
247792225                &      GI Tau                   &      1.28             &       0.04                     &      2.05                     &      0.9                          &      1.34                     &      0.87                      &      1.33                     &      \ldots                   &       \ldots                   &      \ldots                   &      \ldots                  &      1.3                      &      0.4              \\
247799571                &      HK Tau                   &      1.95             &       0.1                      &      2.4                      &      \ldots                  &      2.32                     &      \ldots                   &       3.38                     &      \ldots                   &      \ldots                  &      \ldots                   &      \ldots                   &       2.4                      &      0.3              \\
247805410                &      FX Tau                   &      0.50             &       0.06                     &      0.8                      &      1.1                          &      \ldots                   &      1.08                      &      1.49                     &      \ldots                   &       \ldots                   &      \ldots                   &      \ldots                  &      1.0                      &      0.4              \\
247820821                &      \ldots                   &      -1.35            &       \ldots                   &      \ldots                   &      \ldots                  &      \ldots                   &      \ldots                   &       \ldots                   &      \ldots                   &      \ldots                  &      \ldots                   &      \ldots                   &       0.0                      &      0.0              \\
247885481                &      \ldots                   &      -0.43            &       \ldots                   &      \ldots                   &      \ldots                  &      \ldots                   &      \ldots                   &       \ldots                   &      0.0                      &      \ldots                  &      \ldots                   &      \ldots                   &       0.0                      &      0.0              \\
247935061                &      GO Tau                   &      0.05\tablefootmark{*}                  &      0.09                     &      1.5                      &       1.2                      &      \ldots                   &      1.18                  &      2.44                     &      \ldots                   &       \ldots                   &      \ldots                   &      \ldots                  &      1.6                      &      0.9              \\
248006676                &      JH 223                   &      1.59             &       0.0                      &      1.2                      &      \ldots                  &      \ldots                   &      \ldots                   &       \ldots                   &      \ldots                   &      \ldots                  &      \ldots                   &      \ldots                   &       1.4                      &      0.3              \\
248015397                &      ITG 34                   &      \ldots           &       \ldots                   &      \ldots                   &      \ldots                  &      \ldots                   &      \ldots                   &       \ldots                   &      \ldots                   &      2.6                          &      \ldots                   &      1.77$^2$                  &      2.2                      &      0.6              \\
248029373                &      DK Tau                   &      1.47             &       0.27                     &      0.7                      &      0.8                          &      1.42                     &      0.76                      &      1.4                      &      \ldots                   &       \ldots                   &      \ldots                   &      \ldots                  &      0.7                      &      0.3              \\
248046139                &      GM Tau                   &      -0.28            &       0.26                     &      2.1                      &      \ldots                  &      \ldots                   &      \ldots                   &       \ldots                   &      \ldots                   &      3.54                  &      \ldots                   &      4.34$^2$                 &       2.1                      &      0.3              \\
248047443                &      IS Tau                   &      2.24             &       0.02                     &      2.55                     &      \ldots                  &      \ldots                   &      \ldots                   &       \ldots                   &      \ldots                   &      \ldots                  &      \ldots                   &      \ldots                   &       2.4                      &      0.3              \\
\hline
\end{tabular}
\tablefoot{Only the extinction measured by \cite{Luhman17} with optical or CTTS methods is reported here. The veiling is measured at 7510\,\AA\,   by \cite{Herczeg14}.\\
\tablefoottext{*}{Value discarded because it is inconsistent with the literature.}
}
\tablebib{
(a) \cite{Herczeg14}; (b) \cite{Strom88}; (c) \cite{Gullbring98}; (d) \cite{Kenyon95}; (e) \cite{Strom89}; (f) \cite{Luhman17}; (g) \cite{Mayne12}; (h) \cite{White05}; (i) other refs; (j) this paper; (1) \cite{Hartmann05}; (2) \cite{Guieu07}; (3) \cite{Schaefer12}.
}
\label{tab:Av}
\end{table*}

\section{Notes on individual objects}
\label{sec:app:individuals}
In the following section, remarks about individual objects are presented. Three stars (St 34, HD 285893, and 2MASSJ05023985+2459337) are possible Taurus members \citep{Rebull20}.

\paragraph{HD 285893} According to the HR diagram, the star lies almost on the ZAMS, which means that it is much older than the Taurus population. There are several possibilities to explain this. The star might not be a Taurus member (it is considered a possible member), the spectral type might not be correctly estimated, and/or the star might be seen edge-on, thus its luminosity would be strongly underestimated. The star shows a clear dipper behavior in the light curve, which is unexpected for a spectral type F8 because the higher temperature of the star depletes the inner disk of dust at distances much larger than corotation. The object is poorly studied in the literature and does not appear in more recent spectroscopic surveys. Its behavior appears to be related to the so-called \lq old\rq\ dippers, which can host a debris disk and whose occultations might be caused by disrupting planetesimals \citep{Gaidos19,Tajiri20}.
The dips are narrow, as is commonly the case for aperiodic dippers. The light curve continuum is unstable and aperiodic.

\paragraph{CFHT Tau 12}
No binarity information has been reported for this star. The star is quasiperiodic  \cite[period agrees with][]{Scholz18}, but there is no information about $v\sin{i}$ from which we could derive an inclination. The light curve never reaches a continuum, and no periodicity for this variation can be derived within the time window of the observations. The shape of the dips clearly distinguishes the dipper phenomenon from spots. The folded light curve is noisy.
\paragraph{2MASSJ04384725+1737260} No companion is known in the literature. The star never clearly reaches the continuum; the periodicity is pretty clear, but some dips seem to be aperiodic (see folded light curve in Fig.~\ref{fig:app:phased_lc}). There might also be one or more spots.
\paragraph{2MASSJ04295950+2433078} No binarity is known \citep{Davies14}. It has a high variability amplitude (0.66\,mag). The sinusoidal pattern of the light curve looks rather spot-like, but the star is classified as a dipper based on the irregular shape of the dips. The WPS shows a significant periodicity at $\sim 7.5$\,d, which is much weaker in the periodogram. This might be due to the shape of the light curve, which is well-matched to the Morlet wavelet and is  therefore highly correlated. Folded at 7\,d, a periodic pattern of several dips appears. A harmonic is excluded.
\paragraph{2MASSJ05023985+2459337} No companion is known in the literature. The light curve exhibits both a spot and a dipper pattern. A dedicated discussion is presented in Sec.~\ref{subsec:disc:diff_period}. 
\paragraph{ITG 34} is known to be single \citep{Davies14, Akeson19}. The amplitude of the dips is large (0.39\,mag). According to the light curve, the star either almost never reaches the continuum level or a few bursts occur during the \textit{K2} campaign. The periodogram shows two double-peaked periodicities. The main periodicity is also more stable over time according to the WPS. The second periodicity cannot be a harmonic of the first. The folded light curve is noisy, and no new features emerge when it is folded at a different period. The apparent beat cannot be explained with binarity.
\paragraph{JH 223} is the best example of a quasiperiodic dipper in the sample. The star is a binary with a separation of $2\arcsec$. Both components appear to have a disk \citep{Itoh15}. The star is reported in \cite{Kraus12} as having a disk, but without evidence for accretion. 
\paragraph{IS Tau} is a binary with $0.2\arcsec$ separation \citep{Schafer14}. It is not known which (or if both) component hosts a disk. \cite{Watson09} reported a very high mass fraction (80\%-100\%) of crystalline silicates in the inner disk. The light curve is characterized by a combination of broad and narrow eclipses. The strongest periodicity is given by the broad dips. No \textit{Gaia} parallax is available. The star is also located on the L1495 filament of Taurus, which contains members of both the near and the far population \citep{Fleming19}. In the HR diagram in Fig. \ref{fig:HR_diag}, the possible luminosity of IS Tau is plotted according to mean distances of 130.6 and 160.2\,pc \citep{Fleming19}. The age is more consistent with that of the remaining sample if IS Tau is a member of the far population.
The star shows a double dip; the primary has a larger, quite constant amplitude. For this star, the phase shift of the dips varies more than for other double-peaked dippers.

\paragraph{St 34} is a multiple system (Aab+B) with separation $1.18\arcsec$, surrounded by a circumbinary, transitional disk \citep{Rigliaco15}. The star is reported as SB2 by \cite{Akeson19}. The inner disk appears to be depleted, although some dust is still present \citep{Hartmann05}. Its membership in the Taurus association is controversial \citep{Hartmann05, White05,Dahm11}. \cite{Dahm11} reported an age for St 34 of $\sim$10\,Myr. Its lithium depletion \citep{White05}  would lead to an age of 25\,Myr, thus the authors must assume a distance of only 90\,pc. In the HR diagram of our study (Fig. \ref{fig:HR_diag}), the star appears to have an age of 1-2\,Myr, given the limitations on the luminosity due to photometric variability. The \textit{Gaia} parallax reports a distance of 143\,pc, thus showing that St 34 cannot be a foreground star. Accordingly, the age-derivation based on the lithium depletion timescale apparently does not apply to this star.
The light curve is irregular, and a period forest is present in both the periodogram and WPS. The presented period is uncertain. Some contamination might arise from multiplicity.
\paragraph{V807 Tau} is a multiple system (A + Bab) and a separation of $0.3\arcsec$ for the wide components, $0.04\arcsec$ for the secondary \citep{Schaefer12}. The \textit{Gaia} distance of V807 Tau is 113\,pc, thus it is maybe underestimated. The system is multiple, which means that there might be an issue with the \textit{Gaia}  parallax. The star is located in the B18 region of Taurus \citep{Fleming19}, which has a distance of 110-150\,pc. This value has to be considered with caution, but it is not an outlier for this region.
\cite{Rodriguez17} derived a period of 0.809\,d, which is inconsistent with \textit{K2} data. \cite{Schaefer12} decomposed the brightness of the single components and obtained a primary star 0.5\,mag fainter than the \textit{2MASS J} measurement. This leads to a much smaller stellar radius and higher, more plausible stellar inclination.
The light curve presents an apparently sinusoidal long-term modulation; moreover, it is unclear if the star reaches its maximum brightness at all. The modulation does not appear to be related to binarity because the binary system has a period of 12\,yr. There is only one clear periodicity, and the shape of the dips does not suggest stellar spots.
\paragraph{Haro 6-37} is a triple system. The close binary AB \citep[0.3\arcsec][]{Duchene99} is separated from C by $2.6\arcsec$ \citep{White01}. The light curve presents unusually broad dips for a dipper with high amplitude (0.5\,mag). The periodicity is unclear in the WPS and periodogram. Our luminosity is highly different from that of \cite{Herczeg14} (here $0.8L_{\odot}$, there $0.07 L_{\odot}$). The derived inclination suggests that the period is either incorrect or that the occultations have another origin.
\paragraph{JH 112 A} is separated from its companion JH 112 B by $6.56\arcsec$. JH 112 A is itself a close binary Aa+Ab with a separation of $1.65\arcsec$ \citep{Kraus11}. The star is a transient, quasiperiodic dipper, the dips occurr with a period of 2.21\,d. The dips are accompanied by a general decrease in brightness; it is not clear if the phenomenon is physical or instrumental. The estimated luminosity is different from that of \cite{Herczeg14}. The temperature at corotation is close to 1600\,K, thus requiring higher gas densities for dust to survive \citep{Pollack94}. The star is flagged as a double-peaked dipper. In a few periods, two well-detached peaks are present. The primary peak is always present, and the secondary is strongly variable and sometimes clearly double-peaked.
\paragraph{LkCa15} is a single star \citep{Akeson19} that hosts a transition disk. \cite{Donati19} and \cite{Alencar18} studied the star in detail. \cite{Alencar18} confirmed it as a dipper, with a light curve that does not show large changes in its variability over the years \cite[see also][]{Grankin07}. The \textit{K2} light curve shows a variability amplitude of 0.4\,mag, in accordance with previous literature. The derived inclination of the inner disks suggests that the inner disk is misaligned with the outer disk \citep[$\sim 50\degr$, see][]{vanderMarel15,Thalmann14}.
The photometric period is very close to the period derived from radial velocity and veiling \citep{Alencar18}, thus supporting the scenario of an inclined warp located at the corotation radius. 
\cite{Donati19} reported a slightly lower $v\sin{i}$ than \cite{Alencar18}. The luminosity we derived ($0.96 \pm 0.18 L_{\odot}$) agrees with theirs ($0.8 \pm 0.15 L_{\odot}$) within the error bars. Nevertheless, the spectroscopically measured $T_{\mathrm{eff}}$ is 450\,K higher than the conversion by \cite{Pecaut13}. In this analysis, the directly derived temperature by \cite{Alencar18} is preferred.
The \textit{K2} light curve folded in phase shows a double-peaked dip that is strongly variable over time. The total width of the eclipsing time for this dipper is therefore very large. The light curve exhibits a long-term trend; it is unclear whether the occultations affect the continuum or if it has other origins. 
At least two clear peaks with variable depth and occurrence (i.e., position in phase) are always present in the light curve.
\paragraph{GO Tau} The light curve shows an active star in which bursts also occur. This makes it difficult to define a brightness continuum. The period forest in the periodogram results in no clear periodicity in the WPS. This aperiodic dipper includes both very narrow and wider dips. \cite{Guedel07} derived an upper limit for the rotation period $\le 3.96$\,d  from $v\sin{i}$ .
\paragraph{HK Tau} is a wide binary with a separation of $2.3\arcsec$ \citep{Akeson14}. Both components have disks misaligned with the orbital motion. The light curve does not show a stable continuum. The periodicity is very clear, but the folded light curve is noisy. Some other phenomenon apparently takes place at the stellar surface. \cite{Guedel07} reported a $v\sin{i} = 10\,\mathrm{km~s}^{-1}$, measured by \cite{Hartmann89}, which would lead to a very low inclination that is inconsistent with both a dipper and the outer disk. We consider this value less reliable than the measurement by \cite{Hartmann86} because 10\,km~s$^{-1}$ is the detection limit for $v\sin{i}$.
Instead of being a double-peaked dipper, the star is multi-peaked. The different dips are always present, with a strongly varying amplitude that appears to be uncorrelated between different dips. The Gaussian-shaped tip of the dips is constant, while the dip shapes vary strongly, suggesting that multiple stellar spots might contribute to the photometric variability.
\paragraph{HP Tau} The star does not have a companion \citep{Akeson19}. Its light curve is quasiperiodic, with strong variations in shape and brightness (up to 0.4\,mag). The dips are strongly variable and never appear to be single. The structure of the dips varies from double-peaked to a broad, single peak. \cite{Guedel07} and \cite{Rebull04} found in the literature a period of 5.90\,d, which disagrees with the \textit{K2} period of 4.33\,d. The light curve is strongly variable and the periodicity complex; this means that the period might change over time.  Its luminosity is probably overestimated ($2.3 L_{\odot}$) and the resulting stellar inclination is very low. It is possible that the stellar parameters cannot
be constrained correctly due to the high amplitude of the variability (3\,mag in the $V$ band for different surveys). The luminosity disagrees with that derived by \cite{Herczeg14}, and the star is also highly veiled.
\paragraph{GH Tau} is separated from its companion by $0.3\arcsec$ \citep{Akeson19}. The light curve is irregular, and it is unclear whether there are two types of eclipses, if the star is bursting, or if there is a long-term variation. The folded light curve is very noisy. Two main periodicities at 2.49\,d and 2.94\,d are present, although the latter does not show any significant structure. \cite{Guedel07} indicated a period $\le 3.57$\,d derived from $v\sin{i}$. No \textit{Gaia} parallax was available for this object, but it is located in the B18 group of Taurus \citep{Fleming19}, with an average distance of 127.4\,pc. The star is the fastest rotator in this sample ($v\sin{i} = 30\,\text{km~s}^{-1}$). 
\paragraph{GM Tau} is a single brown dwarf \citep{Akeson19} with a transitional disk. The star is a strong accretor and is highly veiled \citep{Herczeg08,Herczeg14}. The sinusoidal light curve does not show a brightness continuum; the star is considered a dipper because of the irregularity of the dips. 
\paragraph{FX Tau} is a binary with a separation $0.9\arcsec$. Each component hosts a disk \citep{Akeson19}. This aperiodic dipper has alternating wide and narrow dips. A long-term trend is present in the light curve.
\paragraph{GK Tau} is separated from GI Tau by $13.1\arcsec$ \citep{Akeson19}. It is debated in the literature whether these two stars are physically bound. The star is a clear dipper with high amplitude (0.5\,mag). The occultations probably do not allow us to see the brightness continuum. The periodicity is clear, and no other physical phenomena appear to affect the light curve. The period agrees with previous data \citep{Artemenko12,Percy10,Guedel07,Rebull04}. \cite{Percy10} found a long-term variability with a timescale of 2500\,d for this object. 
\paragraph{GI Tau} \cite{Guo18} reported an uncertain spectral type K5-M0 for this star. We adopted a spectral type of M0.4. The inclination of $60^{\circ}$ agrees with the inclination derived here. \cite{Guo18} linked the 7\,d period that we also found to a spot. In their optical monitoring from 2014 to 2016, the folded light curve is clearly sinusoidal, which is far from the observed behavior in the \textit{K2} light curve. \cite{Guo18} observed quasiperiodic dips on timescales that are a multiple of the rotation period. The folded \textit{K2} light curve is highly irregular and contaminated by GK Tau. The first peak in the periodogram is a periodicity related to GK Tau and not to GI Tau. The 7.1\,d period agrees with previous data \citep{Artemenko12,Percy10,Guedel07,Rebull04}. Our observed amplitude of 0.4\,mag is lower than the variability observed by \cite{Guo18}. The folded light curve suggests that small bursts occurred during the \textit{K2} campaign. The star has been classified as dipper, although aperiodic, by \citet{Rodriguez17}.
A comparison with the light curve of GK Tau (not shown here), split into each phase according to the period of GI Tau, shows that the dips are highly affected by the neighbor star. Although the intrinsic variability of GI Tau emerges well, it is difficult to describe the behavior of its multiple dips, if there are any. The complexity of the \textit{K2} light curve does not allow us to clearly distinguish between the two.
\paragraph{DK Tau} is a binary star with a separation of $2.4\arcsec$. Both stars have disks \citep{Akeson19,Akeson14}. The light curve exhibits strong variations over long timescales \citep[see][and references therein]{Grankin07,Rebull20}, and the period varies from one survey to the next. The WPS shows that the period of DK Tau increases from 7.69\,d to $\sim10$\,d during \textit{K2} C13, after a a fading state in lower brightness. This explains the two different, broad peaks in the periodogram. \cite{Percy10} and \cite{Artemenko12} reported a period of 8.18\,d, while \cite{Rebull20} found 7.84\,d. The difference can be easily explained for the first with the complexity of the light curve, and for the second, by the use of a different period-finding algorithm in presence of a broad peak. \cite{Xiao12} reported 4.14\,d, which is not consistent with the \textit{K2} data. \cite{Percy10} also derived a long-term timescale of 2000\,d for the variability of DK Tau.
When the fading state is excluded, the amplitude of the occultations is 0.6\,mag. 

\section{Wavelet power spectra}
\label{sec:app:wps}
\begin{figure*}[htbp]
    \centering
    \includegraphics[trim = 0 30 0 0, clip, width=0.95\linewidth]{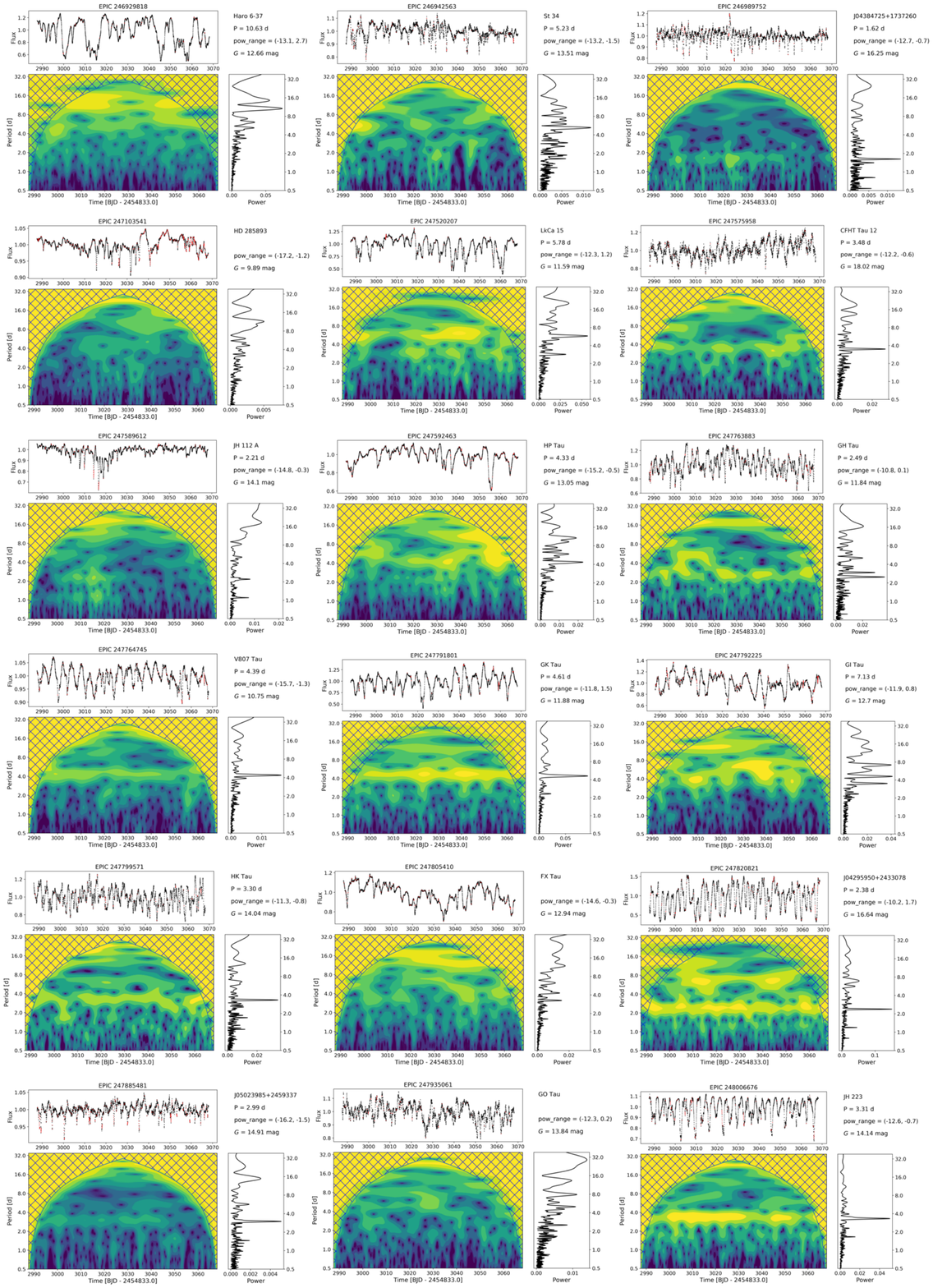}
    \caption{Wavelet power spectra of the dipper sample.}
\end{figure*}
\begin{figure*}[htbp]
    \centering
    \ContinuedFloat
    \includegraphics[trim = 0 560 0 0, clip, width=0.95\linewidth]{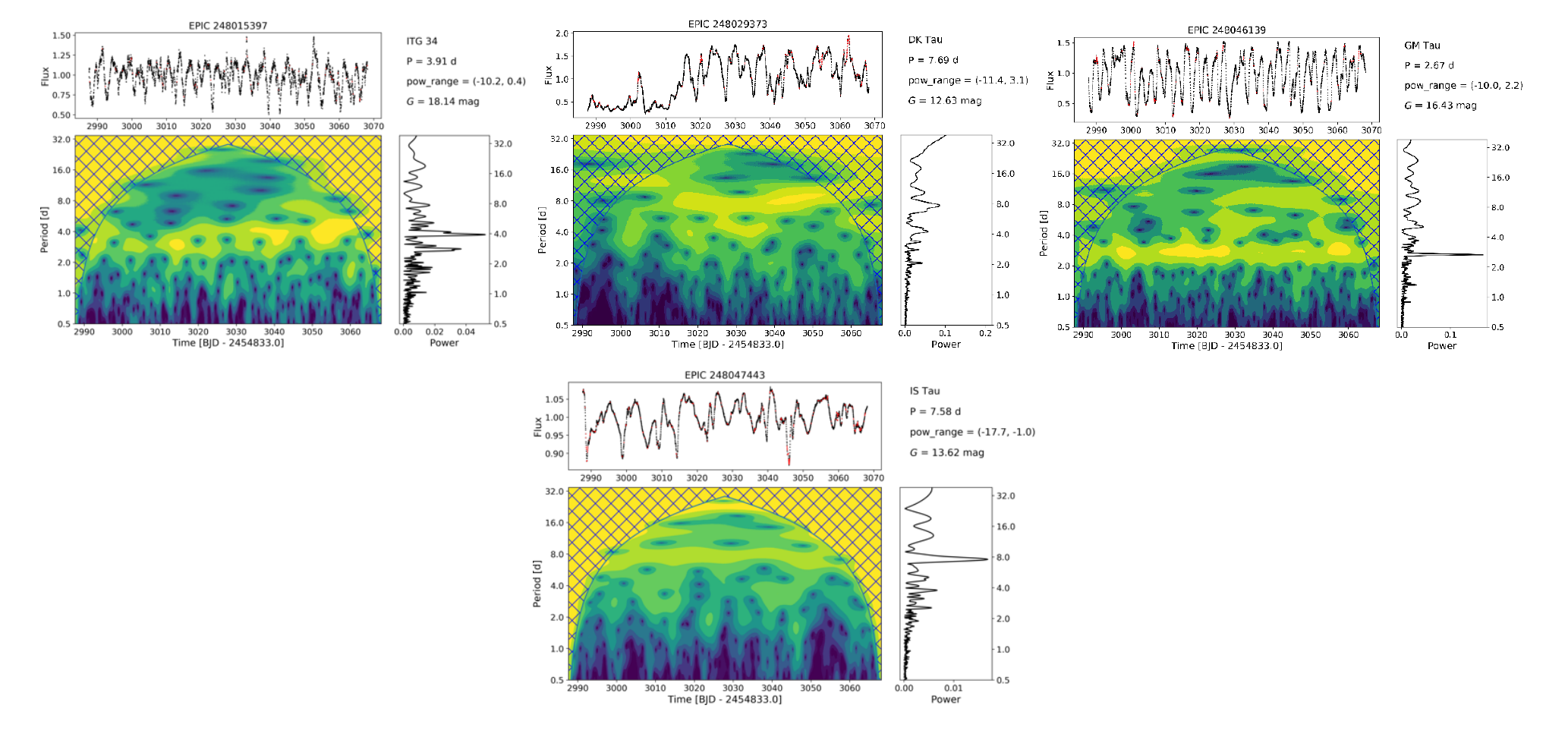}
    \caption{Continued.}
    \label{fig:app:WPS}
\end{figure*}

\section{Folded light curves}
\label{sec:app:phasecurves}
\begin{figure*}[htbp]
    \centering
    \includegraphics[trim = 25 25 50 0, clip, width=0.8\linewidth]{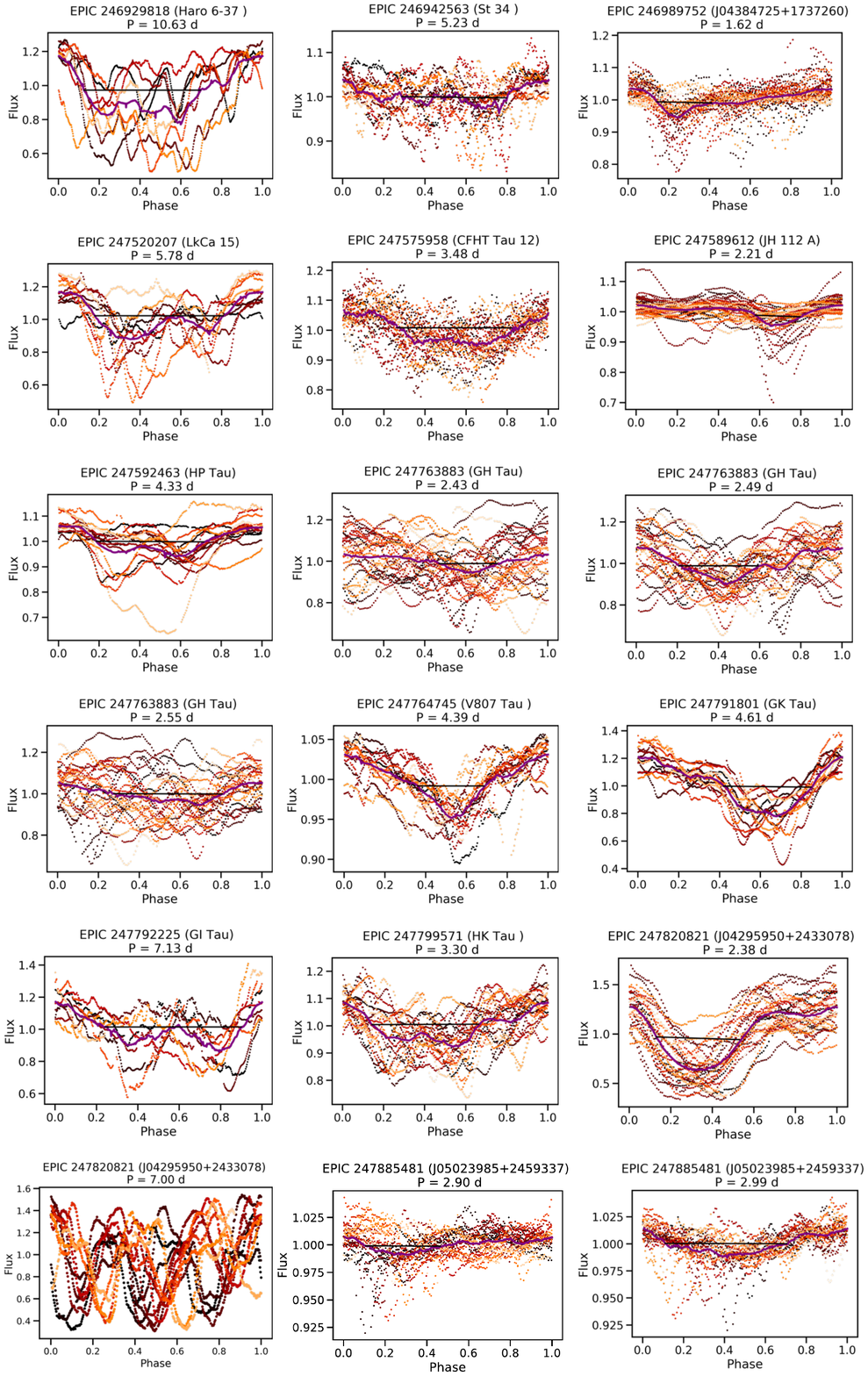}
    \caption{Folded light curves of the dipper sample. Every phase has a different color. The purple line shows the binned light curve. The black line represents the automatically determined dip width. For some stars, more than one possible period is shown. }
\end{figure*}
\begin{figure*}[htbp]
    \centering
    \ContinuedFloat
    \includegraphics[trim = 0 380 0 0, clip,width=0.9\linewidth]{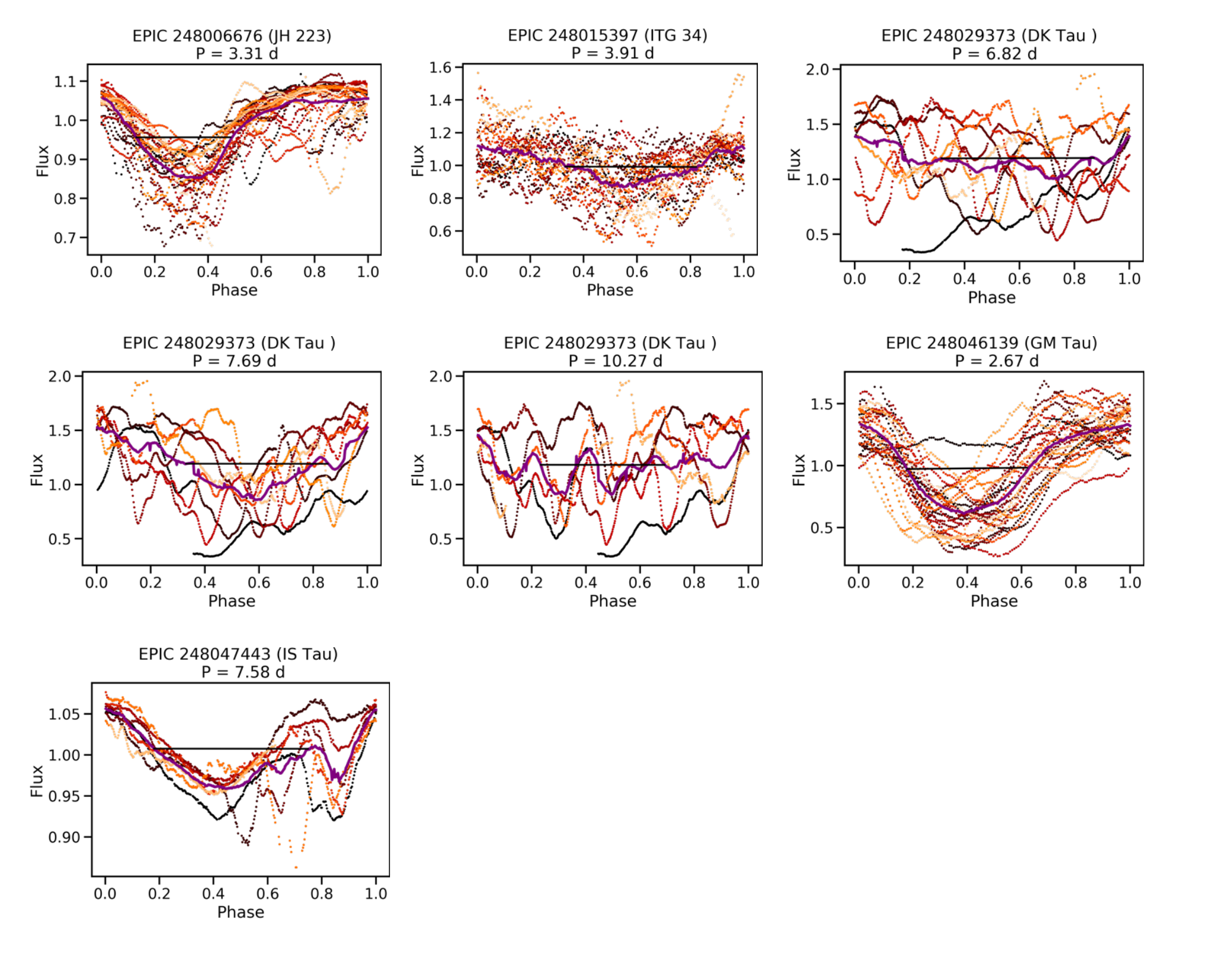}
    \caption{Continued.}
    \label{fig:app:phased_lc}
\end{figure*}

\end{appendix}

\end{document}